\RequirePackage{fix-cm}
\documentclass[smallextended]{svjour3}       
\usepackage{graphicx}

\usepackage{type1cm}
\usepackage{amsmath}
\usepackage{amsfonts}
\usepackage{amssymb,mathrsfs}
\usepackage{upgreek,yfonts}

\usepackage{bm}
\newcommand{\bs}[1]{\boldsymbol{#1}}
\journalname{Nuovo Cimento}
\begin{document}

\title{Hydrodynamic approach to two-dimensional electron systems}


\author{Boris N. Narozhny}


\institute{Boris N. Narozhny \at
              Institut for Theoretical Condensed Matter Physics, Karlsruhe Institute of 
              Technology, 76128 Karlsruhe, Germany 
              \and
              National Research Nuclear University MEPhI (Moscow Engineering Physics Institute),
              115409 Moscow, Russia\\
              \email{boris.narozhny@kit.edu}           
}


\maketitle

\begin{abstract}
  The last few years have seen an explosion of interest in
  hydrodynamic effects in interacting electron systems in ultra-pure
  materials. One such material, graphene, is not only an excellent
  platform for the experimental realization of the hydrodynamic flow
  of electrons, but also allows for a controlled derivation of the
  hydrodynamic equations on the basis of kinetic theory. The resulting
  hydrodynamic theory of electronic transport in graphene yields
  quantitative predictions for experimentally relevant quantities,
  e.g. viscosity, electrical conductivity, etc. Here I review recent
  theoretical advances in the field, compare the hydrodynamic theory
  of charge carriers in graphene with relativistic hydrodynamics and
  recent experiments, and discuss applications of hydrodynamic
  approach to novel materials beyond graphene.

\end{abstract}

\section{Hydrodynamics and condensed matter}\label{sec1}

Collective excitations in solid state physics -- phonons, magnons,
plasmons, etc. -- are often considered in the long-wavelength (small
wavevector) limit with the corresponding observables describing
long-distance properties of matter. One way to develop a macroscopic
theory reflecting such physics \cite{chai} is to combine continuity
equations (manifesting conservation laws) with thermodynamic arguments
to identify how the entropy of the system responds to local density
fluctuations of the conserved quantities. Requiring the total entropy
production rate to be non-negative, one may establish the
``constitutive relations'' between the macroscopic currents and the
external bias. Closing the equations with the help of the
thermodynamic relations one can complete the description of the
long-wavelength dynamics of the system. The resulting theories are
macroscopic since their variables are densities of physical quantities
and the corresponding currents. They are also phenomenological since
they provide no means of calculating the coefficients in the
constitutive relations (i.e., the ``generalized susceptibilities'').
Such approach is justified at distances that are much larger than any
length scales corresponding to the underlying ``microscopic''
scattering processes, the condition that is very often satisfied in
experiments.

The most common equation describing the long-wavelength dynamics in
solids is the diffusion equation \cite{chai}. In the simplest example,
spin diffusion \cite{blo,kad} arises in a system of spin-$1/2$
particles with a velocity- and spin-independent interaction leaving
the total magnetization conserved. This behavior has been
observed experimentally (see, e.g., Ref.~\cite{tuc}) and is generally
expected to be applicable to a wide variety of spin systems (with the
possible exception of one-dimensional integrable models, see
Refs.~\cite{sdme,xxz,sdnat,sdaff}).

Low-temperature charge transport is also often considered to be
diffusive \cite{aar}. In the simplest case, charge carriers are
assumed to be independent and noninteracting, so that their total
number is a conserved quantity, while the dominant relaxation process
is the electron-impurity scattering described by the transport mean
free time, $\tau$. The latter defines both the diffusion constant and
electrical conductivity \cite{ziman} and is still one of the most
important quantities characterizing conductive properties of
experimental samples. The diffusive behavior is commonly expected to
take place in real metals and semiconductors as long as the sample
size is large compared to the mean free path (typically,
$\ell=v_F\tau$ with $v_F$ being the Fermi velocity) \cite{bee} and at
low temperatures, $T\tau\ll1$ \cite{zna} (the units with $\hbar=k_B=1$
are used throughout this paper).

A common feature of the above theories is the decaying (diffusive)
nature of collective modes (defined as the normal modes of the set of
linearized macroscopic equations). In contrast, the collective modes
in conventional fluids, both classical (e.g., water \cite{dau6,lamb})
and quantum (e.g., $^3$He \cite{wolf}), include also sound waves (with
the linear dispersion). This crucial difference can be attributed to
the momentum conservation. Indeed, the usual description of a fluid
(or a gas, see \cite{dau10}) assumes a system of ``particles''
(molecules or atoms) interacting by means of local collisions. In the
simplest case (of a single-component, monoatomic fluid) the collisions
preserve momentum, and hence overall there are three global conserved
quantities -- the number of particles, energy, and momentum. If,
moreover, Galilean invariance is assumed, then the current is defined
by the momentum, which is the key point ultimately leading to the
existence of the sound-like collective mode.

The macroscopic theory describing flows of a conventional fluid --
namely, hydrodynamics -- can be derived in the several ways. One can
follow the above prescription using the continuity equations and
entropy \cite{chai}, one can ``guess'' (or postulate) the constitutive
relations based on the Galilean invariance (or, in the relativistic
case, Lorentz invariance) \cite{dau6}, or one can use the
``microscopic'' kinetic theory \cite{dau10}. The latter approach is
justified, strictly speaking, in a dilute gas, but yields the same set
of hydrodynamic equations as the more phenomenological methods. This
fact is typically attributed to the {\it universality} of the
hydrodynamic approach: the belief that long-distance properties are
largely independent of the short-distance (microscopic) details. As a
result, strongly interacting fluids (such as water) can be
successfully described by the same hydrodynamic theory as an ideal gas
\cite{dau10}.

In condensed matter context, hydrodynamic approaches have been applied
to phonons \cite{gurzhi} (also see the recent experiment \cite{beh}
and references therein) and magnons \cite{hh69}, while applications to
electronic systems \cite{gurzhi1,gurzhi2,dsh} have only recently
attracted widespread attention \cite{rev,luc,pg}. This may appear
surprising, after all the Fermi Liquid theory originally developed for
$^3$He \cite{wolf} has become a dominant paradigm in solid state
physics. In $^3$He, the Fermi Liquid theory can be used to derive the
hydrodynamic equations \cite{akha}, so why cannot the same be done in
solids? Unlike helium atoms, electrons in solids exist in the
environment created by a crystal lattice and can scatter off both
lattice imperfections (or ``disorder'') and lattice vibrations
(phonons). In both cases, their momentum is not conserved. As a result,
the electron motion is typically diffusive \cite{aar}, unless the
sample size is smaller than the mean free path and the system is
``ballistic'' \cite{bee}. For most typical scattering mechanisms in
solids, the mean free path is strongly temperature dependent. In
conventional metals \cite{ziman} electron-impurity scattering
dominates at low temperatures, leading to, e.g., the residual
resistance. At high temperatures, the main scattering mechanism is the
electron–phonon interaction. In many materials, at least one of these
two mechanisms is more effective than electron–electron interaction at
any temperature, leaving no room for hydrodynamic behavior. In terms
of the associated length scales, this statement can be formulated as
$\ell_{ee}\gg\ell_{\rm dis}, \ell_{e-ph}$ (in the self-evident
notation). If a material would exist, where the opposite condition
were satisfied at least in some temperature range, then one would be
justified in neglecting momentum non-conserving processes and applying
the hydrodynamic theory. For a long time such a material was not
known. In recent years, several extremely pure materials became
available bringing electronic hydrodynamics within experimental reach
\cite{geim1,kim1,mac,ihn,goo,gus20,var20,vool,jao,gupta}.

\section{Experimental signatures of hydrodynamic behavior}\label{sec2}

The parameter regime supporting the hydrodynamic behavior can be
readily found in systems where the temperature dependence of key
length scales ($\ell_{ee}$, $\ell_{\rm dis}$, $\ell_{e-ph}$, etc.) is
sufficiently different. This may happen, for example, in
two-dimensional (2D) systems where the electron-electron scattering
length varies with temperature as $\ell_{ee}\sim T^{-2}$ (within the
typical Fermi Liquid description), while the contribution of acoustic
phonon scattering to the electronic mean free path varies as
$\ell_{e-ph}\sim T^{-1}$. At the same time, the low temperature values
of $\ell_{ee}$ are easily surpassed by the mean free path $\ell_{\rm
  dis}$ in ultrapure samples. Hence, 2D systems may offer an
intermediate temperature window \cite{rev,luc,glazman20,adam}, where
electron-electron interaction is the dominant scattering process and
hence appear to be plausible candidates to support the hydrodynamic
behavior. It is then not surprising that many experiments on
electronic hydrodynamics were focusing on 2D systems and especially on
graphene. The latter is a particularly convenient material
\cite{geim1,kim1,geim2,kim2,geim3,geim4,gal} where the mean free path
remains long up to room temperatures,
${{\rm{max}}[\ell_{\rm{dis}},\ell_{e-ph}]>1\,\mu{\rm{m}}}$. At the
same time, at ${T\geqslant150}\,$K the electron-electron scattering
length decreases to ${\ell_{ee}\approx0.1\div0.3\,\mu}$m. Since the
pioneering work on the nonlocal resistance \cite{geim1} and
Wiedemann-Franz law violation \cite{kim1}, several impressive
experiments \cite{geim2,geim3,geim4,gal,young,brar} aimed at
uncovering the hydrodynamic behavior of the electronic system in
graphene. In particular, it was suggested that a viscous hydrodynamic
flow in electronic systems might exhibit enhanced,
higher-than-ballistic conduction \cite{geim2,young,brar}. More
recently, several breakthrough experiments
\cite{ihn,vool,young,brar,fink,halb,zel19,uri,imh,imm,sulp,zel,sulp22}
demonstrated various distinct imaging techniques making it possible to
“observe” the electronic flow in graphene “directly”.

Hydrodynamic flow of electrons in solids should be observable not only
in graphene, but in any material that is clean enough to satisfy the
condition that the electron-electron scattering length is much shorter
than the disorder mean free path. In particular, modern semiconductor
technology allows fabricating ultra-high-mobility heterostructures
\cite{ihn,gus20,gupta,gus18,rai20,gus21}, a noticeable improvement
since the original observation of the Gurzhi effect \cite{mol95}.

At the same time, the hydrodynamic behavior might be observable in a
wide range of novel materials including the 2D metal palladium
cobaltate \cite{mac}, topological insulators (where the conducting
surface states may exhibit hydrodynamic behavior), and Weyl semimetals
\cite{lucnmr,gorbar}. The latter systems have attracted considerable
attention since they exhibit a solid-state realization of the
Adler-Bell-Jackiw chiral anomaly
\cite{jackiw69,adler,jackiw2010,fel16,ong16}. One of the hallmark
manifestations of the anomaly in Weyl systems \cite{lucnmr,spivak13}
is the recently observed negative magnetoresistance
\cite{fel16,fel162}. Observation of relativistic Weyl hydrodynamics in
these systems is the next milestone in the field.

\subsection{Gurzhi effect}\label{sec2.1}

\begin{figure}[t]
\centerline{
\includegraphics[width=0.42\columnwidth]{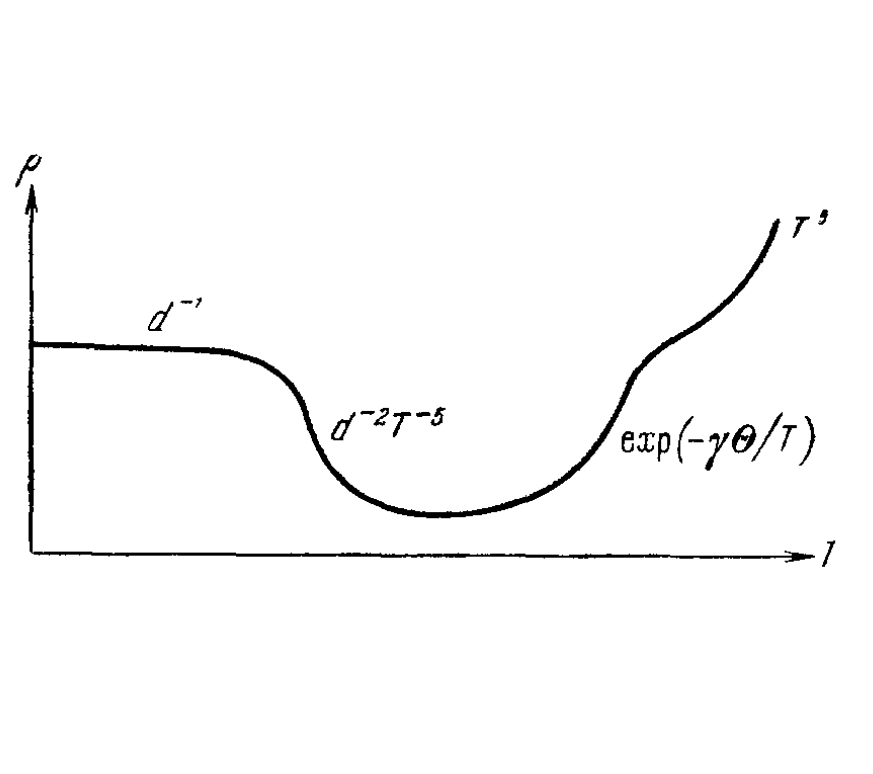}
\qquad
\includegraphics[width=0.42\columnwidth]{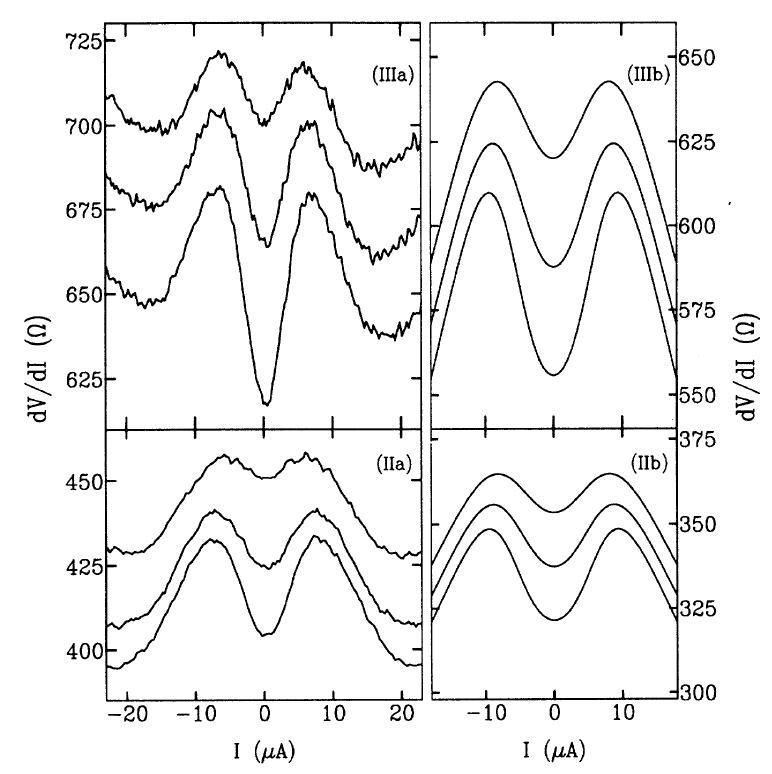}
}
\caption{Gurzhi effect. Left panel: a sketch of the theoretically
  predicted resistance minimum (reprinted with permission from
  Ref.~\cite{gurzhi}; copyright (1968) Uspekhi Fizicheskikh
  Nauk). Right panel: experimental (IIa and IIIa) and theoretical (IIb
  and IIIb) differential resistance $dV/dI$ as a function of the
  current $I$ at the lattice temperatures $T=4.5, \, 3.1, \, 1.8\,$K,
  from top to bottom (reprinted with permission from
  Ref.~\cite{mol95}; copyright (1995) by the American Physical
  Society).}
\label{fig0:gurzhi}
\end{figure}

In his pioneering work \cite{gurzhi,gurzhi1,gurzhi2}, Gurzhi
considered an idealized problem of the electric current flowing in a
thin, clean wire. In this case there are two competing scattering
processes: the electron scattering off the walls of the wire (i.e.,
system boundaries) and the electron-electron interaction, either
direct or effective (e.g., phonon-mediated). Assume that at the lowest
temperatures the electron-electron scattering length is longer than
the width of the wire, $\ell_{ee}\gg d$. Then boundary scattering will
dominate leading to the approximately temperature-independent
resistivity, $\rho\sim1/d$. Now, the electron-electron scattering
length $\ell_{ee}$ is inversely proportional to some power of
temperature (for the direct electron-electron interaction
$\ell_{ee}\propto{T}^{-2}$ \cite{gurzhi1}, while for the
phonon-mediated interaction $\ell_{ee}\propto{T}^{-5}$ \cite{gurzhi2},
see Fig.~\ref{fig0:gurzhi}). As the temperature increases, $\ell_{ee}$
will eventually become smaller than $d$. In the limit $\ell_{ee}\ll
d$, the resistivity will be determined by the electron-electron
scattering, $\rho\sim\ell_{ee}/d^2$ \cite{gurzhi1,gurzhi2} and hence
will decrease with the increasing temperature. This effect can be seen
as the electronic analogy of the crossover between the Knudsen
(molecular) flow and the Poiseuille (viscous) flow in a rarefied gas
driven through a tube \cite{knu}.

The above conclusion crucially depends on the assumption that the
effective mean free path $d^2/\ell_{ee}$ is much smaller than the
length scale $\ell_{\rm dis}$ describing bulk momentum-relaxing
processes (i.e., electron-impurity or electron-phonon scattering).
Then the electronic momentum is approximately conserved and one can
introduce the hydrodynamic description (the expression for $\rho$
follows from the standard expression for the kinematic viscosity,
$\nu=v_F\ell_{ee}/3$ \cite{gurzhi1}).

Once the effective mean free path due to electron-electron interaction
exceeds the disorder scattering length, $d^2/\ell_{ee}\gg\ell_{\rm
  dis}$, the system becomes diffusive and the resistivity resumes its
usual growth with temperature. Hence, $\rho(T)$ is expected to exhibit
a minimum, see Fig.~\ref{fig0:gurzhi}, the result now known as the
Gurzhi effect.

A direct observation of the Gurzhi effect in metals is hindered by
several factors: in addition to the electron-impurity and
electron-phonon scattering, Umklapp scattering, nonspherical Fermi
surface shapes, or Kondo effect may all contribute to the temperature
dependence of the resistivity. An elegant way around these obstacles
was suggested by de Jong and Molenkamp \cite{mol95}. They used 2D
wires defined electrostatically in the two-dimensional electron gas
(2DEG) in semiconductor (GaAs/AlGaAs) heterostructures. Given the
weakness of the electron-phonon coupling in this system, it was
possible to control the electronic temperature selectively without
changing the temperature of the whole sample by passing a {\it dc}
current. The resulting measurement exhibited a minimum in the
differential resistance as a function of the current, see
Fig.~\ref{fig0:gurzhi}, which was argued to be equivalent to the
Gurzhi effect. More recently, the observed decrease of resistivity
with increasing temperature typical of the Gurzhi effect
($\rho\sim{T}^{-2}$) was reported in Ref.~\cite{gus20}.

\subsection{Nonlocal transport measurements}\label{sec2.2}

\begin{figure}[t]
\centerline{\includegraphics[width=0.18\columnwidth]{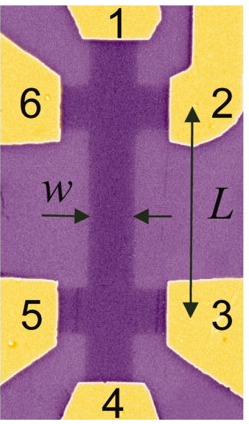}
}
\caption{Hall bar geometry for nonlocal transport measurements.
  Traditional four-terminal measurement involves passing a current
  between leads $1$ and $4$, while measuring the voltage drop between
  leads $2$ and $3$. The resulting resistance
  $R_{23,14}=V_{23}/I_{14}$ is related to the longitudinal
  resistivity, $\rho_{xx}=R_{23,14} W/L$, where $W$ and $L$ are the
  width and length of the Hall bar. In contrast, a nonlocal
  measurement consists of passing a current between, e.g., $2$ and
  $6$, while measuring the voltage between leads $3$ and $5$. In the
  case of usual diffusive transport such voltage should be
  exponentially suppressed \cite{pauw},
  $R_{NL}=R_{35,26}\sim\rho_{xx}\exp(-\pi L/W)$ (From Ref.~\cite{nlr}.
  Reprinted with permission from AAAS).}
\label{fig1:hb}
\end{figure}

The ``modern era'' in electronic hydrodynamics was announced in the
three back-to-back Science papers in 2016 reporting the negative
vicinity resistance \cite{geim1} and Wiedemann-Franz law violation
\cite{kim1} in graphene, as well as hints of the hydrodynamic behavior
in \cite{mac} in PdCoO$_2$. These groundbreaking experiments opened
the door for further studies focusing on unconventional aspects of
electronic transport in ultra-pure materials.

Conventional experiments aimed at uncovering inner workings of solids
often rely on transport measurements \cite{ziman,dau10}, the tool that
proved to be indispensable throughout the history of condensed matter
physics. In a traditional experiment one measures a current-voltage
characteristic and extracts linear response functions determined by
properties of the unperturbed system. A basic quantity that can be
measured in this way is the Ohmic resistance $R$. At the simplest
level, $R$ can be described by the Drude theory \cite{ziman,dru},
which essentially amounts to writing down classical equations of
motion of charge carriers in applied electric and magnetic fields with
a phenomenological friction term.

A more intricate question concerns the spatial distribution of the
electric current density, which is most relevant in small samples
(chips) with multiple leads. Here the current density may exhibit
complex patterns depending on the external bias, electrostatic
environment, chip geometry, and magnetic field. One way to detect such
patterns is provided by nonlocal transport measurements
\cite{skocpol,vwees1,geimnl1,roukes,vonklit,nlgold,geimnl2}, i.e., by
measuring voltage drops between various leads that are spatially
removed from the source and drain, see Fig.~\ref{fig1:hb}. These
techniques were devised to study ballistic propagation of charge
carriers in mesoscopic systems, but recently they were applied to
investigate possible hydrodynamic behavior in ultra-pure conductors
\cite{rev,luc,geim1,geim3,geim4}.

Nonlocal resistance measurements have also been used to study edge
states accompanying the quantum Hall effect
\cite{nlr,mceu,goldman,roth,caza,koma}. While the exact nature of the
edge states has been a subject of debate, the nonlocal resistance,
$R_{NL}$, appears to be an intuitively clear consequence of the fact
that the electric current flows along the edges of the sample. Such a
current would not be subject to exponential decay \cite{pauw}
exhibited by the bulk charge propagation leading to a much stronger
nonlocal resistance.

\subsubsection{Giant nonlocality in magnetic field}\label{sec2.2.1}

While traditional studies of electronic transport tended to focus on
low temperatures, more recent experimental work has been gradually
shifting towards measurements at nearly room temperatures
\cite{geim1,geim3,geim4,nlr,nlgold}. A detailed analysis of the
nonlocal resistance in a wide range of parameters (temperatures,
carrier densities, and magnetic fields) was performed in
Ref.~\cite{nlr} using graphene samples.

\begin{figure}[t]
\centerline{
\includegraphics[width=0.4\columnwidth]{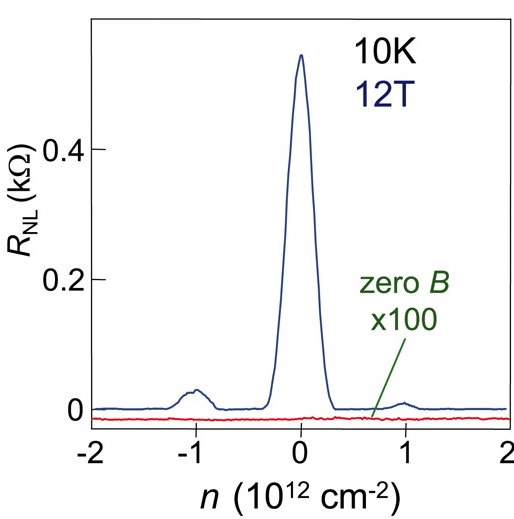}
\qquad
\includegraphics[width=0.4\columnwidth]{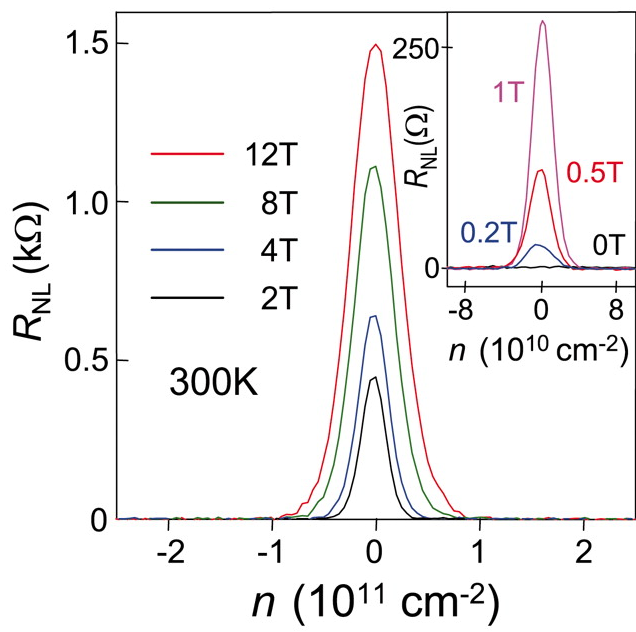}
}
\caption{Nonlocal resistance in graphene. Left panel: QHE regime at
  $T=10\,$K and $B=12\,$T (the red curve indicates that no signal
  could be detected at $B=0$ within the experimental resolution; the
  curve is downshifted for clarity and magnified). Right panel:
  high-temperature regime, $T=300\,$K. (From Ref.~\cite{nlr}.
  Reprinted with permission from AAAS).}
\label{fig2:rnl}
\end{figure}

At low temperatures and in strong magnetic fields, graphene exhibits the
quantum Hall effect (QHE) with well-defined plateaus in Hall
resistivity corresponding to regions of the carrier density where
$\rho_{xx}=0$. At the same densities, the nonlocal resistance also
remains zero, but in between the QHE zeros may reach values as high as
$1\,$k$\Omega$. At high temperatures, all but one such peaks
disappear. The remaining peak at charge neutrality exhibits behavior
that appears to be inconsistent with the QHE interpretation. In
particular, the strong signal persists at near room temperatures, way
beyond the QHE regime with the peak value
${R_{NL}\approx1.5}\,$k$\Omega$ at ${B=12}\,$T and ${T=300}\,$K, three
times higher than that at ${T=10}\,$K, see Fig.~\ref{fig2:rnl}.

The unexpected ``giant'' nonlocality in neutral graphene was
originally explained by diffusion of the mismatched spin-up and
spin-down quasiparticles in the presence of the Zeeman splitting
\cite{nlr}. This interpretation was disputed in Ref.~\cite{chiap}
where the effect was not observed in the nearly parallel field (the
Zeeman splitting is independent of the field direction). Moreover, the
magnitude of the effect proposed in Ref.~\cite{nlr} was disputed in
Ref.~\cite{sven2}, where the residual quasiparticle density due to
Zeeman splitting (at $T=0$ and $B=10\,$T) was estimated to be
$\rho_Q\approx2.2\times10^6\,$cm$^{-2}$ leading to a nonlocal
resistance that is much weaker than the data of Ref.~\cite{nlr}.

The alternative explanation suggested in Ref.~\cite{sven2} was based on
the ``two band'' phenomenology of the electronic system in neutral
graphene \cite{meg,mr1,mrexp}. Indeed, at the charge neutrality point,
the conductance and valence bands in graphene touch. At finite
temperatures, both bands contain mobile carriers leading to a
two-component nature of the electronic system. Given the exact
particle-hole symmetry at neutrality, this system is ``compensated''
and hence there is no classical Hall effect, such that the bulk Hall
conductivity vanishes, $\rho_{xy}=0$, and the longitudinal
conductivity is unaffected by the magnetic field. In contrast, the
same approach yields the nonlocal response that is strongly
field dependent. Indeed, the presence of two types of carriers
(electrons and holes) leads to the existence of two macroscopic
currents: the electric current $\bs{J}$ and the total quasiparticle (or
``imbalance'' \cite{alf}) current $\bs{j}_I$,
\begin{equation}
\label{mc0}
\bs{j} = \bs{j}_e-\bs{j}_h,
\qquad
\bs{j}_I = \bs{j}_e+\bs{j}_h,
\qquad
\bs{J} = e \bs{j},
\end{equation}
where $\bs{j}_{e(h)}$ is the electron (hole) current and $e$ is the
electron charge. In the absence of the magnetic field, the neutral
current $\bs{j}_I$ is decoupled from $\bs{J}$ and is practically
undetectable (it does not couple to the electric field). The electrons
and holes are drifting in parallel, but opposite directions. However,
the magnetic field bends the quasiclassical trajectories of charge
carriers coupling the two currents and turning $\bs{j}_I$ in the
direction that is orthogonal to $\bs{J}$. Now the neutral current can
transport charge carriers to distant parts of the sample, where a
nonlocal response is induced, again, by the magnetic field, see
Section~\ref{sec3.4} for more details.

The arguments of Ref.~\cite{sven2} yield the nonlocal response
capturing the main qualitative features of the effect observed in
Ref.~\cite{nlr}. Quantitatively, these results are consistent with the
rapid decay of the nonlocal signal away from the neutrality point, but
overestimate the magnitude of the effect. The latter discrepancy was
attributed to the simplicity of the model that did not take into
account the effects of electron-electron interaction contributing to
resistivity of neutral graphene, the residual carrier population at
neutrality due to fluctuations of the electrostatic potential
\cite{chiap}, and viscous phenomena, all of which are expected to
suppress $R_{NL}$.

Viscous effects are of particular interest in the context of
electronic hydrodynamics and may also lead to nonlocality. However,
these effects are expected to occur in the absence of magnetic field
as well and in graphene are most pronounced away from charge
neutrality.

\subsubsection{Negative vicinity resistance}\label{sec2.2.2}

Away from charge neutrality, i.e., when the chemical potential exceeds
the temperature, $\mu\gg T$, electrons in graphene are typically
expected to behave similarly to 2DEG in semiconductor
heterostructures. The contribution of the valence band is exponentially
suppressed and the electronic system comprises only the single
component. In that case, a Fermi liquid is expected to behave
hydrodynamically \cite{akha}, the issue with the electronic systems
being whether the material is pure enough.

Assuming the hydrodynamic regime is possible, the single-component
electronic system should obey the Navier-Stokes-like equation
\cite{dau6,navier,stokes,stokes51} with an additional damping due to
disorder scattering \cite{gurzhi2}, as well as the continuity
equation. Within linear response and in the static limit, these
equations can be written as (see, e.g., Section~\ref{sec3})
\begin{equation}
\label{nseqfl}
e\bs{E} = - m \nu \Delta\bs{u} + m \bs{u}/\tau_{\rm dis},
\qquad
\bs{\nabla}\bs{u} = 0,
\end{equation}
where $\bs{u}$ is the hydrodynamic velocity, $\nu$ is the kinematic
viscosity, and $m$ is the effective mass (in graphene this should be
replaced by $\mu/v_g^2$, with $\mu$ being the chemical potential and
$v_g$ the velocity of the Dirac spectrum). The electric current is
expressed in terms of the hydrodynamic velocity as
\begin{equation}
\label{ju}
\bs{j} = n\bs{u},
\end{equation}
where $n$ is the carrier density, see also Eq.~(\ref{mc0}).

The resulting behavior of the current density is determined by the
relative strength of the viscosity and disorder scattering, which can
be expressed in terms of the dimensionless ``Gurzhi number'' (note
that this definition is written in analogy to the Reynolds number
\cite{dau6} and is the inverse of the number defined in
Ref.~\cite{sven1})
\begin{equation}
\label{gu}
{\rm Gu} = \frac{l^2}{\nu\tau_{\rm dis}},
\end{equation}
where $l$ is the typical length scale of the problem. Large values of
${\rm Gu}$ indicate that the disorder scattering dominates (such that
the current density exhibits patterns typical to the traditional
diffusive behavior), whereas small values of ${\rm Gu}$ correspond to
the hydrodynamic viscous flow \cite{pol15,pol16,fl0,sven1,msw2,msw}.

\begin{figure}[t]
\centerline{
\includegraphics[width=0.4\columnwidth]{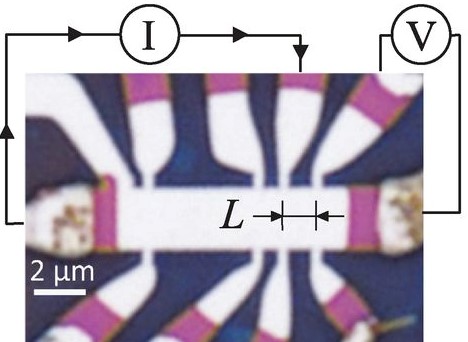}
\qquad
\includegraphics[width=0.5\columnwidth]{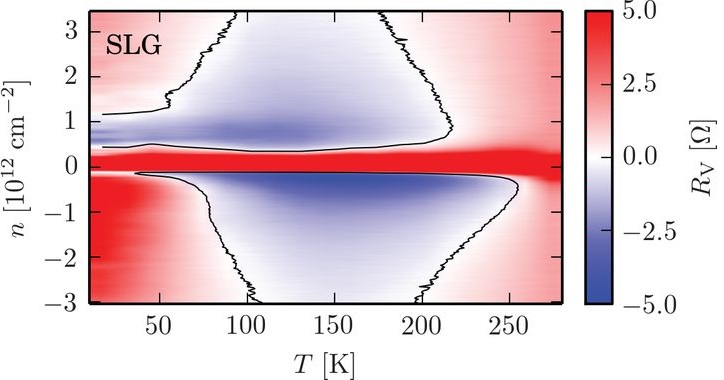}
}
\caption{Negative vicinity resistance in graphene. Left panel:
  multi-lead device with the measurement schematic. Right panel: color
  map showing a wide, intermediate temperature range where the
  vicinity resistance is negative (From Ref.~\cite{geim1}. Reprinted
  with permission from AAAS).}
\label{fig3:nvr}
\end{figure}

\begin{figure}[t]
\centerline{
\includegraphics[width=0.5\columnwidth]{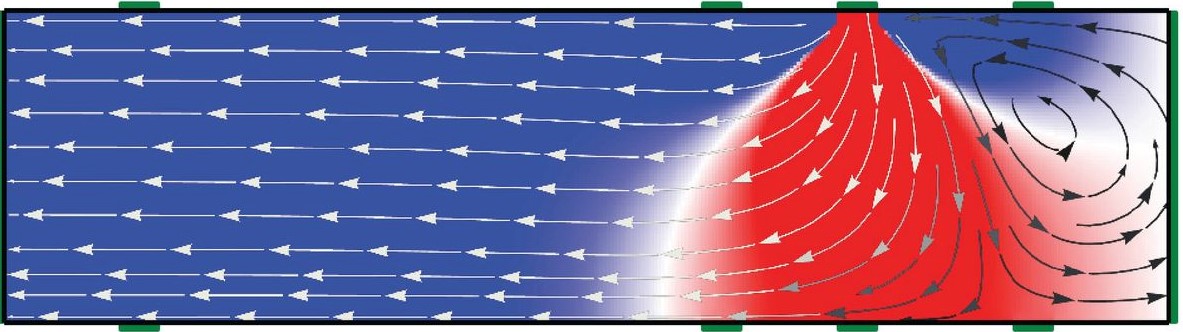}
}
\bigskip
\centerline{
\includegraphics[width=0.98\columnwidth]{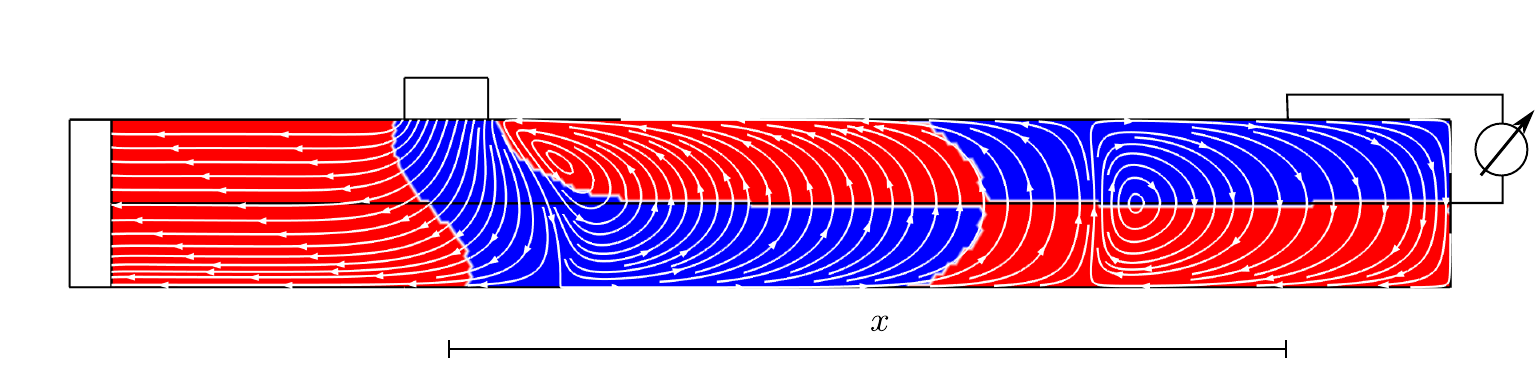}
}
\caption{Vorticity in electronic flows in graphene. Top panel:
  simulated flow in the experimental device shown in
  Fig.~\ref{fig3:nvr} (from Ref.~\cite{geim1}. Reprinted with
  permission from AAAS). Bottom panel: double vortex in a long device
  suggested in Ref.~\cite{sven1}. The red and blue colors indicate the
  alternating sign of the deviation of the electrochemical potential
  from its median value. (Reprinted with permission from IOP
  Publishing).}
\label{fig4:vor}
\end{figure}

In confined geometries, viscous flows may be accompanied by vortices
(or whirlpools) \cite{pol15,pol16,fl0,sven1}, which may be detected by
observing {\it negative} nonlocal resistance by placing the leads on
the opposite sides of a vortex. This idea was realized in the
pioneering experiment of Ref.~\cite{geim1}. Here (unlike the
measurement in Ref.~\cite{nlr}) the leads were placed close to each
other (based on the expected vortex size), see Fig.~\ref{fig3:nvr},
hence the measured quantity was referred to as ``vicinity
resistance''.

In agreement with the expectation that the hydrodynamic behavior
should occur at intermediate temperatures, the measured vicinity
resistance is negative roughly between $70\,$K and $250\,$K (with the
actual range being density dependent), see Fig.~\ref{fig3:nvr}. This
observation was supported in Ref.~\cite{geim1} by a solution to the
above equations (\ref{nseqfl}) showing formation of a vortex close to
the leads. Similar theoretical results were reported in
Refs.~\cite{pol15,pol16,fl0,sven1}, see also Ref.~\cite{stein19}.

Despite the apparent agreement between theory and experiment,
observation of the negative vicinity resistance does not represent the
proverbial ``smoking gun'' proving that the system is in fact in the
hydrodynamic regime. The reason is that ballistic systems may also
exhibit negative nonlocal resistance \cite{imm} as has been shown both
experimentally \cite{nlgold} and theoretically \cite{fl18}. This issue
has been specifically studied in Ref.~\cite{geim3}, where it was shown
that in addition to being negative, the vicinity resistance has to
grow with temperature (the crossover from the ballistic to
hydrodynamic behavior was identified with the minimum in the vicinity
resistance as a function of temperature). More recently,
Ref.~\cite{sven1} reported a numerical solution to the hydrodynamic
equations (\ref{nseqfl}) showing the existence of multiple vortices in
long samples, see Fig.~\ref{fig4:vor}. Since the vorticity of the
adjacent vortices has the opposite sign, placing multiple leads along
the sample and measuring the voltage as a function of distance from
the source electrode should yield a {\it sign-alternating nonlocal
  resistance} which should in principle distinguish the ballistic and
hydrodynamic behavior. Alternatively, one could try to use one of the
novel imaging techniques \cite{imh,imm,sulp,zel} to observe vortices
``directly''.

\subsection{Hydrodynamic flow around macroscopic obstacles}\label{sec2.3}

The collective hydrodynamic flow is expected to differ strongly from
the single-particle ballistic motion in systems with macroscopic
obstacles. Whereas particles tend to scatter off anything they may
encounter -- sample boundaries, other geometrical features, or
long-range potentials, a viscous fluid tends to avoid obstacles by
flowing around them. As a result, the collective flow maybe more
efficient in carrying the constituent particles through the system in
question. In the context of the traditional hydrodynamics of rarefied
gases, this fact has been established already by Knudsen
\cite{knu}. In the context of electronic hydrodynamics, this issue was
first addressed theoretically in Ref.~\cite{fl2} and experimentally in
Ref.~\cite{geim2}.

\subsubsection{Superballistic transport}\label{sec2.3.1}

One of the most common types of ``obstacles'' studied in the context
of electronic transport is a constriction (or a point contact). This
object was extensively studied in mesoscopic physics \cite{bee}, with
the conductance quantization \cite{qpc1,qpc2} being the hallmark
effect. In particular, it was established that ballistic propagation
of charge carriers through a point contact yields the conductance that
is constrained by a fundamental upper bound \cite{sha}.

Quantization of the point contact conductance can be understood by
considering one-dimensional (1D) subbands in the constriction of the
width $W$ (corresponding to the quantized values of the transverse
momentum, $k_y=\pm\pi n/W$). Each subband contributes equally to the
conductance due to the cancellation of the group velocity and the 1D
density of states (DoS) \cite{bee}. Observing that the number of the
occupied subbands is naturally an integer, one finds that the total
conductance is quantized, $G_b=2Ne^2/h$. In the classical limit, the
number of propagating (Landauer) channels in 2D can be estimated as
$N=\left[k_FW/\pi\right]$ (square brackets indicate the integer
value), yielding the upper bound known as the Sharvin limit
\cite{bee,sha}.

\begin{figure}[t]
\centerline{
\includegraphics[width=0.28\columnwidth]{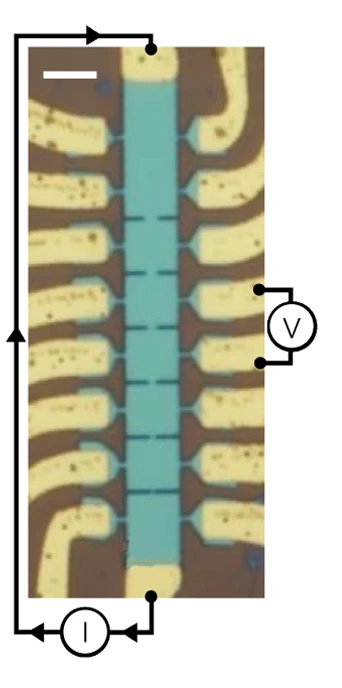}
\qquad
\includegraphics[width=0.4\columnwidth]{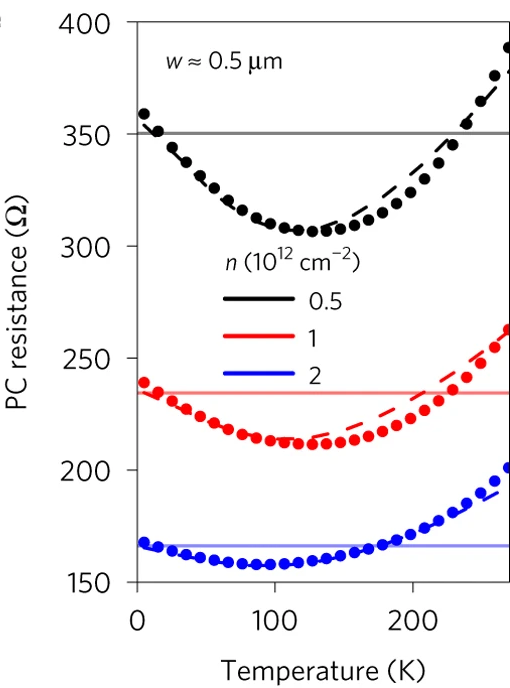}
}
\caption{Superballistic transport in graphene. Left panel: a typical
  measuring device showing multiple point contacts varying in width
  from $W=0.1$ to $W=1.2\,\mu$m. Right panel: point contact resistance
  for a $W=0.5\,\mu$m constriction at representative carrier
  densities. The experimental data are represented by dots, while the
  horizontal lines indicate the Sharvin limit of the maximum classical
  ballistic conductance. Lower-than-the-limit resistance at
  intermediate temperatures is indicative of the collective, viscous
  flow of electrons. (From Ref.~\cite{geim2}. Reprinted with
  permission from Springer Nature).}
\label{fig5:sbc}
\end{figure}

The above argument neglects electron-electron interaction and is
justified when the corresponding scattering length is large compared
to the width of the constriction, $\ell_{ee}\gg W$. In the
hydrodynamic regime, $\ell_{ee}\ll W$, electrons move collectively
avoiding the boundaries and thus may carry the charge through the
point contact more effectively than free fermions (i.e., achieving
conductance higher than $G_b$, see Fig.~\ref{fig5:sbc}). Indeed, the
solution to the hydrodynamic equations describing the electron flow
through a simplest 2D constriction reported in Ref.~\cite{fl2} yields
the conductance
\begin{equation}
\label{superg}
G_h = \frac{\pi e^2 n^2 W^2}{32\eta},
\end{equation}
where $\eta$ is the shear viscosity. Since $G_h$ grows with width
faster than $G_b$, there is a possibility for the ``superballistic''
conduction for wide enough channels.

The theoretical expectation ($G_h>G_b$) was first confirmed in the
experiment of Ref.~\cite{geim2}, see Fig.~\ref{fig5:sbc}, and more
recently corroborated in Ref.~\cite{young}, where a novel imaging
technique was applied to the point contact problem (see
Section~\ref{sec2.4}), see also Ref.~\cite{brar}. The theory of
Ref.~\cite{fl2} was revisited and expanded upon in
Ref.~\cite{glazman20}, where the same hydrodynamic equation was solved
for the current density profile. The authors of Ref.~\cite{glazman20}
also analyzed the intermediate parameter regime where hydrodynamic
flows could be realistically observed. Heating effects in similar
inhomogeneous flows ware analyzed in Ref.~\cite{gor21}.

\subsubsection{Flows around macroscopic obstacles}\label{sec2.3.2}

\begin{figure}[t]
\centerline{
\includegraphics[width=0.9\columnwidth]{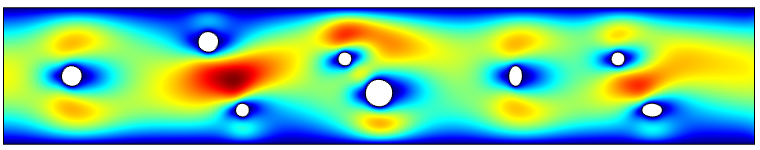}
}
\medskip
\caption{Numerical simulation of the Poiseuille flow in a 2D channel
  with randomly placed macroscopic obstacles (represented by white
  shapes). The color map indicates the magnitude of the flow velocity
  (ranging from zero shown in blue to the maximum shown in dark red).}
\label{fig6:pf}
\end{figure}

The transition from the Ohmic to hydrodynamic flow observed in the
point contact geometry in Refs.~\cite{geim2,young} is similar to the
transition between the Knudsen and Poiseuille flows
\cite{gurzhi,knu,mol95}. The tendency of the viscous flow to avoid
obstacles is well known in hydrodynamics and is illustrated in
Fig.~\ref{fig6:pf}. However, a naive solution of the hydrodynamic
equations in a 2D system with macroscopic obstacles within linear
response leads to the so-called ``Stokes paradox''
\cite{dau6,lamb,gus20,lucas17,guo17}.

The problem of a motion of a spherical object through an otherwise
stationary viscous fluid (or equivalently, viscous flow around a
stationary sphere subject to the condition of constant flow velocity
at infinity) is a classic problem in hydrodynamics
\cite{dau6,lamb,stokes51}. For flows characterized by small Reynolds
numbers, one may neglect the nonlinear term in the Navier-Stokes
equation \cite{dau6,navier,stokes} and solve the resulting system of
linear equations. In 3D the problem can be solved analytically not
only for the sphere but also for several other simple shapes
\cite{lamb}, where one typically calculates the ``drag force'' acting
on the obstacle.

\begin{figure}[t]
\centerline{
\includegraphics[width=0.45\columnwidth]{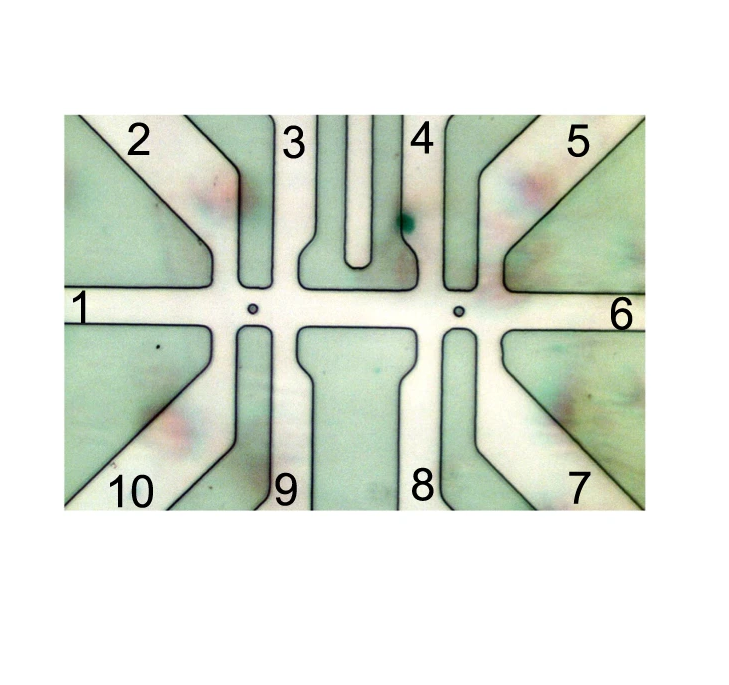}
\quad
\includegraphics[width=0.4\columnwidth]{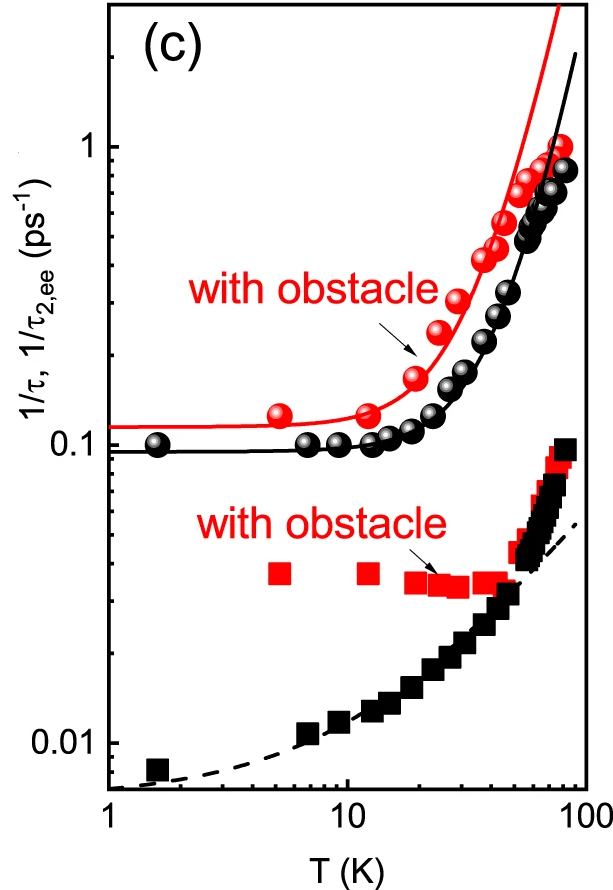}
}
\caption{Stokes flow around an obstacle in GaAs. Left panel: image of
  the Hall bar with two anti-dots used in the experiment
  \cite{gus20}. Right panel: the disorder (squares) and
  electron-electron interaction (circles) scattering rates obtained
  from the experimental data measured in sample with (red) and without
  (black) the obstacle (From Ref.~\cite{gus20}).}
\label{fig7:obst}
\end{figure}

The above simple solution of the linearized hydrodynamic equations
appears to fail if the obstacle has the form of an infinitely long
cylinder (or equivalently, in 2D), the issue known as the ``Stokes
paradox''. The reason for the apparent paradox lies in the
approximation used to linearize the Navier-Stokes equation: the
Reynolds number (i.e., the quantitative expression for the relative
strength of the nonlinear and viscous terms) is scale-dependent and
cannot be assumed small at arbitrary large distances
\cite{dau6,lamb}. Instead of simply neglecting the nonlinear term, one
should linearize it following Oseen \cite{oseen}, whose modified
equation yields a consistent solution (as well as the corrected
expression for the drag force).

In contrast to traditional hydrodynamics, in solid state physics one
is typically interested in linear response properties and has to take
into account momentum relaxation due to weak impurity scattering. The
latter allows one to stabilize the solution, while keeping it within
linear response \cite{lucas17,guo17}. Indeed, in ultra-pure electronic
systems the Gurzhi number (\ref{gu}) may be much larger than the
Reynolds number
\begin{equation}
\label{gure}
\frac{\rm Gu}{\rm Re} = \frac{l^2/(\nu\tau_{\rm dis})}{ul/\nu}
= \frac{l}{u\tau_{\rm dis}},
\end{equation}
justifying the Stokes approximation in the presence of the impurity
scattering.

Stokes flow in the 2D electron system with a circular obstacle was
observed in Ref.~\cite{gus20}. The experiment was performed in a GaAs
heterostructure with the role of the obstacle played by an anti-dot
(or a micro-hole) in the middle of the Hall bar, see
Fig.~\ref{fig7:obst}. The measured resistivity was interpreted using
the macroscopic approach of Refs.~\cite{ale,moo}. The two scattering
mechanisms (one due to impurity scattering and another due to
viscosity) were treated as two parallel channels of momentum
relaxation (based on the fact that the corresponding relaxation rates
can be attributed to the first and second moments of the semiclassical
distribution function). The two contributions can be separated since
they have a different temperature dependence, in particular, the
viscous contribution should exhibit the Gurzhi-like $\rho\sim{T}^{-2}$
behavior. Now, the obstacle does not seem to affect the latter, while
the disorder contribution at low temperatures is significantly
enhanced, see Fig.~\ref{fig7:obst}, which is consistent with the
expectation of the viscous fluid avoiding the obstacle (as opposed to
individual electrons scattering off it).

\subsection{Imaging of electronic flows}\label{sec2.4}

Although traditional linear response measurements may be strongly
affected by the collective, hydrodynamic behavior, interpretation of
such experiments is not straightforward \cite{geim3}. It would be much
easier if one could simply ``watch'' the flow (in a close analogy to
the usual hydrodynamics). Fortunately, in recent years several
``scanning'' or ``imaging'' techniques were suggested allowing one to
do just that even if indirectly.

The basic requirement for any imaging technique is that it should be
non-invasive, i.e., it should not disrupt the flow itself. When
trying to image the flow of electrons, one can rely either on
detecting spatial variation of electric potential \cite{imm,sulp} or
on detecting the local magnetic field induced by the charge motion
\cite{imh}.

\subsubsection{Scanning carbon nanotube single-electron transistor}\label{sec2.4.1}

\begin{figure}[t]
\centerline{
\includegraphics[width=0.8\columnwidth]{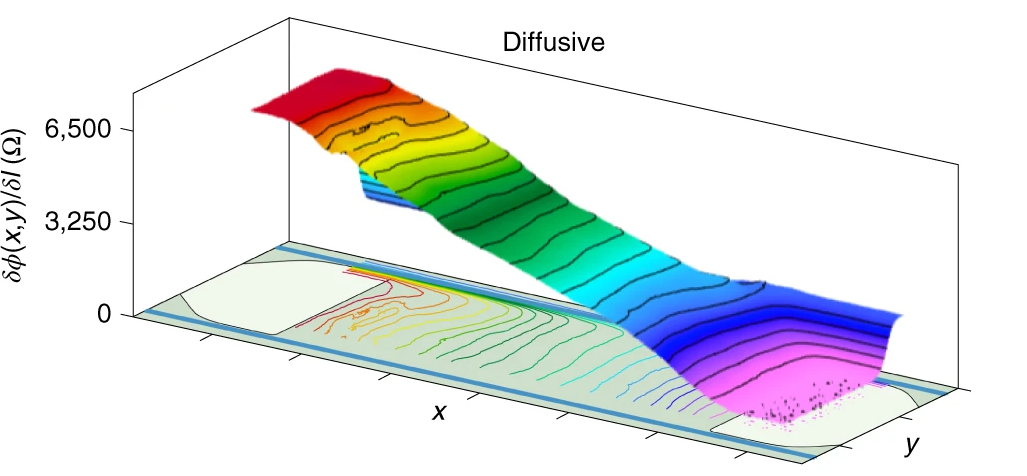}
}
\bigskip
\centerline{
\includegraphics[width=0.8\columnwidth]{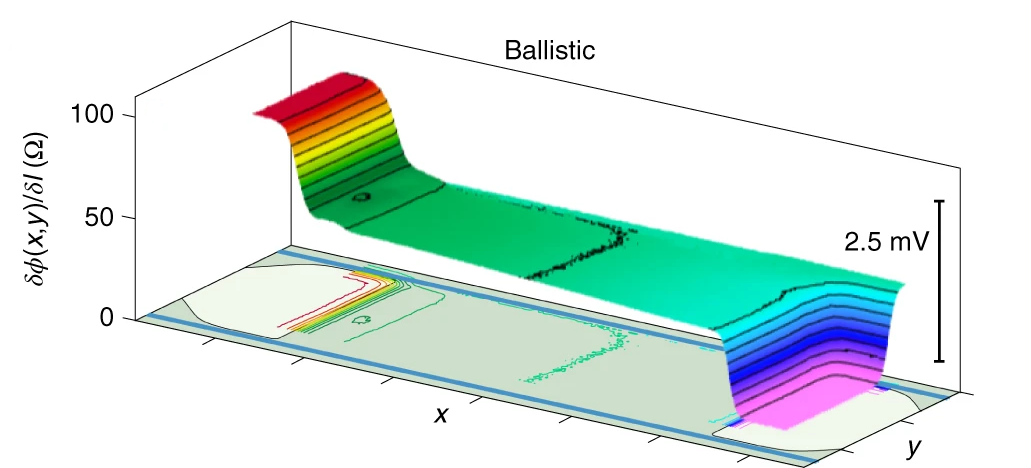}
}
\caption{Spatial imaging of the voltage drop of flowing electrons in
  the diffusive (top) and ballistic (bottom) regimes \cite{sulp}. Both
  plots show the imaged electrostatic potential normalized by the
  total current (yielding a quantity with the units of
  resistance). The data were measured at $T=4\,$K. The diffusive flow
  was observed at charge neutrality (determined by the sharp maximum
  in the two-terminal resistance of the sample), while the ballistic
  behavior was imaged at the hole density of
  $1\times10^{12}\,$cm$^{-2}$. In the latter case, most of the voltage
  drop occurs at the contacts, with the contact resistance approaching
  the ideal Sharvin value \cite{sha,tar}. The bottom plane shows the
  equipotential contours superimposed on the schematic of the graphene
  channel and contacts, indicating the gradual voltage drop in the
  diffusive case contrasted to the flat potential typical of the
  ballistic motion \cite{dat} (From Ref.~\cite{sulp}. Reprinted with
  permission from Springer Nature).}
\label{fig8:diffbal}
\end{figure}

Electric current flowing through a conductor is known to generate a
local change in electrostatic potential (or ``voltage drop''). This
potential can be detected using the capacitive coupling to a local
probe such as the scanning single-electron transistor (SET), see
Fig.~\ref{fig8:diffbal}. In particular, a nanotube SET may exhibit
extreme voltage sensitivity, while the planar probe design could help
minimizing the back action on (or gating) the sample
\cite{sulp}. Moreover, by applying weak perpendicular magnetic field,
the same probe is able to resolve the Hall voltage associated with the
flow, yielding a direct measure of the local current density.

\begin{figure}[t]
\centerline{
\includegraphics[width=0.8\columnwidth]{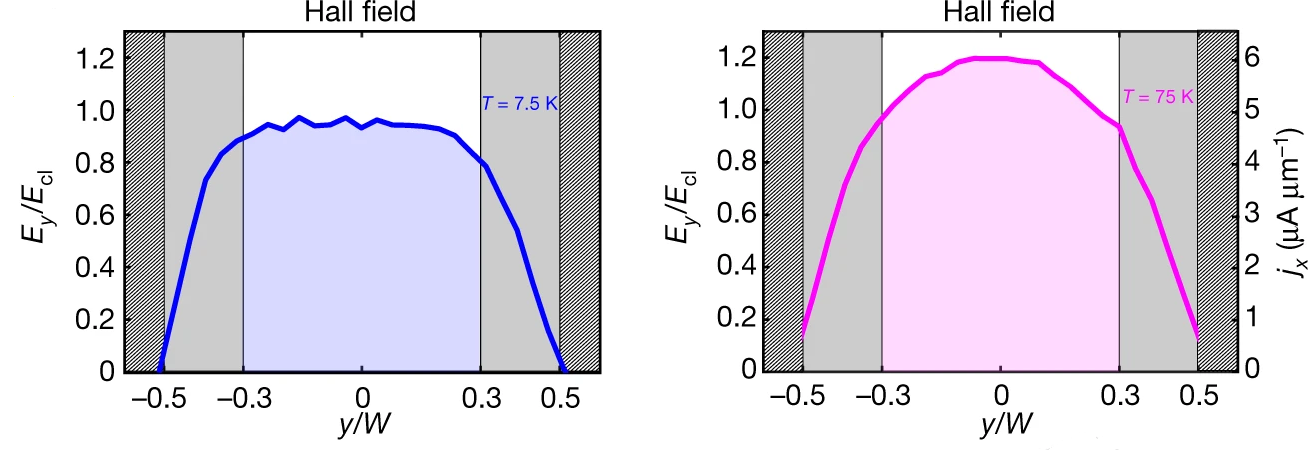}
}
\bigskip
\centerline{
\includegraphics[width=\columnwidth]{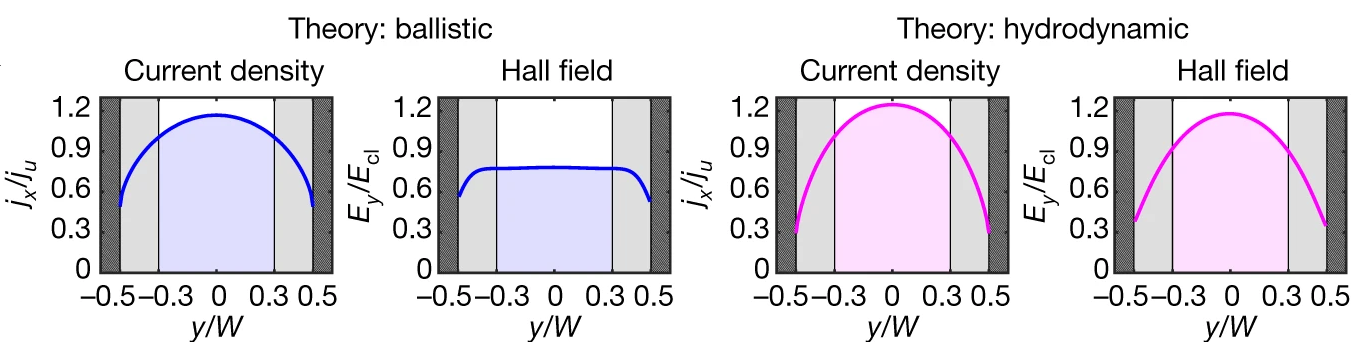}
}
\caption{Spatial imaging of the hydrodynamic flow of electrons in
  doped graphene \cite{imm}. Top: the Hall field $E_y$ as obtained by
  numerical differentiation of the measured Hall voltage with respect
  to $y$, normalized by the classical value, $E_{\rm
    cl}=BJ/(neW)$. The top left panel shows data taken at $T=7.5\,$K,
  $B=\pm12.5\,$mT, and $E_{\rm cl}=91\,$Vm$^{-1}$. The right top panel
  shows data in the presumed hydrodynamic regime at $T=75\,$K,
  $B=\pm18\,$mT, and $E_{\rm cl}=162\,$Vm$^{-1}$. The right vertical
  axis converts the field into the units of the current density by
  scaling with $ne/B$. Bottom: calculated current density $J_x/J_u$
  (with $J_u=J/W$) and Hall field $E_y/E_{\rm cl}$. The numerical
  values were obtained using the parameters corresponding to the
  experimental data in the top panels. (From Ref.~\cite{imm}.
  Reprinted with permission from Springer Nature).}
\label{fig9:hydro_Hall}
\end{figure}

Applying the nanotube SET technique to doped graphene in the
hydrodynamic regime allowed to image the Poiseuille flow of charge
carriers \cite{imm}. Similarly to the case of the Gurzhi effect, see
Sec.~\ref{sec2.1}, the main goal of the experiment was to distinguish
the collective (hydrodynamic) motion from the single-particle
(ballistic) behavior (assuming $\ell_{\rm dis}$ is the largest length
scale in the problem). However, instead of contrasting the temperature
dependence of the sample's resistance \cite{geim3}, here one has to
compare the spatial profile of the current density. In the channel
geometry, see Fig.~\ref{fig8:diffbal}, one studies its dependence on
the lateral coordinate, $\bs{j}=j(y)\bs{e}_x$ (where $\bs{e}_x$ is the
unit vector directed along the channel and $y$ is the coordinate
across the channel). The difficulty is that in contrast to the
textbook diffusive behavior, where the current density is uniform
(except in the narrow regions close to the sample boundaries), both in
the ballistic and hydrodynamic cases $j(y)$ is characterized by a
non-uniform profile with the maximum at the center of the channel
\cite{bee,dat}, making it difficult to distinguish the two regimes
experimentally.

The hydrodynamic Poiseuille flow in a narrow channel is a textbook
problem \cite{dau6}. Taking into account weak impurity scattering and
making the common assumption of the no-slip boundary conditions, one
finds for the electric current density in doped graphene in the
channel geometry
\begin{equation}
\label{pf}
J_x = \sigma E_x \left[1-\frac{\cosh(y/\ell_G)}{\cosh[W/(2\ell_G)]}\right],
\end{equation}
where $\sigma$ in the bulk longitudinal conductivity and $\ell_G$ is
the Gurzhi length \cite{sven1,moo,pol17,mr2,cfl}
\begin{equation}
\label{lg}
\ell_G = \sqrt{\nu\tau_{\rm dis}}.
\end{equation}
Here $\nu$ is the kinematic viscosity, see Eq.~(\ref{nseqfl}). The
parabolic current density profile typical of the standard Poiseuille
flow \cite{dau6,poi} can be recovered by assuming a large Gurzhi
length, ${\ell_G\gg{W}}$. In this limit, the sample resistance is
proportional to the shear viscosity \cite{mr2}, a manifestation of the
Gurzhi effect \cite{gurzhi}.

Introducing more realistic (Maxwell's) boundary conditions with
nonzero slip length \cite{ks19} effectively sets the coordinates where
the catenary curve (\ref{pf}) reaches zero outside of the channel, but
does not significantly affect the current density in the bulk of the
sample. From the experimental viewpoint, however, the resulting curve
is difficult to distinguish from the non-uniform current density in
the ballistic regime, see the bottom panel in
Fig.~\ref{fig9:hydro_Hall} and Sec.~\ref{sec3.2}. As a result, one has
to perform other measurements (e.g., the Hall field, see
Fig.~\ref{fig9:hydro_Hall}) to distinguish the two regimes \cite{imm}.

\subsubsection{Quantum spin magnetometry}\label{sec2.4.2}

\begin{figure}[t]
\centerline{
\includegraphics[width=0.278\columnwidth]{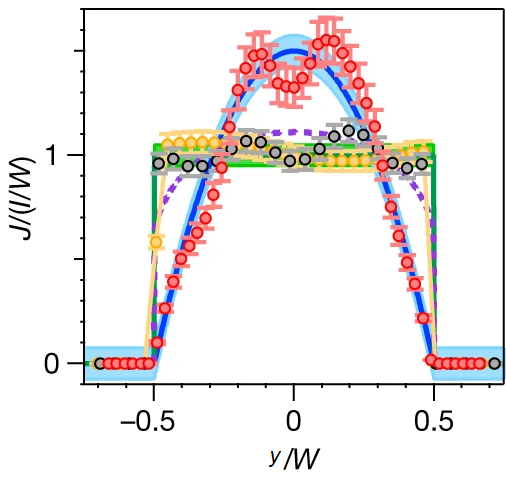}
\quad
\includegraphics[width=0.3\columnwidth]{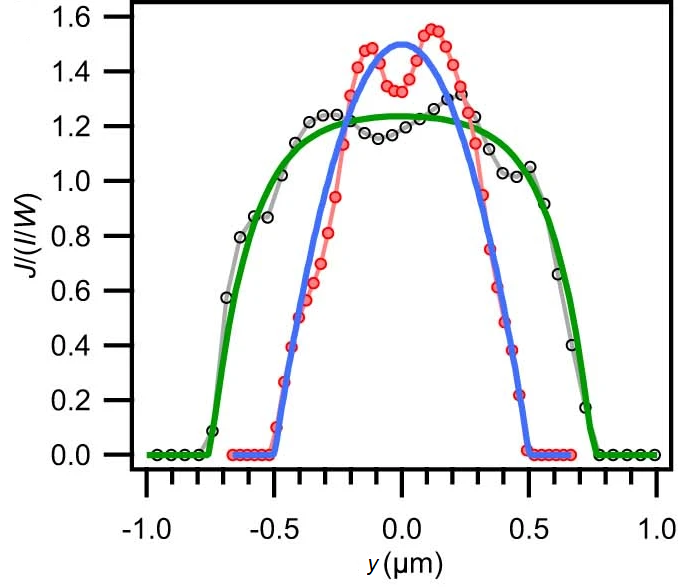}
\quad
\includegraphics[width=0.337\columnwidth]{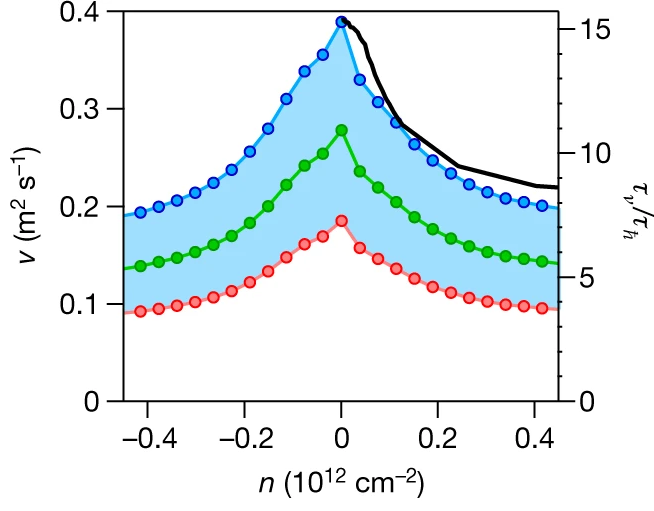}
}
\caption{Spatial imaging of the electric current in neutral graphene
  \cite{imh}. Left: reconstructed current density as a function of the
  lateral coordinate. The current is normalized by the average charge
  carrier flux $I/W$, where $I$ is the total flux and $W=1\,\mu$m is
  the width of the channel. The spatial coordinate $y$ is normalized
  by $W$ and centered on the channel. Red points show data measured in
  neutral graphene, gray points -- in palladium channel, orange points
  -- low-mobility graphene. The curves correspond to idealized
  theoretical expectations: blue -- ideal viscous flow, green --
  uniform current, purple dashed -- the current profile of
  non-interacting electrons with diffusive boundary condition. Center:
  similar measurement for $W=1.5\,\mu$m compared to the data on the
  left. Solid lines are fit to Eq.~(\ref{pf}). Right: bounds on
  kinematic viscosity obtained from fitting the data to
  Eq.~(\ref{pf}). The black curve is the result of a theoretical
  calculation of Ref.~\cite{me2} at $T=300\,$K and no adjustable
  parameters (From Ref.~\cite{imh}. Reprinted with permission from
  Springer Nature).}
\label{fig10:df}
\end{figure}

An alternative technique for imaging the electric current density is
based on the idea of measuring the associated stray magnetic field
\cite{imh}. A sensitive quantum spin magnetometer was realized using
nitrogen vacancy (NV) centers in diamonds \cite{luk}. In contrast to
Ref.~\cite{imm}, the experiment of Ref.~\cite{imh} targeted the
so-called Dirac fluid in neutral graphene and contrasted the presumed
hydrodynamic regime with the diffusive behavior in low-mobility
devices. The latter measurements served as a benchmark and yielded the
standard picture of nearly uniform current (exhibiting a sharp decay
near the channel boundaries, see also Sec.~\ref{sec3.2}) shown in
Fig.~\ref{fig10:df}.

The main result of Ref.~\cite{imh} is the observation (by means of the
scanning NV magnetometry) of a Poiseuille-like flow of the electric
current in neutral graphene described by a catenary curve
(\ref{pf}). Comparing the data to Eq.~(\ref{pf}) the authors have
extracted the kinematic viscosity of the Dirac fluid in graphene (see
the right panel in Fig.~\ref{fig10:df}) showing a good quantitative
agreement with the theoretical calculations of Ref.~\cite{me2}
(without any fitting procedure). Nevertheless, the results of
Ref.~\cite{imh} remain controversial. Within the existing theory of
electronic hydrodynamics, the electric current is related to the
hydrodynamic velocity by Eq.~(\ref{ju}) up to an Ohmic
correction. Precisely at charge neutrality, $n=0$, and Eq.~(\ref{ju})
yields zero, implying that any electric current at charge neutrality
is not hydrodynamic, but is rather given by the Ohmic correction
\cite{me1,hydro1} with the corresponding bulk conductivity determined
by electron-electron interaction \cite{schutt}. The situation is more
involved if the system is subjected to the external magnetic field. In
that case, the Ohmic correction acquires an additional dependence on
the hydrodynamic velocity \cite{me1}, which in particular leads to
positive magnetoresistance \cite{mus,hydro0}. However, a recent
theoretical calculation of the electronic flow in a channel geometry
in neutral graphene based on the direct solution of hydrodynamic
equations (see Sec.~\ref{sec3.4}) yields the so-called
``anti-Poiseuille'' flow \cite{megt2}, with the current density
exhibiting a minimum in the center of the channel -- in contrast to
the maximum in Eq.~(\ref{pf}), see Sec.~\ref{sec4}.

Another feature of the data shown in Fig.~\ref{fig10:df} not accounted
for by the existing theory is that the electric current vanishes at
the channel boundaries. Indeed, the boundary conditions for the Ohmic
correction to Eq.~(\ref{ju}) should be derived from the kinetic theory
similarly to those describing ballistic propagation of electrons
\cite{bee}. In that case, one has to solve the kinetic equation
imposing boundary conditions on the electronic distribution
function. Both extreme limits typically considered in literature,
namely the diffusive and specular boundary conditions, do not lead to
the current vanishing at the boundary. Moreover, the kinetic theory
derivation of the hydrodynamic equations yields the Maxwell's boundary
conditions for the hydrodynamic velocity \cite{ks19}. Finally, there
is strong experimental evidence \cite{zel} for the existence of
classical edge currents in graphene that are not taken into account in
existing theories but casting further doubts on the results shown in
Fig.~\ref{fig10:df}.

\begin{figure}[t]
\centerline{
\includegraphics[width=0.95\columnwidth]{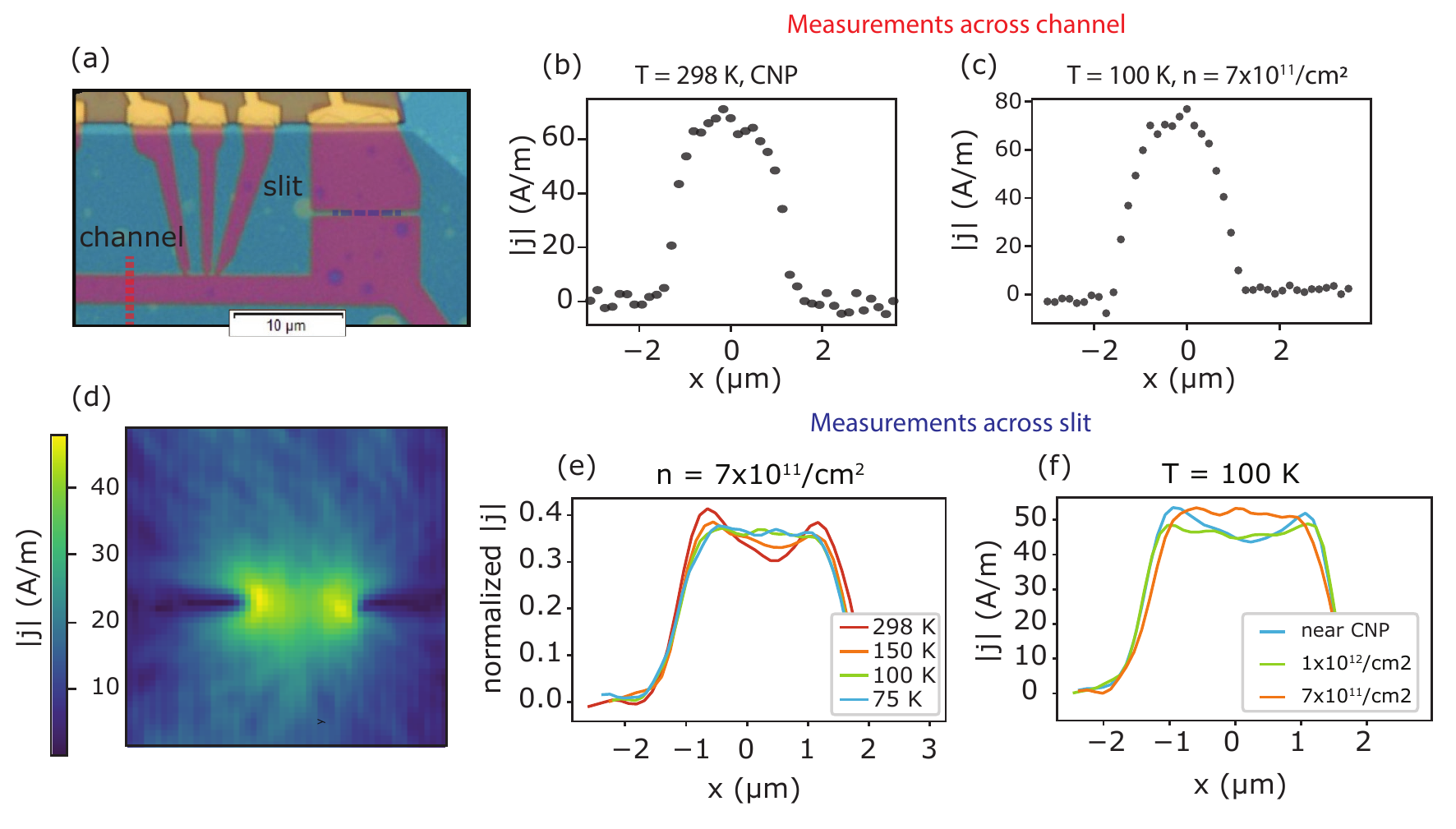}
}
\caption{Spatial imaging of the electric current in neutral graphene
  \cite{young}. (a) Optical image of the graphene device showing the
  locations used in obtaining the current density measurements for the
  channel geometry (b-c), $W=2.7\,\mu$m, and the slit geometry
  (d-f). (b) Current density profile in the channel near the charge
  neutrality point (CNP) at $T=298\,$K. The black dots are the
  reconstructed current density. (c) Measurement of the current
  density profile of the channel at the same position as in (b), but
  at $T=100\,$K and $n=7\times10^{11}\,$cm$^{-2}$. (d) Reconstructed
  current density magnitude at $T=298\,$K, near the CNP, showing the
  characteristic double peaks of Ohmic flow. (d) Temperature
  dependence of the reconstructed $j_y$ at fixed carrier density
  $n=7\times10^{11}\,$cm$^{-2}$ in a line cut through the
  constriction. (e) Carrier density dependence of $j_y$ at fixed
  temperature $T=100\,$K. (From Ref.~\cite{young}. Reprinted with
  permission from the authors).}
\label{fig11}
\end{figure}

The importance of edge physics is further highlighted by the
experiment of Ref.~\cite{young}, where the NV magnetometry was used to
image the flow of charge through a constriction (or a slit) in neutral
graphene. The authors performed measurements in a channel geometry as
well with somewhat contradicting results, see Fig.~\ref{fig11}. While
the channel measurement at nearly room temperature ($T=298\,$K)
yielded the current density profile similar to that reported in
Ref.~\cite{imh}, see Fig.~\ref{fig10:df}, the same profile was
observed at $T=100\,$K implying that the charge flow in the channel is
not very sensitive to the variation of the scattering length. In
contrast, the current density measured in the slit geometry exhibited
Ohmic behavior at room temperature, while at lower temperatures and
finite charge densities the Ohmic double peaks disappeared indicating
the crossover into the hydrodynamic regime. The authors of
Ref.~\cite{young} explained the contradiction between the results in
the channel and slit geometries by fact that the latter is not
affected by the boundary conditions as much as the former. They
conclude that while the edge physics is poorly understood the slit
geometry is better suited to observe the Ohmic-viscous crossover.

\subsubsection{Non-topological edge currents}\label{sec2.4.3}

\begin{figure}[t]
\centerline{
\includegraphics[width=0.7\columnwidth]{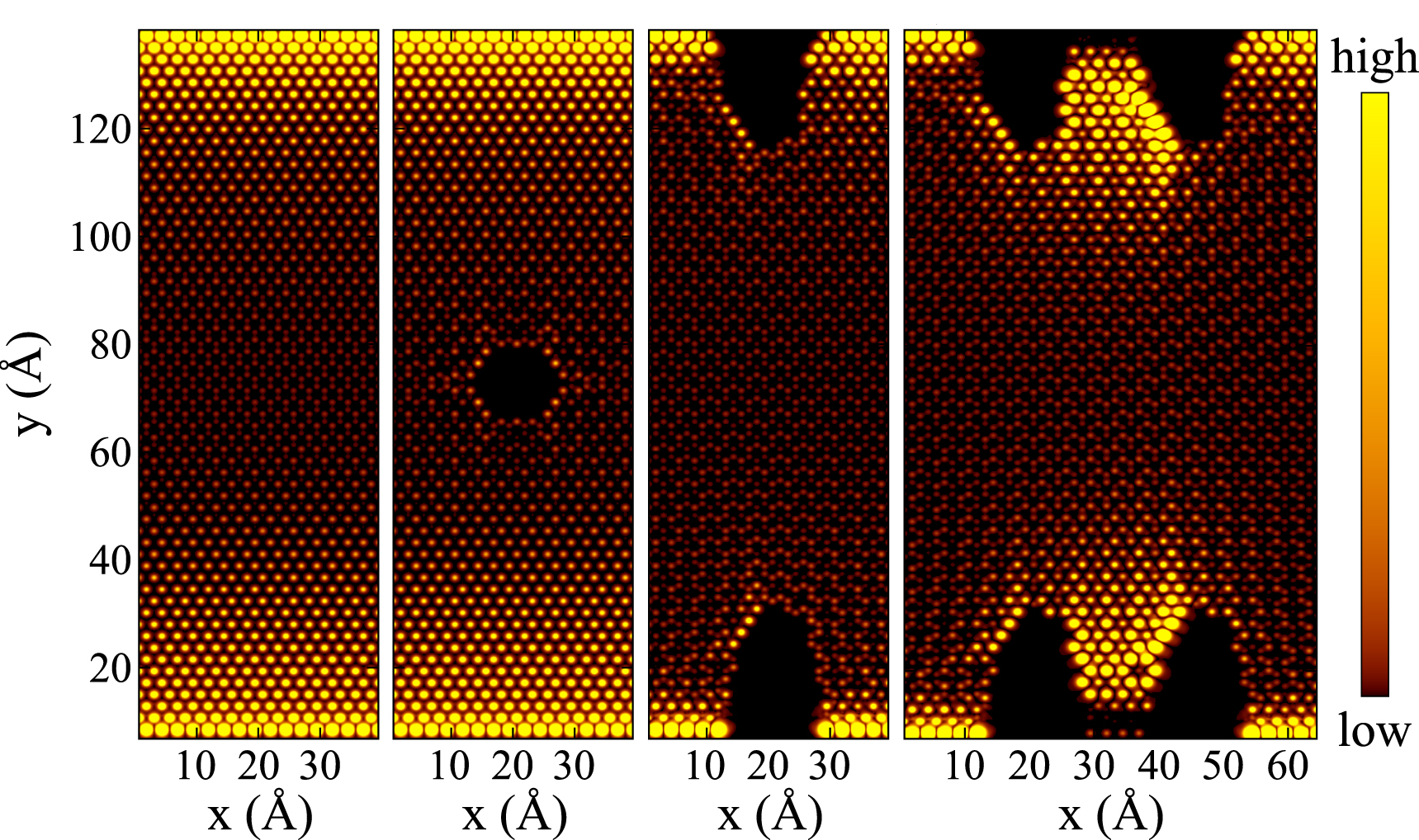}
}
\caption{DFT calculation of the local density of states (LDoS) in a
  graphene flake \cite{nic18}. The enhanced LDoS at the edges
  appears regardless of the shape of the edge and the presence of
  macroscopic defects in the bulk (From Ref.~\cite{nic18}).}
\label{fig12:dos}
\end{figure}

Sample edges play a crucial role in all of the experiments discussed
so far. Yet, understanding of the physics of the edges themselves has
proven somewhat challenging. In traditional condensed matter physics
\cite{ziman}, the focus is typically on bulk behavior and hence a
system is modeled to be infinite. Sample geometry and edge scattering
becomes important in mesoscopic physics \cite{bee,dat}, but most
details are encoded in the boundary conditions. Finally, edge states
are being actively researched in the context of the Quantum Hall
Effect (QHE) \cite{eis,gir} and more generally in the field of topological
insulators \cite{ber}. But even in the latter case, the edge behavior
is dictated by the topological properties of the bulk.  At the same
time, experiments show that sample edges (in particular, in graphene,
see Fig.~\ref{fig12:dos}) may exhibit charge accumulation
\cite{zel19,dgg16,zhit,nic18} and carry non-topological currents
\cite{zel,yac16}.

\begin{figure}[t]
\centerline{
\includegraphics[width=0.452\columnwidth]{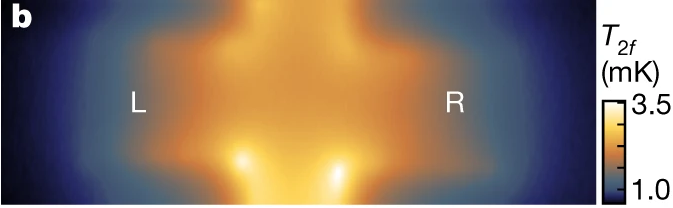}
\quad
\includegraphics[width=0.45\columnwidth]{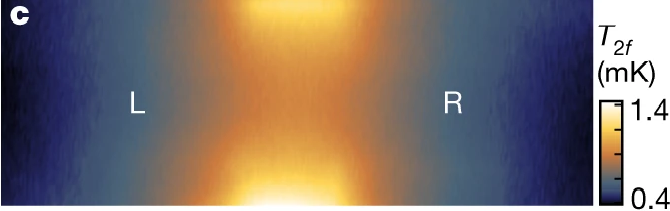}
}
\caption{Thermal imaging of a graphene sample \cite{zel}. Both images
  show the local temperature distribution obtained using the scanning
  SOT at the background temperature $T=4.2\,$K and $B=0$. Left:
  enhanced nonlocality in neutral graphene -- heat dissipation is
  extended into the left and right arms of the Hall bar. Right: Ohmic
  behavior -- heat dissipation is confined to the central region of
  the sample between the source and drain electrodes. (From
  Ref.~\cite{zel}. Reprinted with permission from Springer Nature).}
\label{fig13:thermo}
\end{figure}

\begin{figure}[t]
\centerline{
\includegraphics[width=0.85\columnwidth]{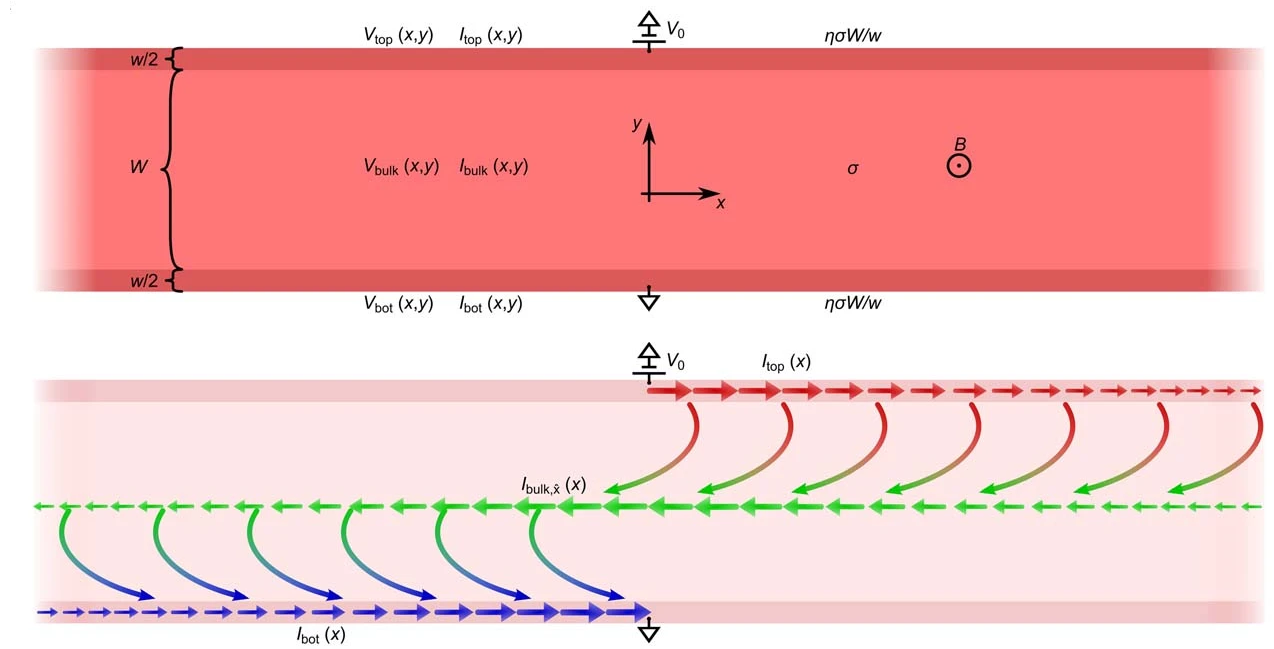}
}
\caption{Classical model mimicking the effects of charge accumulation
  at the sample edges \cite{zel}. Top: the setup -- a strip-like
  sample of width $W$ with bulk conductivity $\sigma$ and narrow edge
  regions (width $w/2$) with the conductivity $\eta\sigma W/w$ with
  $\eta$ being the phenomenological measure of charge
  accumulation. Bottom: nonuniform current density in the presence of
  the magnetic field $B$ featuring the bulk flow in the direction
  opposite to the applied electric field. (From
  Ref.~\cite{zel}. Reprinted with permission from Springer Nature).}
\label{fig14}
\end{figure}

Charge accumulation at the surface is a known phenomenon in
semiconductors \cite{sze} and is a key feature in the traditional
theory of the Schottky barrier \cite{schottky}. Typically, these
effects are associated with ``band bending'' or local,
position-dependent changes in quasiparticle energy levels in the
vicinity of the sample surface (or an interface). The band bending can
also occur in 2D systems. In particular, it has been suggested that in
graphene band bending leads to p-doping of the edges, due to either
intrinsic mechanisms or charged impurities (or defects)
\cite{zel19,yac16,geimedge}. The resulting hole accumulation at the
sample edges has been used in Ref.~\cite{zel} to interpret the highly
unusual nonlocal transport observed by means of SQUID-on-tip (SOT)
thermal imaging and scanning gate microscopy \cite{halb,zel19,halb17}
(for applications of scanning gate microscopy to 2D electron systems
in semiconductor heterostructures see Ref.~\cite{bra18}).

The experiment of Ref.~\cite{zel} provided a deeper insight into the
giant nonlocality observed in neutral graphene subjected to magnetic
field in Ref.~\cite{nlr}, see Sec.~\ref{sec2.2.1}. While confirming
the giant enhancement of the nonlocal resistance at charge neutrality
and in magnetic field, the new data show a number of novel features:
(i) the nonlocality exists even in the absence of magnetic field;
although the observed $R_{NL}$ is much smaller than in the presence of
the field, it is still an order of magnitude stronger that the Ohmic
expectation; (ii) the observed nonlocality is asymmetric with respect
to electron and hole doping; (iii) in magnetic field, the system
exhibits the Hall voltage of the opposite sign (as compared to the
naive expectation); and most importantly, (iv) the observed
nonlocality can be suppressed by applying a potential at the sample
edges. The latter observation represents the key evidence in support
of the interpretation of the data offered in Ref.~\cite{zel}. The
authors argue that the sample edges may carry electric current which
in turn leads to nonlocal resistance. The fact that this current can
be suppressed by a local potential points towards its non-topological
origin (a topological current tends to flow around obstacles
\cite{zel19} such that applying a potential would just ``redefine''
the edge). The existence of the edge current is further corroborated
by the thermal imaging, see Fig.~\ref{fig13:thermo}.

The authors of Ref.~\cite{zel} offer a simple theoretical model to
account for the experimental data. Consider a sample that is infinite
in $x$ direction, while having a width $W$ in the $y$
direction. Without charge accumulation at the edges, the sample can be
assumed to host a uniform charge density, while the current density
can be found using the Ohm's law and the continuity equation. Consider
now a different situation, where the charge density in narrow regions
close the sample edge exceeds the bulk density. Now, the same
equations have to be solved separately in the edge and bulk regions
leading to the complicated behavior shown in Fig.~\ref{fig14}.

The classical model accounts for the unexpected inversion of the Hall
voltage and edge currents observed in the experiment, but does not
explain the physical origin of these effects at a microscopic
level. Some of these features appear to be rather general for the
usual transport equations in the strip geometry. For example, current
flows against the direction of the applied electric fields have also
been reported in Ref.~\cite{cfl}, where the hydrodynamics-like
phenomenology was used to define distinct edge regions where charge
carriers react to the applied magnetic field differently than carriers
in the bulk of the sample, see also Ref.~\cite{mr1}.

Although implications of the results of Ref.~\cite{zel} are not fully
understood at the time of writing, it is clear that the boundary
effects play a very important role in the observed behavior of small
graphene samples. This presents a clear challenge for the theory which
so far was focusing on bulk systems, see Sec.~\ref{sec3}. In
particular, the existing solutions of the hydrodynamic equations in
the strip geometry (similar to Fig.~\ref{fig14}) were
found under the simplest model assumptions of either the no-slip or
Maxwell's boundary conditions, see Sec.~\ref{sec3}.

One could try to avoid the issue of the boundary conditions (except
for the boundaries with the source and drain electrodes \cite{fal19})
by utilizing the Corbino disk geometry \cite{corbino}. Due to
inherently inhomogeneous current flow (even in the Ohmic regime), the
Corbino disk was suggested as a potential device to measure electronic
viscosity \cite{corb}. More recently, hydrodynamic behavior in this
setting was reported in the imaging experiment of Ref.~\cite{sulp22}.

\begin{figure}[t]
\centerline{
\includegraphics[width=0.45\columnwidth]{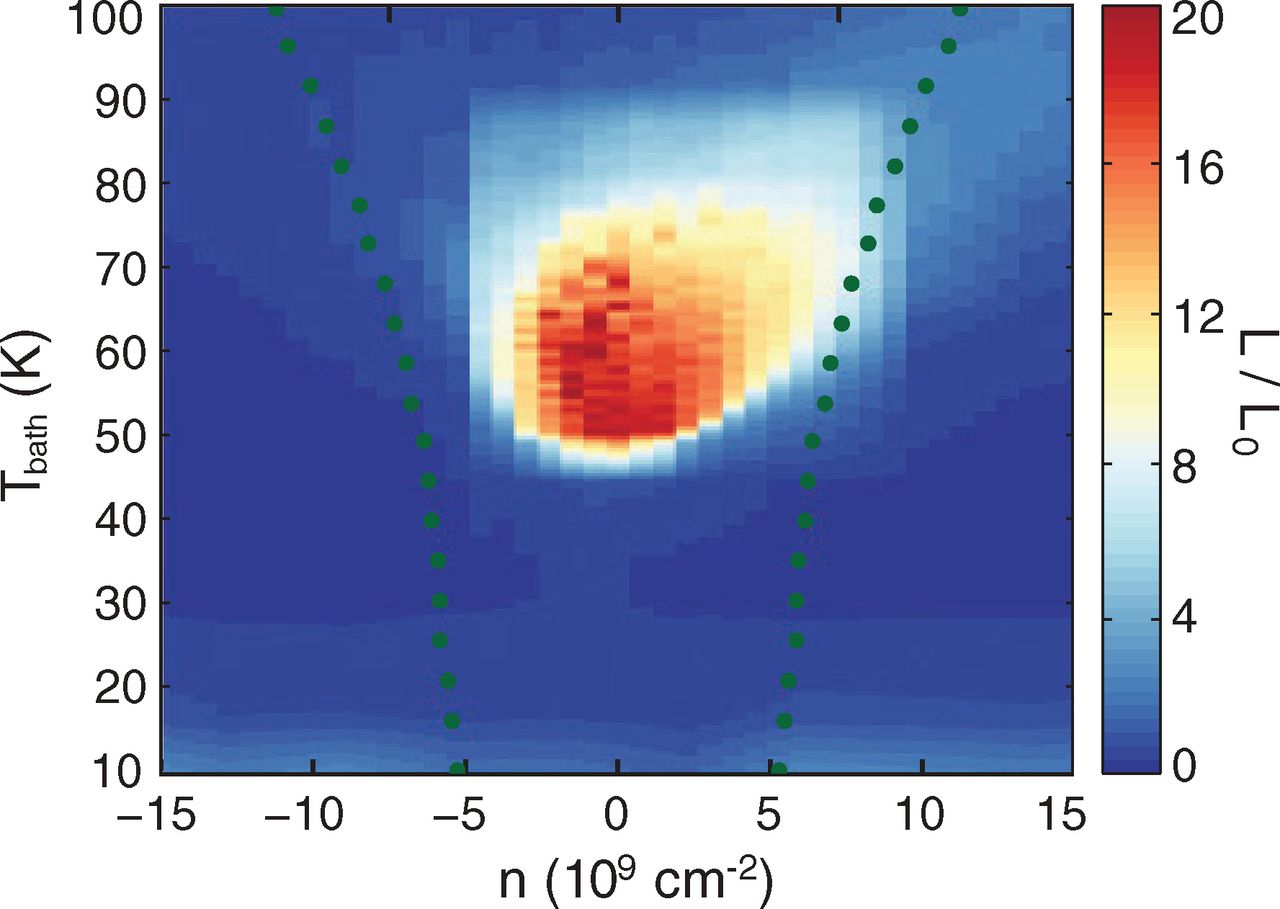}
}
\caption{Wiedemann-Franz law violation in neutral graphene
  \cite{kim1}. The color scheme shows the Lorenz number as a function
  of the charge density and bath temperature. The unusually large
  Lorenz number is observed in the vicinity of charge neutrality and
  in a temperature window above the disorder-dominated regime, but
  below the onset of electron-phonon coupling (From
  Ref.~\cite{kim1}. Reprinted with permission from AAAS).}
\label{fig15:wfl}
\end{figure}

\subsection{Wiedemann-Franz law violation}\label{sec2.5}

Unconventional charge transport properties exhibited by electronic
systems presumed to be in the hydrodynamic regime may be accompanied
by unusual heat transport leading to strong violations of the
Wiedemann-Franz law \cite{wfl,ll}. Initially an empirical observation,
the Wiedemann-Franz law can be readily understood within the standard,
single-particle transport theory \cite{ziman}. Qualitatively, if both
charge and heat are carried by the same excitations and affected by
the same scattering mechanisms (as is the case for noninteracting
electron models), then the only difference between the electric and
thermal conductivities is the dimensionality, leading to the famous
expression
\begin{equation}
\label{wflaw}
\frac{\kappa}{\sigma} = {\cal L} T,
\qquad
{\cal L}={\cal L}_0=\frac{\pi^2}{3e^2}.
\end{equation}
Here $\sigma$ and $\kappa$ are the electric and thermal conductivities
and the coefficient ${\cal L}$ is known as the Lorenz number, while
the ``universal'' value ${\cal L}_0$ corresponds to free electrons.
Now, electrons in solids are typically not free and hence there is no
reason for Eq.~(\ref{wflaw}) to be universally valid. In conventional
metals, the Wiedemann-Franz law is approximately obeyed, for example
the Lorenz number in copper exhibits deviations from ${\cal L}_0$ up
to a factor of $2$ at intermediate temperatures (depending on sample
purity) \cite{hust}. Consequently, a strong violation of the
Wiedemann-Franz law almost certainly an indication of unconventional
physics, that in the context of electronic systems may include
hydrodynamic behavior.

\subsubsection{Large Lorenz number in neutral graphene}\label{sec2.5.1}

Unconventional thermal transport in neutral graphene was reported
already in early experiments of Refs.~\cite{kim0,ong}. The
Wiedemann-Franz law was then studied in detail in Ref.~\cite{kim1}
where it was interpreted as evidence for the hydrodynamic ``Dirac
fluid''. An observation of the related phenomenon of giant thermal
diffusivity in a Dirac fluid was reported in Ref.~\cite{kopp21}.

In hindsight, strong violation of the Wiedemann-Franz law in graphene
should have been expected on the basis of the two celebrated features
(see also Ref.~\cite{har}) -- the linear spectrum \cite{wal,sem,dvm}
and ``quantum'' conductivity
\cite{schutt,geimqhe,mfs,mfss,mish1,mish}. The latter indicates that
the unusual feature of the electrical conductivity at charge
neutrality is not its value, but rather the scattering mechanism
behind it -- electron-electron interaction. In contrast, the former
ensures that the electron-electron interaction does not relax the
energy current (since it is equivalent to the momentum flux, see
Sec.~\ref{sec3}), which implies that the thermal conductivity is
determined by disorder scattering. As a result, the Lorenz number is
expected to be proportional to the ratio of the disorder mean free
time to the electron-electron scattering time, which in the
hydrodynamic regime (or otherwise in ultra-clean graphene in the
appropriate temperature interval) is assumed to be large,
${\cal{L}}\propto\tau_{\rm{dis}}/\tau_{ee}\gg1$, see
Fig.~\ref{fig15:wfl}.

\begin{figure}[t]
\centerline{
\includegraphics[width=0.305\columnwidth]{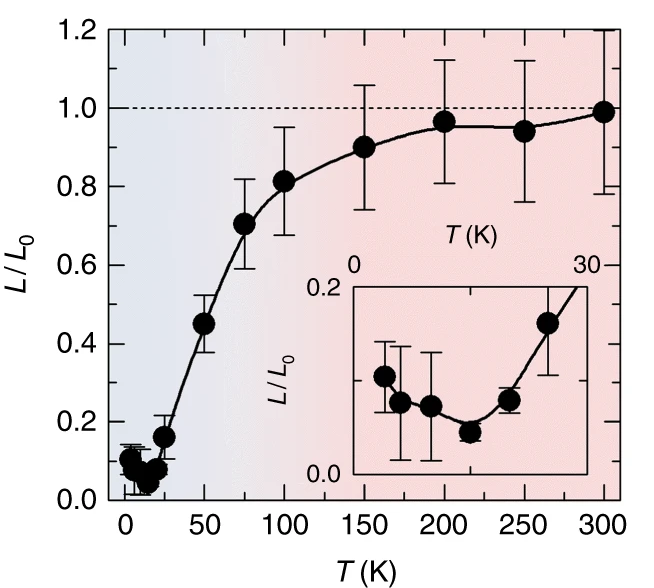}
\quad
\includegraphics[width=0.28\columnwidth]{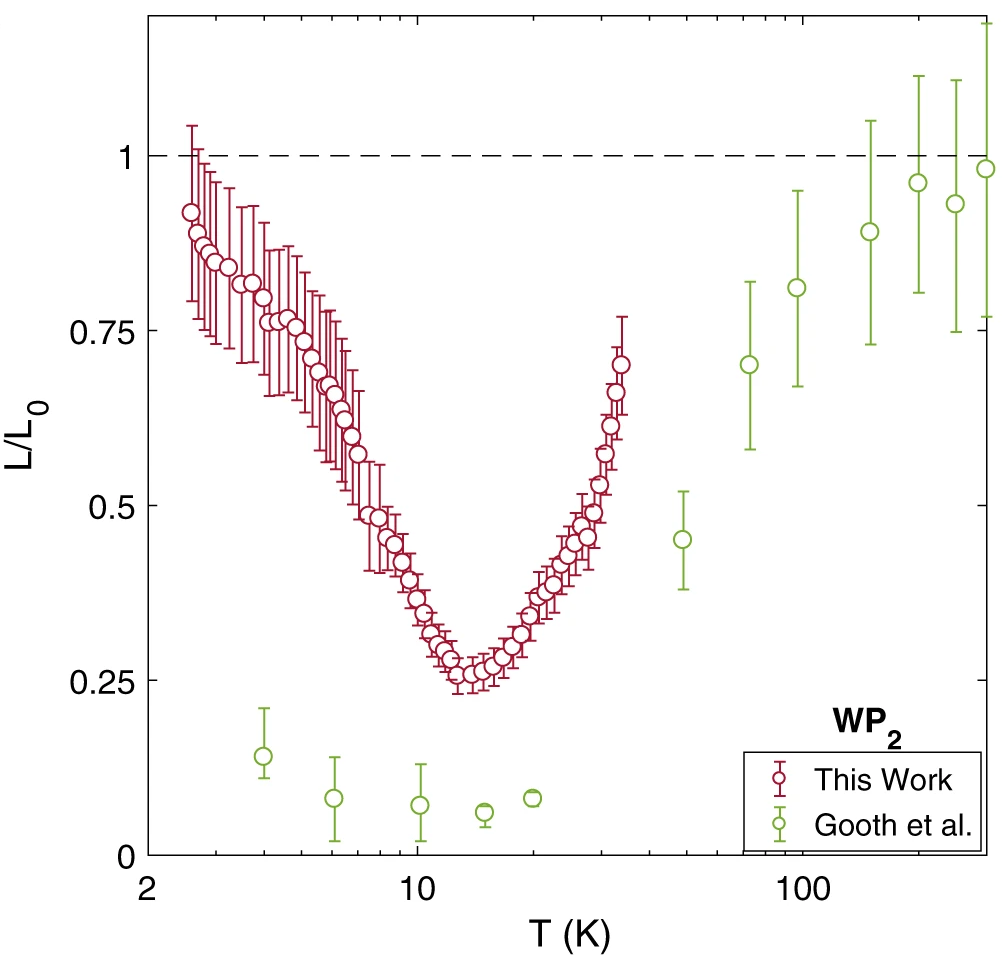}
\quad
\includegraphics[width=0.323\columnwidth]{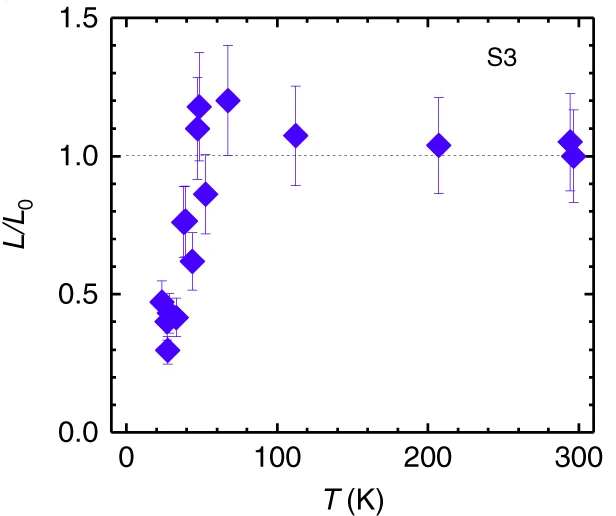}
}
\caption{Wiedemann-Franz law violation in topological materials. Left:
  the Lorenz number extracted from measurements of the electrical and
  thermal conductivities in a $WP_2$ micro-ribbon (width $2.5\,\mu$m)
  (Reprinted from Ref.~\cite{goo18}). The inset shows the zoomed-in
  low-temperature region of the same data.  Central: the Lorenz number
  in bulk (mm-sized) single crystals of $WP_2$ (Reprinted from
  Ref.~\cite{goo}). Green dots show the data from the left
  plot. Right: the Lorenz number in $MoP$ (Reprinted from
  Ref.~\cite{Kumar19}). }
\label{fig16:top}
\end{figure}

The intermediate nature of the hydrodynamic regime suggested by the
data in Fig.~\ref{fig15:wfl} is corroborated by the results of the
experiments on the thermoelectric power \cite{kim2}. Here it
manifested itself in the failure to uncover the ideal hydrodynamic
limit where (in the absence of disorder) the thermopower equals the
thermodynamic entropy per carrier charge \cite{alf,fos16}. Still, the
observed thermopower at relatively high temperatures significantly
exceeded the standard Mott relation indicating the hydrodynamic
behavior \cite{kim2}.

Interestingly, the hydrodynamic theory predicts the Wiedemann-Franz
law violation even in doped graphene (in the Fermi-liquid regime)
\cite{luc,vig15} (for a detailed discussion of the Wiedemann-Franz law
violation in Fermi liquids in general see Ref.~\cite{lds2}), but now
the Lorenz number is predicted to be small (and in fact vanish in the
limit of large densities, see Sec.~\ref{sec3}). The effect can not be
clearly seen in Fig.~\ref{fig15:wfl}, presumably due to relatively low
densities explored in the data shown. This prediction suggests a
possible relation with the small Lorenz number observed in topological
materials, which has not been fully addressed so far.

\subsubsection{Small Lorenz number in topological materials}\label{sec2.5.2}

Recently, hints of electronic hydrodynamics have been observed in the
topological material $WP_2$ \cite{goo,goo18}, where the
Wiedemann-Franz law is also strongly violated, see
Fig.~\ref{fig16:top}. The measured thermal and electrical
conductivities in $WP_2$ exhibit features that are significantly
different from those observed in graphene. The presumed hydrodynamic
regime is limited to temperatures below $20\,$K (as determined by the
electron-phonon scattering dominating transport properties at higher
temperatures). Here, the measured Lorenz number turns out to be small,
${{\cal L}\ll{\cal L}_0}$, the result that was attributed to the
existence of the hydrodynamic regime (confirmed by the extremely large
measured values of the typical length scale describing
momentum-relaxing scattering properties). Interestingly enough,
similar effects have been observed in a different topological
material, $MoP$ \cite{Kumar19}.

The precise microscopic nature of the proposed hydrodynamic state and
especially its relation to the hydrodynamic regime in graphene remains
unclear. An interesting proposal on the experimental measurement of
one of the relevant length scales, the ``momentum-relaxing'' length
(e.g., $\ell_{\rm dis}$), which together with the
``momentum-conserving'' length $\ell_{ee}$ determines whether the
sample is in the hydrodynamic, ballistic, or Ohmic regime, was
suggested in Ref.~\cite{moll21}. The authors used Sondheimer
oscillations \cite{sond} to extract $\ell_{\rm dis}$ even in the
ballistic case $\ell_{\rm dis}\gg L$ (where $L$ is the typical system
size) and suggested that this effect can be used as an effective
quantitative probe for identifying scattering processes in ultra-clean
materials.

\section{Electronic hydrodynamics}\label{sec3}

Hydrodynamic description of interacting particles (or excitations) has
long been part of the theoretical toolbox used (in addition to
traditional fluid mechanics \cite{dau6}) in a wide range of fields
including many-body theory \cite{kad}, superfluids \cite{wolf,akha},
quark-gluon plasma \cite{shu}, or interstellar matter \cite{sho}. The
underlying general idea allowing to develop the hydrodynamic theory
suitable to such different circumstances is the universality of the
long-time, long-wavelength behavior, i.e., the assumption that
macroscopic (long-distance) physics is independent of microscopic
details and is governed by symmetries, which can be expressed in
terms of continuity equations.

The most common symmetry assumed in physics is time translation
invariance leading to energy conservation. The corresponding continuity
equation reads
\begin{subequations}
\label{ce00}
\begin{equation}
\label{ece0}
\partial_t n_E + \bs{\nabla}\!\cdot\!\bs{j}_E=0,
\end{equation}
where $n_E$ is the energy density and $\bs{j}_E$ is the energy current.

The second conservation law typically assumed in the context of
electronic systems is the particle number (or charge) conservation
(manifesting gauge invariance) described by the continuity equation
\begin{equation}
\label{ce0}
\partial_t n + \bs{\nabla}\!\cdot\!\bs{j}=0.
\end{equation}
Here $n$ and $\bs{j}$ are the particle number and current densities
while the charge and electric current densities differ by a factor of
the electric charge, see also Eq.~(\ref{mc0}).

Supplementing equations (\ref{ece0}) and (\ref{ce0}) by the
thermodynamic equation of state and the entropy balance equation
\cite{chai} one may arrive at the macroscopic theory describing the
long-distance properties of the system and find the spectrum of the
collective modes. The resulting behavior is diffusive (i.e.,
equivalent to the standard Drude-like approach to electronic transport
\cite{ziman}).

In contrast, conventional fluids are additionally assumed to be
translationally invariant which implies momentum conservation described
by the continuity-like equation for the momentum density,
$\bs{n}_{\bs{k}}$,
\begin{equation}
\label{mce0}
\partial_t n^\alpha_{\bs{k}} + \nabla^{\beta} \Pi_E^{\alpha\beta} = 0.
\end{equation}
\end{subequations}
Here $\Pi_E^{\alpha\beta}$ is the momentum flux (or stress-energy)
tensor. Introducing momentum conservation has a drastic effect on the
collective modes of the system leading to the appearance of a mode
with the linear dispersion, i.e., the sound mode \cite{chai}. The
existence of the latter is the crucial distinction between
hydrodynamics and other macroscopic, long-wavelength theories
(although a more general interpretation of the term ``hydrodynamics''
is also used in literature, see, e.g., Ref.~\cite{chai}).

The explicit form of the hydrodynamic equations can be obtained by
supplementing the continuity equations (\ref{ce00}) by the so-called
``constitutive relations'' reducing the amount of independent
variables and turning Eqs.~(\ref{ce00}) into a closed set. This is
typically done under the assumption of local equilibrium
\cite{dau6}. Moreover, the form of the stress-energy tensor in the
moving fluid is often obtained by relating to the properties of the
stationary fluid (that are assumed to be known). To do that,
one needs to change the reference frame to the rest frame of the
fluid. Consequently, traditional hydrodynamics \cite{dau6}
distinguishes the two cases of Galilean- and Lorentz-invariant fluids,
i.e., the classical and relativistic hydrodynamics. While early
applications of the hydrodynamic approach to electronic transport were
based on the classical theory \cite{gurzhi,dsh,stein}, it is the
possibility of realization of relativistic hydrodynamics in graphene
that ignited the current interest in the field.

\subsection{Relativistic hydrodynamics in a solid state laboratory}\label{sec3.1}

The discovery of graphene and Dirac fermions in it \cite{geimqhe} has
provided a unique opportunity to study relativistic effects in a solid
state laboratory \cite{kats}. In particular, early work on collective
electronic flows attempted to adapt relativistic hydrodynamics in
$(2+1)$ dimensions to Dirac fermions in graphene \cite{har,job}.

\subsubsection{Ideal relativistic fluid}\label{sec3.1.1}

Standard equations of relativistic hydrodynamics \cite{dau6} are encoded
in the relation 
\begin{equation}
\label{rhe0}
\frac{\partial T_i^k}{\partial x^k} = 0,
\end{equation}
where $T^{ik}$ is the relativistic stress-energy tensor (in graphene,
this is a $3\times3$ tensor in the $(2+1)$-dimensional space-time)
\begin{subequations}
\label{tik00}
\begin{equation}
\label{tik0}
T^{ik} = w u^i u^k - p g^{ik},
\end{equation}
with $w$ and $p$ being the enthalpy and pressure, respectively, in the
local rest frame.

For the purposes of this review, it will be instructive to write down
the explicit form of the individual components of $T^{ik}$: the energy
density
\begin{equation}
\label{t00}
T^{00} = \frac{w}{1-u^2/v_g^2} - p,
\end{equation}
the momentum density (here we adopt the usual practice of denoting the
space components by Greek indices, while the Roman indices refer to
the space-time)
\begin{equation}
\label{t0a}
T^{0\alpha} = \frac{wu_\alpha}{v_g\left(1-u^2/v_g^2\right)},
\end{equation}
and finally the momentum flux density 
\begin{equation}
\label{tab}
T^{\alpha\beta} = \frac{wu_\alpha u_\beta}{v_g^2\left(1-u^2/v_g^2\right)}+p\delta_{\alpha\beta}.
\end{equation}
\end{subequations}
The energy flux density is proportional to the momentum density and is
given by $v_gT^{0\alpha}$. This fact will be explored in more detail below.

The relativistic generalization of the Euler equation \cite{euler} can
be obtained by projecting Eq.~(\ref{rhe0}) onto the direction
perpendicular to the $3$-velocity $u^i$ \cite{dau6}. This yields
\begin{subequations}
\label{releul}
\begin{equation}
\label{ree0}
\frac{w}{1-u^2/v_g^2} \left[\frac{\partial}{\partial t} + \bs{u}\!\cdot\!\bs{\nabla}\right]\bs{u}
+ v_g^2 \bs{\nabla}p + \bs{u} \frac{\partial p}{\partial t} = 0.
\end{equation}
Supplementing the Euler equation (\ref{ree0}) by the relativistic
continuity equation
\begin{equation}
\label{rce0}
\frac{\partial \left(nu^k\right)}{\partial x^k} = 0,
\end{equation}
and the thermodynamic equation of state
\begin{equation}
\label{est}
w=3p,
\end{equation}
one can quickly convince oneself that the ideal flow described by
Eq.~(\ref{ree0}) is isentropic
\begin{equation}
\label{rceen0}
\frac{\partial \left(su^k\right)}{\partial x^k} = 0.
\end{equation}
\end{subequations}
Equations (\ref{releul}) represent the closed set of hydrodynamic
equations describing an ideal (non-dissipative) flow of a
single-component relativistic fluid in a $(2+1)$-dimensional
space-time with the velocity $v_g$ playing the role of the speed of
light. This theory possesses a collective mode
\cite{luc,hydro1,schutt,cosmic,lev13,fog,svin,ldsp,ks20,fat,megt} with
the linear dispersion relation
\begin{equation}
\label{cs}
\omega = \frac{v_gq}{\sqrt{2}}.
\end{equation}
In the literature, this mode has been referred to as the ``cosmic
sound'' \cite{cosmic} or the ``second sound'' \cite{ks20}.

\subsubsection{Electronic fluid in graphene}\label{sec3.1.2}

The ideal hydrodynamic theory outlined in the previous Section can be
considered a purely phenomenological since it is based on an implicit
assumption of equilibrium in the local rest frame without discussing
the physical processes responsible for the equilibration. In the case
of graphene, that has to be electron-electron interaction, which is
the {\it classical, three-dimensional} Coulomb interaction. The latter
point refers to the fact that although graphene is atomically thin so
that the electron motion is restricted to two dimensions, the electric
field induced by the electron charges is not. The former point refers
to the orders of magnitude difference between the electron velocity
and the speed of light, $v_g\ll c$, preventing the above hydrodynamic
theory and electromagnetic fields to be transformed by the same
Lorentz transformation. This issue was addressed in detail in
Ref.~\cite{har}.

Coulomb interaction can be included in the hydrodynamic description by
re-writing the relativistic Euler equation (\ref{rhe0}) in the form 
\begin{equation}
\label{fj}
\frac{\partial T^k_i}{\partial x^k} = \frac{e}{c}F_{ik}j^k.
\end{equation}
Notice, that in the right-hand side of this equation one has to write
the speed of light, which is inconsistent with the use of the velocity
$v_g$ in the stress-energy tensor (\ref{tik00}). A possibility to
resolve this issue was suggested in Ref.~\cite{har}. Indeed,
redefining the electromagnetic field tensor $F_{ik}$ and the current
$j^k$ as
\begin{equation}
\label{fikv}
F_{ik}=
\begin{pmatrix}
0 & (c/v_g) E_x & (c/v_g) E_y \cr
-(c/v_g) E_x & 0 & -B \cr
-(c/v_g) E_y & B & 0
\end{pmatrix},
\end{equation}
\begin{equation}
\label{jiv}
j^k = 
\begin{pmatrix}
v_g n, & \bs{j}
\end{pmatrix},
\end{equation}
one may remove the inconsistency from Eq.~(\ref{fj}) turning it into
the standard form of the relativistic Euler equation. However, this is
only a partial solution since the redefined field tensor (\ref{fikv})
leaves only two Maxwell's equations intact,
\begin{equation}
\label{meqs}
\bs{\nabla}\times\bs{B} = 0, 
\qquad
\bs{\nabla}\!\cdot\!\bs{E} = 4\pi en,
\end{equation}
while the other two are violated leaving the above approach questionable.

Even if the modified equation (\ref{fj}) can be accepted for those
problems that do not involve the two violated Maxwell's equations
(e.g., a description of stationary currents), there are other issues
that prevent one from treating electronic flows in graphene as truly
relativistic. As already mentioned above, there are other scattering
processes in graphene (and in any other solid) affecting the behavior
of charge carriers. These may include electron-phonon and disorder
scattering, Auger processes, and three-particle collisions, none of
which are Lorentz-invariant. Moreover, typical currents studied in
present-day experiments are small enough, such that the hydrodynamic
velocity is small as well, $u\ll v_g$. As a result, one would be
interested in the non-relativistic limit of the hydrodynamic equation
(\ref{fj}) anyways. Now, the non-relativistic form of hydrodynamics
can also be derived within the kinetic theory approach (see the next
Section), where all of the above issues can be consistently
taken into account. In the absence of dissipative processes, the
generalized Euler equation for the hydrodynamic electronic flows in
graphene obtained from the kinetic theory does indeed closely resemble
Eq.~(\ref{ree0}), while containing additional terms taking into
account scattering processes that were not considered so far. In
addition, introducing dissipative processes within the
phenomenological approach involves defining new parameters, such as
electrical conductivity and viscosity, that can only be determined in
an experiment. While the kinetic theory provides a method to
``calculate'' these parameters, the accuracy of these calculations may
be limited depending on the initial assumptions allowing one to
formulate the kinetic equation in the first place. The form of the
dissipative corrections remains the same in both approaches providing
a useful checkpoint.

\subsection{Kinetic theory approach}\label{sec3.2}

Kinetic approach has been used to describe electronic transport in
solids for decades \cite{ziman}. While applicability of the kinetic
theory to quantum many-body systems remains an active area of research
\cite{aka}, it is often assumed that at least at high enough
temperatures electrons behave semiclassically such that the kinetic
theory is applicable. At the same time, this implies that
quasiparticle excitations are long-lived, the assumption that might
not be valid in strongly correlated or hydrodynamic regimes. Strictly
speaking, the kinetic equation can only be applicable in weakly
interacting electronic systems. This might be a problem in graphene,
where the effective coupling constant in an idealized model is
$\alpha_g=e^2/v_g\approx2.2$ (which may be reached in suspended
graphene) and while an insulating substrate may reduce this value (by
a factor of the dielectric constant), the resulting $\alpha_g$ is not
small (typically, $\alpha_g\approx0.2\div0.3$ \cite{gal,sav}).
Consequently, derivation of the hydrodynamic equations has to rely on
universality: one assumes that the form of the equations is
independent of the interaction strength (similarly to how the
Navier-Stokes equation derived from the kinetic theory of rarefied
gases \cite{dau10} can be used to describe properties of water, where
the kinetic equation is not applicable). Calculation of kinetic
coefficients then has to rely on the renormalization group procedure
\cite{shsch,me2} treating $\alpha_g$ as a running coupling constant
\cite{msf,paconpb,paco99,julia}. One renormalizes the theory to the
parameter regime, where the coupling constant is small, solves the
kinetic equations, and then renormalizes back to realistic parameter
values. For a more microscopic approach to deriving the hydrodynamic
equations based on the nonequilibrium Keldysh technique, see
Ref.~\cite{aa17}. This paper provides a proper microscopic treatment
of inelastic electron-electron scattering that is responsible for
establishing the local equilibrium that is the central assumption of
the kinetic approach discussed below.

\subsubsection{Quasiclassical Boltzmann equation}\label{sec3.2.1}

At high enough temperatures (where the hydrodynamic behavior is
observed \cite{geim1,geim2,geim3,geim4}), the quasiparticle spectrum
in monolayer graphene \cite{can} comprises two bands of carriers (the
``conductance'' and ``valence'' bands) that touch in the corners of
the hexagonal Brillouin zone, i.e., at the ``Dirac points'' (multilayer
graphene was discussed in Ref.~\cite{sim20}). In the vicinity of the
Dirac points the spectrum can be approximately considered to be linear
(logarithmic renormalization due to electron-electron interaction
\cite{paconpb}, see also Ref.~\cite{adam18}, is observed at much lower
temperatures \cite{vgeim1}). The linearity of the Dirac spectrum leads
to two important kinematic effects: (i) the suppression of Auger
processes \cite{alf,svin18} and hence approximate conservation of the
number of particles in each band independently \cite{rev,luc,alf,bri};
and (ii) the so-called ``collinear scattering singularity''
\cite{schutt,mus,hydro0,mfss,msf,bri,kash,drag2}. The former
represents an additional conservation law that is not taken into
account in the above phenomenological hydrodynamics. The latter is
justified by the smallness of the effective coupling constant and
allows for a nonperturbative solution of the Boltzmann equation
(recall that the Boltzmann approach itself is justified in the weak
coupling limit, $\alpha_g\to0$).

Consider now the two-band model of low-energy quasiparticles in
graphene. Within the kinetic approach, the quasiparticles can be
described by a distribution function, $f_{\lambda\bs{k}}$, where each
quasiparticle state is characterized by the band index (or chirality),
$\lambda=\pm1$, and 2D momentum, $\bs{k}$. The spectrum is assumed to
be linear,
\begin{subequations}
\label{eg}
\begin{equation}
\epsilon_{\lambda\bs{k}} = \lambda v_g k,
\end{equation}
with the straightforward relation between velocities and momenta,
\begin{equation}
\label{vg}
\bs{v}_{\lambda\bs{k}}=\lambda v_g \frac{\bs{k}}{k}, \qquad
\bs{k} = \frac{\lambda k}{v_g}\bs{v}_{\lambda\bs{k}}
=\frac{\epsilon_{\lambda\bs{k}}\bs{v}_{\lambda\bs{k}}}{v_g^2}.
\end{equation}
\end{subequations}
The distribution function satisfies the kinetic (Boltzmann) equation
\begin{subequations}
\label{be}
\begin{equation}
\label{ke}
{\cal L}f_{\lambda\bs{k}} = {\rm St}_{ee} [f_{\lambda\bs{k}}] + {\rm St}_{R} [f_{\lambda\bs{k}}] 
+ {\rm St}_{\rm dis} [f_{\lambda\bs{k}}],
\end{equation}
where the left-hand side (LHS) is defined by the Liouville's operator
\begin{equation}
\label{l}
{\cal L} = \partial_t + \bs{v}\!\cdot\!\bs{\nabla}_{\bs{r}} + 
\left(e\bs{E} \!+\! \frac{e}{c} \bs{v}\!\times\!\bs{B}\right)\!\cdot\!\bs{\nabla}_{\bs{k}},
\end{equation}
\end{subequations}
and the right-hand side (RHS) represents the collision integral.

In the simplest Golden-Rule-like approximation, different scattering
processes contribute to the collision integral in the additive
fashion; hence, the form of the RHS in Eq.~(\ref{ke}). In the
hydrodynamic regime, the electron-electron interaction (described by
${\rm St}_{ee}$) is the dominant scattering process responsible for
equilibration of the system. Consequently, {\it local equilibrium} is
described by the distribution function that nullifies ${\rm St}_{ee}$
\cite{dau10}
\begin{equation}
\label{le}
{\rm St}_{ee}\left[f^{(le)}_{\lambda\bs{k}}\right]=0,
\qquad
f^{(le)}_{\lambda\bs{k}} =
\left\{
1\!+\!\exp\left[\frac{\epsilon_{\lambda\bs{k}}\!-\!\mu_\lambda(\bs{r}) \!-\! 
\bs{u}(\bs{r})\!\cdot\!\bs{k}}{T(\bs{r})}\right]\!
\right\}^{\!-1}\!,
\end{equation}
where ${\mu_\lambda(\bs{r})}$ is the local chemical potential and
$\bs{u}(\bs{r})$ is the hydrodynamic (or ``drift'') velocity. The
local equilibrium distribution function (\ref{le}) allows for
independent chemical potentials in the two bands, which can be
expressed in terms of the ``thermodynamic'' and ``imbalance''
\cite{alf} chemical potentials
\begin{equation}
\label{mui}
\mu_\lambda = \mu + \lambda\mu_I.
\end{equation}
In global equilibrium (i.e., for stationary fluid)
\begin{equation}
\label{ff}
f^{(0)} = f^{(le)}_{\lambda\bs{k}}(\mu_I=0, \bs{u}=0).
\end{equation}

In addition, two more scattering processes need to be taken into
account. Even ultra-pure graphene samples contain some degree of
(weak) disorder. Scattering on impurities violates momentum
conservation leading to a weak decay term in the generalized Euler
equation \cite{rev,luc,me1,hydro1}. This process (as well as other
momentum-relaxing processes) is described in Eq.~(\ref{ke}) by
${\rm{St}}_{\rm dis}$. At the same time, electron-phonon interaction
may lead not only to the loss of electronic momentum (which is already
taken into account in ${\rm{St}}_{\rm dis}$), but also to the loss of
energy. Consequently, despite being subdominant in the hydrodynamic
regime the electron-phonon interaction should be taken into account as
one of the dissipative processes. However, due to the linearity of the
Dirac spectrum, lowest order scattering on acoustic phonons is
kinematically suppressed. Instead, it is a higher order process, the
so-called disorder-assisted electron-phonon scattering \cite{srl} or
``supercollisions'' \cite{ralph13,betz,tik18,kong}, that plays the
most important role in the hydrodynamic regime. Indeed,
supercollisions violate not only energy conservation, but also
conservation of the number of particles in each band. As a result,
continuity equations for energy and single-band particle numbers also
acquire weak decay terms. In the kinetic equation (\ref{ke}), these
effects are described by ${\rm{St}}_R$ (the subscript ``R'' here
stands for ``recombination'', see
Refs.~\cite{meg,mr1,mrexp,alf,mr2,prin2020,mr3,drag12}).

Within the kinetic theory, conservation laws are manifested in the sum
rules for the collision integrals. There are four conservation laws to
consider: energy, momentum, and particle number in the two bands. The
latter can be expressed in terms of the ``charge'' and ``total
quasiparticle'' (or imbalance) numbers similarly to Eq.~(\ref{mui})
\begin{equation}
\label{nui}
n_\lambda = \frac{1}{2} \left(\lambda n + n_I\right).
\end{equation}
The continuity equation (\ref{ce0}) representing global charge
conservation can be obtained by summing the kinetic equation
(\ref{ke}) over all quasiparticle states. During this procedure, all
three collision integrals in Eq.~(\ref{ke}) vanish \cite{dau10}
\begin{subequations}
\label{syms}
\begin{equation}
\label{gsym}
N\!\sum_\lambda\!\int\!\!\frac{d^2k}{(2\pi)^2} {\rm St}_{ee} [f_{\lambda\bs{k}}]
=
N\!\sum_\lambda\!\int\!\!\frac{d^2k}{(2\pi)^2} {\rm St}_{R} [f_{\lambda\bs{k}}]
=
N\!\sum_\lambda\!\int\!\!\frac{d^2k}{(2\pi)^2} {\rm St}_{\rm dis} [f_{\lambda\bs{k}}]
=0.
\end{equation}
Moreover, electron-electron and disorder scattering also conserve the
number of particles in each band, such that 
\begin{equation}
\label{imsym1}
N\sum_\lambda\!\int\!\!\frac{d^2k}{(2\pi)^2} \lambda {\rm St}_{ee} [f_{\lambda\bs{k}}]
=
N\sum_\lambda\!\int\!\!\frac{d^2k}{(2\pi)^2} \lambda {\rm St}_{\rm dis} [f_{\lambda\bs{k}}]
=0,
\end{equation}
whereas supercollisions lead to a decay term in the continuity
equation for the imbalance density
\begin{eqnarray}
\label{cephir}
N\sum_\lambda \!\int\!\!\frac{d^2k}{(2\pi)^2} \lambda \, {\rm St}_{R} [f_{\lambda\bs{k}}] 
\approx
-
\mu_I n_{I,0} \lambda_Q
\approx
-
\frac{n_I\!-\!n_{I,0}}{\tau_R}.
\end{eqnarray}
Here $n_{I,0}$ is the imbalance density at global equilibrium, see
Eq.~(\ref{ff}), i.e., for $\mu_I=0$ and $\bs{u}=0$. The first equality
in Eq.~(\ref{cephir}) was suggested in Ref.~\cite{alf} and serves as
the definition of the dimensionless coefficient $\lambda_Q$, while the
second (valid to the leading order) was suggested in
Refs.~\cite{mr1,me1} and offers the definition of the ``recombination
time'' $\tau_R$ (see also Ref.~\cite{prin2020}).  The two expressions
are equivalent since ${n_I\!-\!n_{I,0}\propto\mu_I}$.

Similarly, both electron-electron and disorder scattering conserve
energy, hence the corresponding collision integrals vanish upon
summation over all quasiparticle states with an extra factor of energy
\begin{equation}
\label{ciee}
N\sum_\lambda\!\int\!\!\frac{d^2k}{(2\pi)^2} \epsilon_{\lambda\bs{k}} {\rm St}_{ee} [f_{\lambda\bs{k}}] 
=
N\sum_\lambda\!\int\!\!\frac{d^2k}{(2\pi)^2} \epsilon_{\lambda\bs{k}} {\rm St}_{\rm dis} [f_{\lambda\bs{k}}] 
=0.
\end{equation}
Integrating the ``recombination'' collision integral one finds \cite{meig1}
\begin{equation}
\label{cepher}
N\sum_{\bs{k}}\epsilon_{\lambda\bs{k}} {\rm St}_{R} [f_{\lambda\bs{k}}]
=
-
\mu_I n_{E,0} \lambda_{QE}
\approx
-
\frac{n_E\!-\!n_{E,0}}{\tau_{RE}}.
\end{equation}
The equivalence of the two forms of the decay term stems from
${n_E\!-\!n_{E,0}\propto\mu_I}$ assuming the electrons and holes are
characterized by the same temperature.

Supercollisions contribute differently to recombination and energy
relaxation. Recombination typically implies scattering between the
quasiparticle states in different bands only. At the same time,
supercollisions may also take place within a single band \cite{srl}.
This process does not affect the number of particles in the band, but
is accompanied by the energy loss as the electron scatters from a
higher energy state into a lower energy state (losing its momentum to
the impurity). Consequently, this process provides an additional
contribution to energy relaxation. Thus, the time scales $\tau_R$ and
$\tau_{RE}$ should be quantitatively different, although of the same
order of magnitude (at least at charge neutrality and in the
hydrodynamics regime).

Now, other processes may contribute to $\tau_R$ and $\tau_{RE}$,
including direct electron-phonon scattering
\cite{meg,alf,hydro0,drag2,srl,bistr,tse}, scattering on optical
phonons \cite{fos16,lev19}, three-particle collisions
\cite{luc,lev19}, and Auger processes \cite{rev,luc,alf,bri}. Taking
into account these effects does not change the functional form of the
continuity equations leaving the integrated collision integrals
(\ref{cephir}) and (\ref{cepher}) intact, but may affect the
theoretical estimates of the values of $\tau_R$ and $\tau_{RE}$ (see
Refs.~\cite{srl,meig1}). Given the approximate nature of such
calculations, one may treat these parameters as phenomenological
taking into account all relevant scattering processes.

Finally, electron-electron interaction conserves momentum and hence
\begin{equation}
\label{ksym1}
N\sum_\lambda\!\int\!\!\frac{d^2k}{(2\pi)^2} \bs{k} \, {\rm St}_{ee} [f_{\lambda\bs{k}}]
= 0.
\end{equation}
On the other hand, weak disorder scattering leads to a weak decay term
that should be included in Eq.~(\ref{mce0}). Within the simplest
$\tau$-approximation \cite{dau10,me1} 
\begin{equation}
\label{mdis}
N\sum_\lambda\!\int\!\!\frac{d^2k}{(2\pi)^2} \bs{k} \, {\rm St}_{\rm dis} [f_{\lambda\bs{k}}] 
= \frac{\bs{n}_{\bs{k}}}{\tau_{\rm dis}}.
\end{equation}
\end{subequations}
The remaining collision integral ${\rm{St}}_R$ also does not conserve
momentum, but given the phenomenological nature of $\tau_{\rm dis}$
\cite{gal} (a better version of the disorder collision integral in
graphene should involve the Dirac factors suppressing backscattering
\cite{kash18} which would lead to the similar approximation but with
the transport scattering time, which in graphene differs by a factor
of 2), the contribution of the next-order supercollisions (involving
both disorder and phonons) may be considered to be included in
$\tau_{\rm{dis}}$ (similarly to the above discussion of $\tau_R$ and
$\tau_{RE}$).

\subsubsection{Continuity equations in graphene}\label{sec3.2.2}

Using the above properties of the collision integrals, one can easily
obtain the continuity equations in graphene \cite{rev,luc,me1} by
integrating the kinetic equation (\ref{ke}). In comparison to the
``phenomenological'' continuity equations (\ref{ce00}), the resulting
equation will contain extra terms due to the weak decay processes
(discussed in the previous Section) and external electromagnetic
fields. Hence the only true symmetry of the electronic fluid in a
solid is gauge invariance that manifests itself by means of the
continuity equation (\ref{ce0})
\begin{subequations}
\label{ces}
\begin{equation}
\label{ce}
\partial_t n + \bs{\nabla}\!\cdot\!\bs{j}=0,
\end{equation}
where the kinetic definitions of the ``charge'' density and current
are [cf. Eq.~(\ref{nui})]
\begin{equation}
\label{n}
n = n_+ - n_-, \qquad
n_+=N\!\int\!\!\frac{d^2k}{(2\pi)^2} f_{+,\bs{k}}, \qquad
n_-=N\!\int\!\!\frac{d^2k}{(2\pi)^2} \left(1-f_{-,\bs{k}}\right),
\end{equation}
and [cf. Eq.~(\ref{mc0})]
\begin{equation}
\label{j}
\bs{j} = \bs{j}_+ - \bs{j}_- = 
N\!\int\!\!\frac{d^2k}{(2\pi)^2} 
\left[\bs{v}_{+,\bs{k}} f_{+,\bs{k}} - \bs{v}_{-,\bs{k}}\left(1\!-\!f_{-,\bs{k}}\right)\right].
\end{equation}

In the two-band model of graphene, the number of particles in each band
is approximately conserved (see above). Hence, in addition to
Eq.~(\ref{ce}), one finds a continuity equation for the ``imbalance
density'', see Eq.~(\ref{nui}),
\begin{equation}
\label{cei}
\partial_t n_I + \bs{\nabla}\!\cdot\!\bs{j}_I = - \frac{n_I\!-\!n_{I,0}}{\tau_R},
\end{equation}
where
\begin{equation}
\label{ni}
n_I = n_+ - n_-, \qquad
\bs{j} = \bs{j}_+ + \bs{j}_-,
\end{equation}
and the RHS in Eq.~(\ref{cei}) comes from integrating the collision
integral, see Eq.~(\ref{cephir}).

The continuity equation for the energy density is obtained by
multiplying the kinetic equation (\ref{ke}) by
$\epsilon_{\lambda\bs{k}}$ and summing over all quasiparticle
states,
\begin{equation}
\label{ece}
\partial_t n_E + \bs{\nabla}\!\cdot\!\bs{j}_E
=
e\bs{E}\!\cdot\!\bs{j} -
\frac{n_E\!-\!n_{E,0}}{\tau_{RE}},
\end{equation}
where $n_E$ and $\bs{j}_E$ are defined as
\begin{equation}
\label{ne}
n_E = N\sum_\lambda\!\int\!\!\frac{d^2k}{(2\pi)^2} \epsilon_{\lambda\bs{k}} f_{\lambda\bs{k}}
\end{equation}
and
\begin{equation}
\label{nk}
\bs{j}_E = Nv_g^{2}\sum_\lambda\!\int\!\!\frac{d^2k}{(2\pi)^2} \bs{k} f_{\lambda\bs{k}}
=
v_g^{2} \bs{n}_{\bs{k}}.
\end{equation}
The last equality represents the fact that in graphene the momentum
density is proportional to the energy density [due to the properties
  of the Dirac spectrum Eq.~(\ref{eg})]. The two terms in the RHS in
Eq.~(\ref{ece}) come from the Lorentz term in the Liouville's operator
(\ref{l}) and the integrated collision integral, see
Eq.~(\ref{cepher}). The former physically represents Joule's heat.

Finally, the continuity equation representing momentum conservation is
obtained by multiplying the kinetic equation (\ref{ke}) by $\bs{k}$
and summing over all states. In contrast to the ``phenomenological''
equation (\ref{mce0}), the resulting equations contains extra terms
stemming from the effect of the electromagnetic field and weak
disorder (\ref{mdis})
\begin{equation}
\label{mce}
\partial_t n_{\bs{k}}^\alpha + \bs{\nabla}^\beta\Pi_E^{\alpha\beta} - enE^\alpha
-\frac{e}{c}\left[\bs{j}\!\times\!\bs{B}\right]^\alpha 
= - \frac{n_{\bs{k}}^\alpha}{\tau_{\rm dis}}.
\end{equation}
Here $\bs{n}_{\bs{k}}$ is given by Eq.~(\ref{nk}) and the momentum
flux tensor is defined as
\begin{equation}
\label{pie}
\Pi_E^{\alpha\beta} = N\sum_\lambda\!\int\!\!\frac{d^2k}{(2\pi)^2} 
k^\alpha v_{\lambda\bs{k}}^\beta f_{\lambda\bs{k}}.
\end{equation}
\end{subequations}

\subsubsection{Constitutive relations}\label{sec3.2.3}

Continuity equations represent the global conservation laws and are
valid without any further assumptions. Hydrodynamics, however, assumes
that the set of continuity equations can be closed by expressing the
vector and tensor quantities (i.e., the currents and stress-energy
tensor) in terms of the ``velocity field'' $\bs{u}(\bs{r})$. Such
expressions are known as ``constitutive relations''.
Phenomenologically, they can be derived using the Galilean or (in the
relativistic case) Lorentz invariance \cite{dau6}. However, neither is
valid for Dirac fermions in graphene (the former due to the linear
spectrum and the latter due to the classical nature of the Coulomb
interaction, see Sec.~\ref{sec3.1.2}). Instead, one can derive the
constitutive relations from the kinetic theory under the assumption of
local equilibrium \cite{rev,me1}. Indeed, substituting the local
equilibrium distribution function into the definitions of the three
currents (\ref{j}), (\ref{ni}), and (\ref{nk}) yields the expected
relations
\begin{subequations}
\label{crs0}
\begin{equation}
\label{js0}
\bs{j} = n\bs{u}, \qquad \bs{j}_I = n_I\bs{u}, \qquad 
\bs{j}_E = {\cal W}\bs{u},
\end{equation}
where ${\cal W}$ is the enthalpy density. This thermodynamic quantity
can also be evaluated using the local equilibrium distribution
function, which yields the ``equation of state''
\begin{equation}
\label{eqst}
{\cal W} = n_E + P = \frac{3n_E}{2\!+\!u^2/v_g^2},
\end{equation}
where $P$ is the thermodynamic pressure. Both of these quantities
appear in the explicit expression of the momentum flux tensor
\begin{equation}
\label{pi0}
\Pi_E^{\alpha\beta} = P \delta^{\alpha\beta} + \frac{\cal W}{v_g^2}u^\alpha u^\beta.
\end{equation}
\end{subequations}

Combining Eqs.~(\ref{crs0}) with the continuity equation for momentum
density (\ref{mce}), one may generalize the Euler equation
\cite{euler} to Dirac quasiparticles in graphene
\begin{equation}
\label{euler}
{\cal W}(\partial_t+\bs{u}\!\cdot\!\bs{\nabla})\bs{u}
+
v_g^2 \bs{\nabla} P
+
\bs{u} \partial_t P 
+
e(\bs{E}\!\cdot\!\bs{j})\bs{u} 
=
v_g^2 
\left[
en\bs{E}
+
\frac{e}{c} \bs{j}\!\times\!\bs{B}
\right]
-
\frac{{\cal W}\bs{u}}{\tau_{{\rm dis}}}.
\end{equation}
It is instructive to compare Eq.~(\ref{euler}) to the relativistic
version of the Euler equation, Eq.~(\ref{ree0}). Formally, the first
three terms in the LHS of Eq.~(\ref{euler}) coincide with the three
terms of Eq.~(\ref{ree0}). The rest of the terms -- the Joule's heat,
Lorentz force, and weak decay due to disorder -- have not been
considered in the relativistic theory and are explicitly not
Lorentz-invariant. Even though the first three terms in
Eq.~(\ref{euler}) have the same form as Eq.~(\ref{ree0}), there is a
subtle difference: the pressure $p$ in Eq.~(\ref{ree0}) is the
thermodynamic pressure in the local rest frame, while $P$ in
Eq.~(\ref{euler}) is the pressure in the laboratory frame. The latter
is evaluated with the distribution function (\ref{le}) and hence is a
function of the velocity $\bs{u}$, while $p=P(\bs{u}=0)$. This point
is the only difference between the relativistic equation of state
(\ref{est}) and Eq.~(\ref{eqst}) as well.

The generalized Euler equation (\ref{euler}) together with the
continuity equations (\ref{ce}), (\ref{cei}), and (\ref{ece}) describe
the ``ideal'' flow of the electronic fluid. In conventional
hydrodynamics ``ideal'' means ``in the absence of dissipation'', which
is not quite the case here, since weak disorder scattering,
quasiparticle recombination, and energy relaxation are already taken
into account. However, none of these processes are due to
electron-electron interaction and hence are absent in the conventional
theory \cite{dau6}.

\subsubsection{Dissipative corrections}\label{sec3.2.4}

In its simplest form, conventional hydrodynamics \cite{dau6,dau10}
considers a system of particles (atoms, molecules, etc.) with the
contact (short-range) interaction, such that individual scattering
processes are almost literally ``collisions''. These collisions
represent the physical process responsible for equilibration: if the
system is driven out of equilibrium, they tend to restore it. In the
process the system is bound to lose energy, hence the collisions are
responsible for dissipation.

In graphene (and other solids, see below), the situation is slightly
more involved, but the main idea remain the same -- physical processes
responsible for equilibration lead to dissipation that is described by
``kinetic coefficients''. This can be described as follows
\cite{dau6,dau10}. Nonequilibrium states are characterized by nonzero
macroscopic current. In the process of equilibration the currents
relax (their values are being reduced towards zero). Hence, the
quasiparticle currents (\ref{js0}) acquire additional terms -- the
dissipative corrections \cite{rev,luc,me1,megt}
\begin{subequations}
\label{js}
\begin{equation}
\bs{j} = n\bs{u} + \delta\bs{j}, \qquad 
\bs{j}_I = n_I\bs{u} + \delta\bs{j}_I.
\end{equation}
In the absence of magnetic field, the dissipative corrections are
related to external bias by means of a ``conductivity matrix''
\cite{me1,alf,lev19}
\begin{equation}
\label{djs}
\begin{pmatrix}
\delta\bs{j} \cr
\delta\bs{j}_I
\end{pmatrix}
=\widehat\Sigma
\begin{pmatrix}
e\bs{E} - T\bs{\nabla}(\mu/T) \cr
-T\bs{\nabla}(\mu_I/T)
\end{pmatrix}.
\end{equation}
At charge neutrality ${\mu=\mu_I=0}$ the matrix $\widehat\Sigma$ is
diagonal. In the absence of disorder, the upper diagonal element
defines the ``quantum'' or ``intrinsic'' conductivity
\cite{rev,luc,alf,me1,lev19}
\begin{equation}
\label{squ}
\sigma_Q = e^2 \Sigma_{11}(0).
\end{equation}
\end{subequations}
The third current $\bs{j}_E$ does not acquire a dissipative
correction since it is proportional to the momentum density, see
Eq.~(\ref{nk}), and electron-electron interaction conserves momentum.
This point represents the key difference between electronic
hydrodynamics in graphene (or any semimetal with linear spectrum) from
conventional fluid mechanics of systems with parabolic
(Galilean-invariant) spectrum. In the latter case, it is the particle
number (or mass) current $\bs{j}$ that is proportional to the momentum
density. As a result, the energy current gets a dissipative correction
described by the thermal conductivity $\varkappa$ that is determined
by interparticle collisions. In the hydrodynamic theory of graphene,
the role that is equivalent to that of $\varkappa$ is played by the
elements of the matrix $\widehat\Sigma$. The matrix nature of
$\widehat\Sigma$ reflects the band structure of graphene. In the case
of strong recombination, the imbalance mode becomes irrelevant and one
is left with the single dissipative coefficient $\sigma_Q$, see
Ref.~\cite{luc}. Now, the thermal conductivity in graphene arises
purely due to weak disorder scattering that is already taken into
account in the Euler equation (\ref{euler}). This is the reason for
the strong violation of the Wiedemann-Franz law in neutral graphene,
see Sec.~\ref{sec2.5.1}.

The kinetic coefficients $\widehat\Sigma$ can be found by solving the
kinetic equation (\ref{ke}) perturbatively using the standard
procedure \cite{dau10,luc,hydro1,har}. In a bulk system and in the
absence of magnetic field, this calculation was performed in detail in
Ref.~\cite{me1}, where a $3\!\times\!3$ matrix was considered [i.e.,
  adding the energy current and its relaxation due to weak disorder to
  Eq.~(\ref{djs})]. The following $2\!\times\!2$ matrix was introduced
in Ref.~\cite{megt}. In both cases, one expresses the matrix
$\widehat\Sigma$ as a linear combination of the interaction and 
disorder contributions
\begin{subequations}
\label{sigma3}
\begin{equation}
\label{sigmam}
\widehat\Sigma=
\widehat{\textswab{M}}\,
\widehat{\textswab{S}}_{xx}^{-1}
\widehat{\textswab{M}},
\qquad
\widehat{\textswab{S}}_{xx} = \frac{\alpha_g^2T^2}{2{\cal T}^2}
\widehat{\textswab{T}}
+
\frac{\pi}{{\cal T}\tau_{\rm dis}}
\widehat{\textswab{M}},
\end{equation}
where (following the $2\!\times\!2$ notation)
\begin{equation}
\label{mh}
\widehat{\textswab{M}}\!=\!
\begin{pmatrix}
1\!-\!\frac{2\tilde{n}^2}{3\tilde{n}_E}\frac{T}{\cal T} &
\frac{xT}{\cal T} \!-\! \frac{2\tilde{n}\tilde{n}_I}{3\tilde{n}_E}\!\frac{T}{\cal T}\cr
\frac{xT}{\cal T} \!-\!
\frac{2\tilde{n}\tilde{n}_I}{3\tilde{n}_E}\frac{T}{\cal T} & 
1\!-\!\frac{2\tilde{n}_I^2}{3\tilde{n}_E}\frac{T}{\cal T}
\end{pmatrix}\!,
\end{equation}
with dimensionless densities (in self-evident notation; $Li_n(z)$ is the polylogarithm)
\begin{eqnarray}
\label{tiln}
&&
\tilde{n} = {\rm Li}_2\left(-e^{-x}\right) \!-\! {\rm Li}_2\left(-e^x\right)\!,
\;\;
\tilde{n}_I=\frac{x^2}{2}\!+\!\frac{\pi^2}{6},
\;\;
\tilde{n}_E = -{\rm Li}_3\left(-e^x\right) 
\!-\! {\rm Li}_3\left(-e^{-x}\right)\!,
\nonumber\\
&&
\nonumber\\
&&
x=\mu/T,
\quad
{\cal T} = 2T\ln\left[2\cosh(x/2)\right],
\end{eqnarray}
and dimensionless scattering rates
\begin{equation}
\label{taum}
\widehat{\textswab{T}}
=
\begin{pmatrix}
  t_{11}^{-1} & t_{12}^{-1}  \cr
  t_{12}^{-1} & t_{22}^{-1}
\end{pmatrix},
\qquad
t_{ij}^{-1}=\frac{8\pi{\cal T}}{\alpha_g^2NT^2} 
 \tau_{ij}^{-1}.
\end{equation}
\end{subequations}
Here $\tau_{ij}^{-1}$ represent the integrated collision integral
appearing while solving the kinetic equation within the three-mode
approximation \cite{hydro1,me1,mfss,ks20}. The fact that the collision
integrals can be represented by the effective scattering rates
$\tau_{ij}^{-1}$ is not equivalent to the simplest $\tau$
approximation that was employed above for the collision integrals
${\rm St}_{\rm dis}$ and ${\rm St}_R$. Instead, this is simply a
manifestation of the dimensionality of a collision integral (that is
inverse time).

Numerical values of the scattering rates (\ref{taum}) were
discussed in Ref.~\cite{me3}. In particular, at charge neutrality the
off-diagonal elements vanish, $t_{12}^{-1}(0)=0$. The diagonal element
$t_{11}^{-1}(0)$ determines the ``intrinsic'' or ``quantum''
conductivity matrix, $\sigma_Q$. For small $x\ll1$ the dimensionless
``scattering rates'' $t_{ij}$ have the form \cite{me3}
\begin{subequations}
\label{tausdp}
\begin{equation}
\label{tau11dp}
\frac{1}{t_{11}} =\frac{1}{t_{11}^{(0)}} 
+ x^2\!\left(\frac{1}{t_{11}^{(2)}}\!-\!\frac{1}{8\ln2}\frac{1}{t_{11}^{(0)}}\right)\! 
+ {\cal O}(x^3),
\end{equation}
\begin{equation}
\label{tau12dp}
\frac{1}{t_{12}} = \frac{x}{t_{12}^{(1)}} + {\cal O}(x^3),
\end{equation}
\begin{equation}
\label{tau22dp}
\frac{1}{t_{22}} = \frac{1}{t_{22}^{(0)}} 
+ x^2\!\left(\frac{1}{t_{22}^{(2)}}\!-\!\frac{1}{8\ln2}\frac{1}{t_{22}^{(0)}}\right)\! 
+ {\cal O}(x^3).
\end{equation}
\end{subequations}
For unscreened Coulomb interaction, the dimensionless quantities
$t_{ij}^{(0,1,2)}$ are independent on any physical
parameter. Numerically, one finds the values \cite{megt} (neglecting the
small exchange contribution \cite{kash}):
\[
\left(t_{11}^{(0)}\right)^{-1} \approx 34.63,
\quad
\left(t_{11}^{(2)}\right)^{-1} \approx 5.45,
\]
\[
\left(t_{12}^{(1)}\right)^{-1} \approx 5.72,
\quad
\left(t_{22}^{(0)}\right)^{-1} \approx 19.73,
\quad
\left(t_{22}^{(2)}\right)^{-1} \approx 5.65.
\]
In the case of screened interaction, the quantities $t_{ij}^{(0,1,2)}$
depend on the screening length.

The above values for the effective scattering rates yield the
following value for the intrinsic conductivity
\begin{equation}
\label{sq}
\sigma_Q = {\cal A}e^2/\alpha_g^2,
\qquad
{\cal A}\approx0.12.
\end{equation}
The quantity $\sigma_Q$ was studied by multiple authors
\cite{luc,me1,hydro1,schutt,mus,har,mfs,mfss,kash} and is a
temperature-dependent constant. This temperature dependence appears
due to the logarithmic renormalization of the coupling constant
$\alpha_g$ \cite{shsch}.

The above theoretical values can be related to the experimental data
of Ref.~\cite{gal}. Using the value of the coupling constant
$\alpha_g\approx0.23$ that is consistent with measurements at charge
neutrality, the dimensionfull scattering rates at a typical
temperature $T=267\,$K have the following values 
\[
\tau_{11}^{-1}\approx 7.35 \, {\rm THz},
\qquad
\tau_{22}^{-1}\approx 4.17 \, {\rm THz}.
\]
The disorder scattering rate at $T=267\,$K can be estimated as
\[
\tau_{\rm dis}^{-1} \approx 0.8  \, {\rm THz}.
\]

In the opposite limit of strongly doped graphene, $x\gg1$, all
elements of the matrix (\ref{taum}) coincide approaching the value
\cite{me1,megt,me3}
\begin{subequations}
\label{tijfl}
\begin{equation}
t^{-1}_{ij}(\mu\gg T) \rightarrow \frac{8\pi^2}{3}.
\end{equation}
The reason for this is the exponentially small contribution of the
second band in which case the two currents $\bs{j}$ and $\bs{j}_I$
coincide. In this limit, the corresponding dimensionfull rate vanishes
\begin{equation}
\tau_{11}^{-1}\approx \frac{\pi N \alpha_g^2 T^2}{3 \mu},
\end{equation}
\end{subequations}
leading to the vanishing dissipative corrections to the quasiparticle
currents
\begin{equation}
\label{djsfl}
\delta\bs{j} = \delta\bs{j}_I \rightarrow 0.
\end{equation}
As a result, electric current has the hydrodynamic form (\ref{ju})
leading to the use of the hydrodynamic approach to electronic
transport in doped graphene, both theoretically
\cite{fl2,pol16,fl0,fl1,fl18} and experimentally
\cite{geim1,geim3,geim4,imm}

In the presence of magnetic field or in confined geometries the
dissipative corrections to quasiparticle currents are more
complicated. External magnetic field entangles all three modes and
hence the corrections to quasiparticle currents acquire a dependence
on the hydrodynamic velocity $\bs{u}$ \cite{me1}. In confined
geometries, the coordinate dependence of the distribution function
becomes important and as a result the dissipative corrections
(\ref{js}) become non-uniform \cite{megt2}. In that case, the usual
local conductivity may become poorly defined \cite{zel,megt2}, but the
issue remains insufficiently explored.

\subsubsection{Electronic viscosity}\label{sec3.2.5}

Dissipative processes also contribute a correction to the momentum
flux (or stress-energy) tensor (\ref{pie}). In the non-relativistic
limit, one writes the dissipative correction to $\Pi_E^{\alpha\beta}$
[here $\Pi_{E,0}^{\alpha\beta}$ denotes the tensor given in
  Eq.~(\ref{pi0})]
\begin{subequations}
\label{dpdef}
\begin{equation}
\Pi_{E}^{\alpha\beta}=\Pi_{E,0}^{\alpha\beta}+\delta\Pi_{E}^{\alpha\beta},
\end{equation}
to the leading order in gradient expansion as
\begin{equation}
\label{etadef}
\delta\Pi_{E}^{\alpha\beta} = \eta^{\alpha\beta\gamma\delta} \nabla^\gamma u^\delta,
\end{equation}
where $\eta^{\alpha\beta\gamma\delta}$ is the rank-four viscosity
tensor \cite{dau6}. In a fully rotationally-invariant system the
explicit form of the viscosity tensor is dictated by symmetry and in
2D is given by
\begin{equation}
\label{etasim}
\eta^{\alpha\beta\gamma\delta} = 
\eta 
\left(\delta^{\alpha\gamma}\delta^{\beta\delta}
+
\delta^{\alpha\delta}\delta^{\beta\gamma}\right) + (\zeta-\eta) \delta^{\alpha\beta}\delta^{\gamma\delta},
\end{equation}
where $\eta$ and $\zeta$ are the shear and bulk viscosity,
respectively.

In graphene, the bulk viscosity vanishes, at least to the leading
approximation \cite{rev,luc,pol17,hydro1,mfs}, similarly to the
situation in ultrarelativistic systems \cite{dau10,kha} and Fermi
liquids \cite{akha,sykes} (although it may appear in disordered
systems in magnetic field \cite{bur21}). As a result, the leading term
of the gradient expansion of the dissipative stress tensor has the
form \cite{dau6,me2,me1}
\begin{equation}
\label{visdefB0}
\delta\Pi^{\alpha\beta}_E=-\eta\textswab{D}^{\alpha\beta},
\end{equation}
where
\begin{equation}
\textswab{D}^{\alpha\beta}=
\nabla^\alpha u^\beta+\nabla^\beta u^\alpha-\delta^{\alpha\beta}\bs{\nabla}\!\cdot\!\bs{u}.
\end{equation}
In the presence of magnetic field, the shear viscosity acquires a
field dependence \cite{ale,moo,stein} and the correction to the stress
tensor gains an additional contribution
\begin{equation}
\label{visdefB}
\delta\Pi^{\alpha\beta}_E=-\eta(B)\textswab{D}^{\alpha\beta}
+\eta_H(B) \epsilon^{\alpha ij} \textswab{D}^{i\beta} e_B^j,
\end{equation}
\end{subequations}
where ${\bs{e}_B=\bs{B}/B}$ and $\eta_H(B)$ is the Hall
\cite{ale,moo,pol17,me1,stein,read,brad,grom,ady} viscosity. While the
sign of $\eta$ is fixed by thermodynamics \cite{dau6,dau10}, the sign
of $\eta_H$ is not. Equation (\ref{visdefB}) follows
Ref.~\cite{geim4}: the Hall viscosity is positive for electrons
\cite{me1} (and negative for holes).

Electronic viscosity can be calculated in two different ways. As a
linear response function relating stress to strain \cite{read,julia1},
the viscosity tensor can be found using a Kubo formula
\cite{read,julia1,poli16} (that can be related to the usual Kubo
formula for conductivity \cite{read}). Such calculations are mostly
perturbative and were used to evaluate viscosity in strongly doped
graphene \cite{poli16} and in the high-frequency (collisionless)
regime \cite{julia1}, as well as in disordered 2D electron systems
beyond the hydrodynamic regime \cite{bur19}. A further extension of
this approach yields higher order corrections, such as ``drag
viscosity'' \cite{gal20} (by analogy to Coulomb drag
\cite{dragrev}). Alternatively, one can proceed with the solution of the
kinetic equation (\ref{ke}) following the standard procedure
\cite{dau10,me1,hydro1}. For arbitrary carrier density this yields a
somewhat cumbersome expression that can only be analyzed numerically
\cite{me2}, but simplifies in the limiting cases of neutral and
strongly doped graphene.

At charge neutrality and in the absence of magnetic field, the only
energy scale in the problem is the temperature $T$ and hence the shear
viscosity has the form \cite{msf}
\begin{equation}
\label{eta0}
\eta(\mu\!=\!0, B\!=\!0) = {\cal B}\frac{T^2}{\alpha_g^2v_g^2}.
\end{equation}
The coefficient ${\cal B}$ has been evaluated in Ref.~\cite{msf} to
have the value ${\cal B}\approx0.45$. This result was later confirmed
in Ref.~\cite{me1}. In both cases, the numerical value was obtained
with the simplest model of unscreened Coulomb interaction, which is
valid for small $\alpha_g$, i.e., in the regime of formal validity of
the kinetic approach (as well as the three-mode approximation allowing
for nonperturbative results). At realistic parameter values one has
to supplement kinetic calculations by the renormalization group (RG)
approach treating $\alpha_g$ as a running coupling constant
\cite{shsch,msf,paconpb,paco99,julia}. However, the product
$\alpha_gv_g$ remains constant along the RG flow \cite{msf,kash}, such
that Eq.~(\ref{eta0}) represents the correct form of shear viscosity
in graphene at low temperatures and $B=0$ \cite{shsch}.

Experimentally, a measurement of the shear viscosity is nontrivial
\cite{corb}. However, nonlocal resistance measurements \cite{geim1}
yield an estimate of a related quantity, the kinematic viscosity, see
Eq.~(\ref{nseqfl}). In graphene, the kinematic viscosity is defined as
\begin{equation}
\label{nudef}
\nu = \frac{v_g^2\eta}{\cal W}.
\end{equation}
The appearance of the enthalpy density in this definition is a
manifestation of the fact that the hydrodynamic flow in graphene is
the energy flow, see Eq.~(\ref{nk}). At charge neutrality, the
kinematic viscosity is determined by the ratio of the velocity and
coupling constant rather than their product \cite{me2}
\begin{equation}
\label{nu0}
\nu(\mu\!=\!0, B\!=\!0) \propto \frac{v_g^2}{\alpha_g^2T},
\end{equation}
and hence is renormalized along the RG flow. In doped graphene, the
dominant temperature dependence of the kinematic viscosity can be
estimated as \cite{me2}
\begin{equation}
\label{nufl}
\nu(\mu\gg1, B\!=\!0) \propto \frac{v_g^2\mu}{\alpha_g^2T^2}\frac{1}{1\!+\!T^2/\mu^2}.
\end{equation}
This expression disregards additional temperature dependence arising
from the RG and extra logarithmic factors \cite{luc,me2,poli16}.

\begin{figure}[t]
\centerline{
\includegraphics[width=0.5\columnwidth]{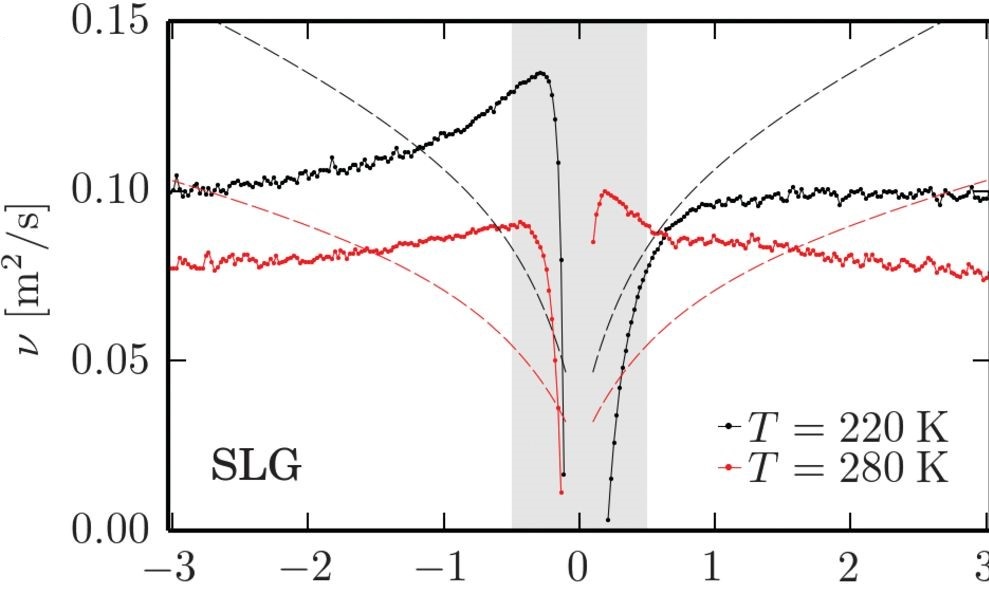}
\quad
\includegraphics[width=0.45\columnwidth]{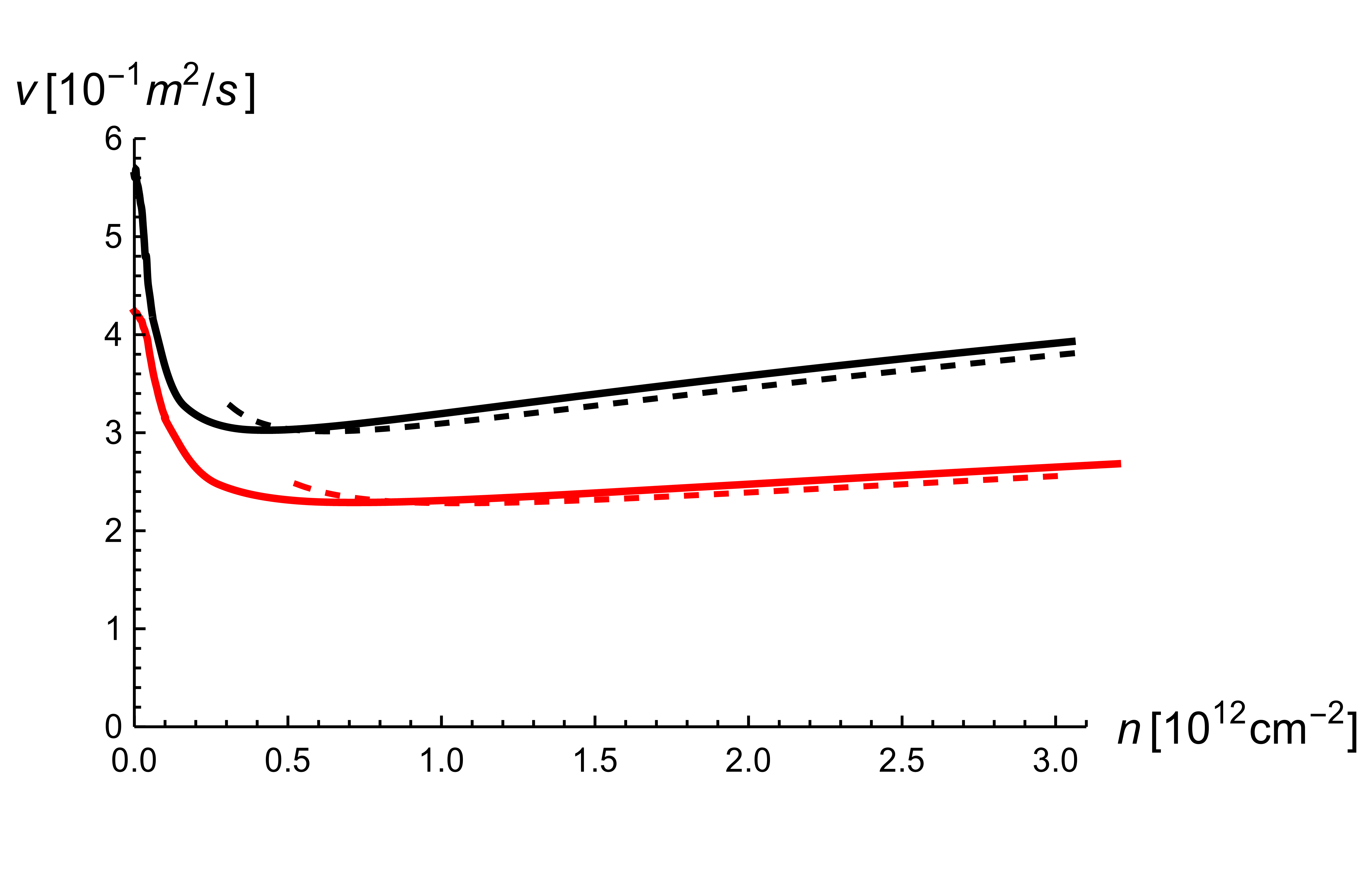}
}
\caption{Kinematic viscosity in monolayer graphene. Left: experimental
  data of Ref.~\cite{geim1} obtained by means of vicinity resistance
  measurements, see Sec.~\ref{sec2.2.2} (From Ref.~\cite{geim1}.
  Reprinted with permission from AAAS). Right: theoretical result of
  Ref.~\cite{me2} obtained using the kinetic theory and
  renormalization group techniques (Reprinted with permission from
  Ref.~\cite{me2}. Copyright (2019) by the American Physical
  Society). }
\label{fig17:eta}
\end{figure}

Taking into account renormalization and screening effects, one can
reach a quantitative estimate of the kinematic viscosity that is of
the same order of magnitude as the experimental data reported in
Ref.~\cite{geim1}, see Fig.~\ref{fig17:eta}. Close to charge
neutrality, the theoretical results show excellent agreement with the
data reported in Ref.~\cite{imh} as shown in Fig.~\ref{fig10:df} (see,
however, Sec.~\ref{sec2.4.2} for the discussion of the controversial
nature of that data).

The field dependence of the shear viscosity was discussed
semiclassically in Refs.~\cite{geim4,ale,moo,stein} in the context of
a single-component Fermi liquid or strongly doped graphene (where only
one band contributes to low-energy physical properties). The resulting
behavior is similar to the conventional magnetoconductivity
\cite{ziman}
\begin{subequations}
\label{etaflB}
\begin{equation}
\eta(B; \mu\gg T) = \frac{\eta(B=0; \mu\gg T)}{1+\Gamma_B^2},
\end{equation}
\begin{equation}
\label{etahflB}
\eta_H(B;\mu\gg T)=\eta(B=0; \mu\gg T)\frac{\Gamma_B}{1+\Gamma_B^2},
\end{equation}
where
\begin{equation}
\label{gb}
\Gamma_B
=
2 \omega_B \tilde\tau_{11},
\quad
\omega_B=|e|v_g^2B/(\mu c).
\end{equation}
\end{subequations}
The kinetic approach \cite{me2} allows one to identify the scattering
rate $\tilde\tau_{11}$ appearing in Eqs.~(\ref{etaflB}). Indeed, this
rate should be distinguished \cite{hydro1,poli16} from the transport
scattering rate \cite{schutt,drag} that determines the electrical
conductivity and the ``quantum'' scattering rate \cite{schutt} that
determines the quasiparticle lifetime. At the same time, the kinetic
theory yields the field dependence of the shear viscosity at charge
neutrality as well \cite{me2}
\begin{equation}
\label{vdpb}
\eta(B; \mu=0) = 
\frac{T^2}{\alpha_g^2v_g^2}
\frac{{\cal B}+{\cal B}_1\gamma_B^2}{1+{\cal B}_2\gamma_B^2},
\end{equation}
where 
\begin{equation}
\label{gammab}
\gamma_B = \frac{|e|v_g^2B}{\alpha_g^2cT^2},
\end{equation}
where ${{\cal B}_1\approx 0.0037}$ and ${{\cal B}_2\approx 0.0274}$.
In contrast to the Fermi liquid results, the shear viscosity at
$\mu=0$ does not vanish in the limit of classically strong field.

Frequency-dependent viscosity was analyzed in
Refs.~\cite{read,poli16,ale18}. In particular, Ref.~\cite{ale18}
suggested an existence of a resonance in strong magnetic fields (as
well as the corresponding plasmon damping). Momentum-dependent
viscosity in Fermi liquids (due to head on collisions
\cite{lai92,gurzhi95}) was suggested in Ref.~\cite{levitov19} (for an
alternative approach to viscosity in Fermi liquids see
Ref.~\cite{ale20}).

Beyond graphene, in anisotropic Dirac systems \cite{julia,brad} one
has to consider the full viscosity tensor (these are the systems where
two Dirac cones merge in momentum space \cite{chus}; this may be
relevant to the organic conductor $\alpha$-(BEDT-TTF)$_{2}$I$_{3}$
under pressure \cite{ad1}, the heterostructure of the $5/3$
TiO$_{2}$/VO$_{2}$ supercell \cite{ad2,ad3}, surface modes of
topological crystalline insulators with unpinned surface Dirac cones
\cite{fafu}, and quadratic double Weyl fermions \cite{hua}). In the
absence of magnetic field, the viscosity matrix contains six
independent components (in accordance with the Onsager reciprocity
\cite{dau6,dau10}), which scale differently with temperature
\cite{julia}. In particular, one of the six components vanishes at
lowest temperatures violating the famous (conjectured) bound for the
shear viscosity to entropy density ratio \cite{kov}. As a result, the
authors of Ref.~\cite{julia} proposed a generalization of the bound to
anisotropic 2D systems, see Sec.~\ref{sec5}. An alternative view on
anisotropic Dirac semimetals taking into account spectrum topology
(i.e., the Berry curvature) has been developed in
Ref.~\cite{sur19}. Hall viscosity in the quantum Hall regime in such
systems was discussed in Ref.~\cite{sur202}. More complicated spectra
can be encountered in 3D Luttinger semimetals \cite{lut56} where the
long-screened nature of the Coulomb interaction leads to a
scale-invariant, non-Fermi-liquid ground state \cite{ab74}.
Hydrodynamic behavior in such systems was considered in
Ref.~\cite{julia20}.

\subsubsection{Hydrodynamic equations in graphene}\label{sec3.2.6}

Taking into account the dissipative corrections in the continuity
equations (\ref{ces}), one finds the generalization of the
Navier-Stokes equation \cite{dau6,navier,stokes} in graphene
\begin{subequations}
\label{heqs}
\begin{eqnarray}
\label{nseq}
&&
{\cal W}(\partial_t+\bs{u}\!\cdot\!\bs{\nabla})\bs{u}
+
v_g^2 \bs{\nabla} P
+
\bs{u} \partial_t P 
+
e(\bs{E}\!\cdot\!\bs{j})\bs{u} 
=
\\
&&
\nonumber\\
&&
\qquad\qquad\qquad\qquad
=
v_g^2 
\left[
\eta \Delta\bs{u}-\eta_H \Delta\bs{u}\!\times\!\bs{e}_{\bs{B}}
+
en\bs{E}
+
\frac{e}{c} \bs{j}\!\times\!\bs{B}
\right]
-
\frac{{\cal W}\bs{u}}{\tau_{{\rm dis}}}.
\nonumber
\end{eqnarray}
The full set of the hydrodynamic equations contains also the
continuity equations
\begin{equation}
\label{ce2}
\partial_t n + \bs{\nabla}\!\cdot\!\bs{j}=0,
\end{equation}
and
\begin{equation}
\label{cei2}
\partial_t n_I + \bs{\nabla}\!\cdot\!\bs{j}_I = - \frac{n_I\!-\!n_{I,0}}{\tau_R},
\end{equation}
and the thermal transport equation \cite{meig1}
\begin{eqnarray}
\label{eqen}
&&
T\left[\frac{\partial s}{\partial t}
+
\bs{\nabla}\!\cdot\!
\left(s\bs{u}-\delta\bs{j}\frac{\mu}{T}-\delta\bs{j}_I\frac{\mu_I}{T}\right)\right]
=
\delta\bs{j}\!\cdot\!
\left[e\bs{E}\!+\!\frac{e}{c}\bs{u}\!\times\!\bs{B}\!-\!T\bs{\nabla}\frac{\mu}{T}\right]
-
\nonumber\\
&&
\nonumber\\
&&
\qquad\qquad\qquad
-
T\delta\bs{j}_I\!\cdot\!\bs{\nabla}\frac{\mu_I}{T}
+
\frac{\eta}{2}\left(\nabla^\alpha u^\beta \!+\! \nabla^\beta u^\alpha
\!-\! \delta^{\alpha\beta} \bs{\nabla}\!\cdot\!\bs{u}\right)^2-
\nonumber\\
&&
\nonumber\\
&&
\qquad\qquad\qquad
-
\frac{n_E\!-\!n_{E,0}}{\tau_{RE}}
+
\mu_I \frac{n_I\!-\!n_{I,0}}{\tau_R}
+
\frac{{\cal W}\bs{u}^2}{v_g^2\tau_{\rm dis}},
\end{eqnarray}
\end{subequations}
where $s$ denotes the entropy density. The equation (\ref{eqen})
replaces the continuity equation for the energy density (\ref{ece}) as
is common in hydrodynamics \cite{dau6}. The hydrodynamic equations are
supplemented by the constitutive equations for the quasiparticle
currents (\ref{js}) and the generalized conductivity matrix
$\widehat\Sigma$, as well as Maxwell's equations for the
electromagnetic field, in other words, Vlasov self-consistency
\cite{dau10,rev,me1,hydro1}.

\subsubsection{Boundary conditions}\label{sec3.2.7}

The state of a conventional fluid is described by the velocity vector
and two thermodynamic quantities, such as density and pressure. The
hydrodynamic equations are differential equations containing spatial
and time derivatives of these variables. Hence, to find a solution to
these equations one has to specify the boundary conditions.

The conventional Navier-Stokes equation \cite{dau6,navier,stokes}
greatly simplifies for an incompressible fluid. In this case, the
fluid density is a constant, while the pressure gradient can be
excluded by applying the curl operation to the equation. The resulting
equation is a differential equation for the velocity only.

If a viscous fluid is flowing near a solid, stationary boundary, a
simple ``no-slip'' boundary condition is often assumed \cite{dau6}
(due to the molecular forces acting between the fluid and the
boundary). On the other hand, a boundary between a fluid and a gas can
be characterized by the ``no-stress'' boundary condition, where the
tangential stress is continuous at the interface. The two conditions
can be ``unified'' as limiting cases of a more general condition due
to Maxwell \cite{max}
\begin{equation}
\label{maxbc}
u^\alpha_t\Big|_S
=
\left.
\ell_S \, e_n^\beta \, \frac{\partial u^\alpha_t}{\partial x^\beta}\right|_S,
\end{equation}
where $\bs{e}_n$ is the unit vector normal to the surface,
$\bs{u}_t=\bs{u}-(\bs{u}\!\cdot\!\bs{n})\bs{n}$ is the tangential
velocity, and $\ell_S$ is the so-called ``slip length''. The no-slip
boundary condition, $\bs{u}=0$ (the normal component of the velocity
has to vanish at any solid boundary by obvious reasons) corresponds to
$\ell_S=0$, while the limit $\ell_S\rightarrow\infty$ describes the
no-stress case.

In electronic systems, the boundary condition (\ref{maxbc}) was
studied in detail in Ref.~\cite{ks19} based on the kinetic approach.
Solving the kinetic equation in the presence of a boundary requires
boundary conditions for the distribution function. The latter are
well studied \cite{falk83}, especially in the context of mesoscopic
physics \cite{bee}. Analytic calculations are possible in the two
limiting cases of specular and diffusive scattering at the boundary.
Boundary conditions in the presence of magnetic field were studied in
Ref.~\cite{ady2}. Recently, the issue of the boundary conditions and
the slip length in the magnetic field was discussed in
Ref.~\cite{rai22prb}.

Specular scattering refers to ideally smooth boundaries such that the
incidence and reflection angles (of the quasiparticle velocity)
coincide. In that case, the distribution function obeys the simple
boundary condition
\begin{equation}
\label{bcf}
f(\varphi)\Big|_S = f(-\varphi)\Big|_S,
\end{equation}
where $\varphi$ is the angle between the quasiparticle (microscopic)
velocity $\bs{v}$ and the boundary. Experimental feasibility of smooth
boundaries was recently explored in Ref.~\cite{adam21}.

\begin{figure*}[t]
\centerline{\includegraphics[width=0.4\textwidth]{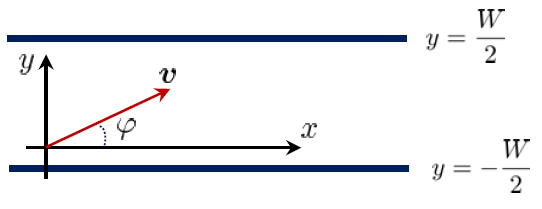}
}
\caption{Channel geometry: the electronic fluid is confined to a
  channel (along the $x$-direction) of the width $W$; $\bs{v}$ is
  the quasiparticle velocity directed at the angle $\varphi$ to the
  channel boundary.}
\label{fig18:ch}
\end{figure*}

In the diffusive case, the boundary is assumed to be sufficiently
rough, such that the incoming quasiparticle can scatter off the
boundary in any direction with equal probability (independent of the
incidence angle). This can be expressed by a more complex
condition. In a channel geometry (see Fig.~\ref{fig18:ch}) the
corresponding condition has the form \cite{bee}
\begin{subequations}
\label{dbc}
\begin{equation}
\label{bcfd1}
f(W/2, -\pi\!<\!\varphi\!<\!0)\ = \frac{1}{2}\int\limits_0^\pi\! d\varphi'
\sin\varphi'f(W/2, \varphi'),
\end{equation}
\begin{equation}
\label{bcfd2}
f(-W/2, 0\!<\!\varphi\!<\!\pi)|_S = \frac{1}{2}\int\limits_{-\pi}^0\! d\varphi'
\sin\varphi'f(-W/2, \varphi').
\end{equation}
\end{subequations}

The resulting slip length is strongly influenced by the choice of the
boundary conditions for the distribution function \cite{ks19}. The
authors of Ref.~\cite{ks19} express $\ell_S$ in terms of the
electron-electron scattering length
\begin{equation}
\label{slip}
\ell_S = g(\kappa) \ell_{ee},
\qquad
\kappa = \frac{h_1^2 h_2^{d-1}}{\lambda^{d+1}},
\qquad
g(\kappa) \rightarrow
\begin{cases}
g_0/\kappa, & \kappa\ll1, \cr
g_\infty, & \kappa\rightarrow\infty,
\end{cases}
\end{equation}
where $h_1$ and $h_2$ are the mean height and correlation length
describing the boundary roughness \cite{falk83}, $\lambda$ is the
(temperature dependent) electron wavelength, and $d$ is the spatial
dimensionality. The precise value of $g(\kappa)$ varies dramatically,
but at experimentally relevant temperatures one finds
$\ell_S\approx0.5\,\mu$m, the value that agrees with experimental
observations, see Ref.~\cite{imm}.

Full solution to the hydrodynamic equations in electronic systems
requires also boundary conditions for thermodynamic quantities. In
electronic systems, these are most conveniently expressed in terms of
electrochemical potentials.

Traditional transport theory is based on a single-electron approach,
where the main mechanism of electrical resistance -- and hence,
dissipation -- is the electron-impurity and electron-phonon
scattering. In this case, the bulk system is characterized by a local
conductivity, while contact interfaces -- by the contact
resistance. The latter appears due to equilibration of (originally
mismatched) electrochemical potentials in the two interfacing
materials \cite{mard}. The bulk and contact resistances could be seen
as independent parts of the overall electrical circuit. If the bulk
system is diffusive, the contribution of the contacts is typically
negligible. On the contrary, in ballistic systems there is almost no
dissipation in the bulk, such that most of the voltage drop occurring
in the contacts, see Fig.~\ref{fig8:diffbal}.

In the context of ideal (inviscid) hydrodynamics in nearly neutral
graphene, boundary conditions taking into account contact resistance
were considered in Ref.~\cite{alf}. Assuming the leads are represented
by a disordered, particle-hole symmetric metal, the electron and hole
currents are given by the difference of the electrochemical potentials
across the interface divided by the contact resistance. If no electric
current is allowed in the system (as is appropriate for measurements
of thermal conductivity \cite{luc,alf}), this leads to a boundary
condition relating the imbalance chemical potential $\mu_I$ and
current $\bs{j}_I$.

An alternative situation was considered in Ref.~\cite{fal19}. In this
paper the authors have considered an idealized situation where a clean
(disorder-free), but viscous electron fluid is contacted by an ideal
conductor with an ideal interface characterized by the vanishing
reflection coefficient \cite{landauer}. The absence of disorder
implies the lack of Ohmic dissipation in the bulk, while the ideal
contacts do not provide any contact resistance. In that case the bulk
dissipation due to viscosity has to be compensated by the work done by
current source. If both the bulk and the contacts are disorder-free,
then the electric potential exhibits a sharp inhomogeneity (on the
hydrodynamic scale - a jump) in a narrow region close to the
interface, which translates into a viscosity-dependent contribution to
the contact resistance that can be positive or negative depending
on the contact curvature sign.

Real samples are likely to exhibit all of the above effects and
moreover may host additional localized charges at the sample edges
leading to classical (nontopological) edge currents \cite{zel}, see
Sec.~\ref{sec2.4.3}. The appropriate boundary conditions then strongly
depend on sample geometry and the specific measurement scheme. For
example, the authors of Ref.~\cite{corb} suggest using the Corbino
disk geometry to measure electronic viscosity. In their setup, the
outer edge of the Corbino disk is isolated, implying the vanishing
radial component of the electric current. In addition, they required
the azimuthal momentum component to diffuse radially, such that the
off-diagonal component of the viscous stress tensor vanishes at both
edges of the disk. Interestingly enough, the authors of
Ref.~\cite{corb} considered the no-slip boundary conditions as well
and found no qualitative difference with the above approach.

\subsection{Hydrodynamic collective modes and plasmons}\label{sec3.3}

Hydrodynamic collective modes have been considered by many authors
\cite{luc,hydro1,schutt,cosmic,lev13,fog,svin,ldsp,ks20,fat,megt,lucas16,falk18}.
The point of consensus is that the ideal (neglecting dissipative
processes) electronic fluid in neutral graphene is characterized by a
sound-like collective mode (sometimes referred to as the ``cosmic
sound'' \cite{cosmic} or the ``second sound'' \cite{ks20}) with the
linear dispersion relation
\begin{equation}
\label{cs0}
\omega = v_gq/\sqrt{2}.
\end{equation}
In a way, this result justifies the claim that the electronic fluid
behave hydrodynamically, see Sec.~\ref{sec1}.

Dissipative processes damp the sound mode (\ref{cs0}). In contrast to
traditional hydrodynamics this happens since dissipation due to
``external'' scattering (e.g., disorder and electron-phonon
scattering) appears already in the description of an ``ideal'' (i.e.,
inviscid) electronic fluid, see Eqs.~(\ref{cei}), (\ref{ece}), and
(\ref{euler}). Another issue is the regime of applicability of the
dispersion relation (\ref{cs0}) or its damped counterparts. The point
is that hydrodynamics is based on the gradient expansion valid at
length scales that are much larger than $\ell_{ee}$ (representing the
energy and momentum conserving interaction responsible for
equilibration). At smaller length scales other, more conventional
collective excitations, such as plasmons
\cite{hydro1,lev13,fog,ldsp,ks20,fat,Giuliani,hill09,prin11,fei12,chen12,bas,kop17,kop18,polkop20,pol20pl,kop20p1,kop20p2,mach19,per20,mach20,falko20}, may be identified.

\subsubsection{Electronic ``sound'' in neutral graphene}\label{sec3.3.1}

Collective excitations in the electronic system in graphene have been
recently studied in detail in Ref.~\cite{megt}. At charge neutrality
and in the absence of magnetic field, the sound mode (\ref{cs0})
damped by the dissipative processes has the dispersion relation
\begin{equation}
\label{csv}
\omega = 
\sqrt{\frac{v_g^2q^2}{2} 
- \frac{1}{4}\left(\frac{1\!+\!q^2\ell_G^2}{\tau_{{\rm dis}}} - \frac{1}{\tau_{RE}}\right)^2}
- i \frac{1\!+\!q^2\ell_G^2}{2\tau_{{\rm dis}}} - \frac{i}{2\tau_{RE}},
\end{equation}
where $\ell_G$ is the Gurzhi length (\ref{lg}). Although
Eq.~(\ref{csv}) can be straightforwardly derived by linearizing the
hydrodynamic equations (\ref{heqs}), the damping in Eq.~(\ref{csv})
can be seen as exceeding the accuracy of the hydrodynamic
regime. Indeed, the gradient expansion in neutral graphene is
justified for momenta smaller than a certain scale defined by the
electron-electron interaction
\begin{equation}
\label{hydro}
q\ell_{\rm hydro}\ll1, \quad
\ell_{\rm hydro}\sim\frac{v_g}{\alpha_g^2\bar{T}}.
\end{equation}
Assuming a clean system $\tau_{\rm dis}\rightarrow\infty$ (energy
relaxation due to supercollisions \cite{meig1} may be also neglected,
$\tau_{RE}\gg\tau_{\rm dis}$), the expression under the square root in
Eq.~(\ref{csv}) can be expanded for small $q$ as
\[
\frac{v_g^2q^2}{2} - \frac{\left(1\!+\!q^2\ell_G^2\right)^2}{4\tau^2_{{\rm dis}}}
\rightarrow
\frac{v_g^2q^2}{2}\left[1 \!-\! A q^2 \ell_{\rm hydro}^2
\!- {\cal O}(\tau_{\rm dis}^{-1})\right]\!,
\]
where $A$ is a numerical coefficient. Hence, within the hydrodynamic
approach, the viscous contribution to damping should be neglected,
leaving one with the simpler dispersion \cite{hydro1}
\begin{equation}
\label{cm2}
\omega = 
\sqrt{\frac{v_g^2q^2}{2} - \frac{1}{4\tau^2_{{\rm dis}}}}
- \frac{i}{2\tau_{{\rm dis}}}.
\end{equation}
Now, the peculiar nature of the Dirac spectrum in graphene leads to
the fact that the linearized version of the hydrodynamic equations is
justified in a wider parameter region than Eqs.~(\ref{heqs})
themselves \cite{hydro1,hydro0,drag2,me3} (due to the ``collinear
scattering singularity'' \cite{rev,luc,hydro1,mfss}). In
the weak coupling limit, the linear response theory is valid at much
larger momenta
\begin{equation}
\label{scasep}
q\ell_{\rm coll}\ll1, \quad
\ell_{\rm coll}\sim\frac{v_g}{\alpha_g^2\bar{T}|\ln\alpha_g|}\ll\ell_{\rm hydro},
\end{equation}
formally providing one with a justification to extend Eq.~(\ref{csv})
beyond the hydrodynamic regime. However, already at
$q\ell_{\rm{hydro}}\sim1$ the imaginary part of the sound dispersion
becomes comparable to the real part, at which point the dispersion is
no longer observable.

The nature of the sound mode (\ref{cs0}) [or Eq.~(\ref{csv})] becomes
clear if one takes into account the fact that in neutral graphene in
the absence of magnetic field the electric charge is decoupled from
the hydrodynamic energy flow. Indeed, at charge neutrality $n=0$ so
that the electric field does not enter the linearized Navier-Stokes
equation (\ref{nseq}), while the ``conductivity matrix'' in
Eqs.~(\ref{js}) is diagonal. Hence, the energy flow is described by
the Navier-Stokes equation (\ref{nseq}), while charge transport is
described by the Ohmic relation (\ref{djs}), together with the Vlasov
self-consistency. The latter can be expressed using the Poisson's
equation
\begin{subequations}
\label{vlasov}
\begin{equation}
\label{v3d}
\bs{E}_V = - e\bs{\nabla}\!\int\! d^2r' \frac{\delta n(\bs{r}')}{|\bs{r}\!-\!\bs{r}'|}.
\end{equation}
In gated structures \cite{mr1,ash}, this can be simplified to
\begin{equation}
\label{v2dg}
\bs{E}_V = - \frac{e}{C}\bs{\nabla}\delta n(\bs{r}),
\end{equation}
\end{subequations}
where ${C=\varepsilon/(4\pi d)}$ is the gate-to-channel capacitance
per unit area, $d$ is the distance to the gate, and $\varepsilon$ is
the dielectric constant. This approximation neglects the long-ranged
(dipole-type) part of the screened Coulomb interaction and is
justified while the charge density $n(\bs{r})$ varies on length scales
exceeding $d$.

The charge sector of the theory is characterized by an overdamped
collective mode with the dispersion
\begin{equation}
\label{cmdp}
\omega =-iD_0q^2\!\left[1\!+\!eV_{s}(q)\frac{\partial n}{\partial\mu}\right]\!,
\quad
D_0 = \frac{1}{2}\frac{v_g^2\tau_{11}\tau_{\rm dis}}{\tau_{11}\!+\!\tau_{\rm dis}}.
\end{equation}
In a gated structure, the mode is diffusive (with the Vlasov
self-consistent potential $V_s=e/C$ providing a correction to the
diffusion coefficient). For long-range Coulomb interaction (here
$V_s=2\pi e/q$), the dispersion remains purely imaginary with
$\omega\sim iq$ at small $q$.

\subsubsection{Electronic ``sound'' in doped graphene}\label{sec3.3.2}

In doped graphene, the charge and energy modes are coupled by the
Vlasov self-consistency \cite{megt}. To the leading order in (weak)
energy relaxation this leads to a sound mode similar to
Eq.~(\ref{csv}) and a diffusive mode that in a gated structure has the
dispersion
\begin{equation}
\label{zeromode}
\omega=-\frac{i}{\tau_{RE}}
\frac{\varkappa v_g^2q^2}{(\varkappa\!+\!2\pi C)v_g^2q^2\!+\!4\pi C\tau_{RE}^{-1}\tau_{\rm dis}^{-1}},
\end{equation}
where the Thomas-Fermi screening length is given by
\begin{equation}
\label{tfs}
\varkappa = N \alpha_g k_F = Ne^2\mu/v_g^2.
\end{equation}
For long-range Coulomb interaction, the factor $2\pi C$ should be
replaced with $q$. Physically, the mode (\ref{zeromode}) describes
energy diffusion appearing due coupling of the charge and energy
fluctuations by Vlasov self-consistency.

For a gated structure, the sound mode coincides with the ``cosmic
sound'' (\ref{cs0}) at the lowest momenta, albeit with the sound
velocity modified by screening. In the case of long-range Coulomb
interaction the dispersion is no longer sound-like. In the limit
$q\rightarrow0$ (and $\mu\gg T$), one finds the spectrum similar to
the usual 2D plasmon \cite{zna,hydro1}
\begin{equation}
\label{cm30}
\omega(q\ll\varkappa) = - \frac{i}{2\tau_{{\rm dis}}}
+
\sqrt{\frac{1}{2}v_g^2q\varkappa-\frac{1}{4\tau^2_{{\rm dis}}}}.
\end{equation}
The expression (\ref{cm30}) is valid when
\[
q\ell_G\ll1, \quad q\ll\varkappa, \quad v_g^2\varkappa q \tau^2_{\rm dis} \gg 1.
\]
These conditions are consistent with the applicability condition of
the hydrodynamic approach if
\[
v_g\varkappa\tau_{\rm dis}\gg1
\quad\Rightarrow\quad
N\alpha_g\mu\tau_{\rm dis}\gg1,
\]
\[
\ell_G\ll v_g^2\varkappa \tau^2_{\rm dis}
\quad\Rightarrow\quad
N^2\alpha^4_g\mu\tau_{\rm dis}(\bar{T}\tau_{\rm dis})^2\gg1.
\]
The above conditions provide a possibility to observe the dispersion
(\ref{cm30}) in a parametrically defined range of wavevectors.

\subsubsection{Hydrodynamic modes and plasmons}\label{sec3.3.3}

The above sound-like modes have to be distinguished from plasmonic
excitations in electronic systems. The latter are well studied, also
in graphene
\cite{hydro1,lev13,ldsp,ks20,fat,Giuliani,hill09,prin11,fei12,chen12,bas,kop17,kop18,polkop20,pol20pl,kop20p1,kop20p2,mach19,per20,mach20,falko20}. In
a degenerate electron gas in 2D, the plasmon dispersion (neglecting
impurity scattering, i.e., $\tau_{\rm{dis}}\rightarrow\infty$) has the form \cite{Giuliani}
\begin{equation}
\label{2Dplasmon}
\omega = \sqrt{2e^2\mu q}\left(1+\gamma \frac{q}{\varkappa}\right),
\end{equation}
where $\gamma$ is a numerical coefficient, that can be evaluated
either within the random phase approximation (i.e., by computing the
Lindhard function; this leads to $\gamma=3/4$ \cite{Giuliani}), or
using a macroscopic (hydrodynamic-like) theory. The latter approach
yields a different value of $\gamma$ which is typically attributed to
the fact that hydrodynamics is applicable at small momenta
($q\ell_{\rm hydro}\ll1$) and frequencies, while plasmons are
nonequilibrium excitations that belong to higher momenta
\cite{Giuliani}. Based on this argument one might expect that the
hydrodynamic collective modes and plasmons simply have nothing to do
with each other \cite{fog}. Yet, given the same leading momentum
dependence in Eqs.~(\ref{cm30}) and (\ref{2Dplasmon}), the relation
between the two is worth investigating.

In graphene, the possibility of discussing momenta exceeding
$1/\ell_{\rm hydro}$ is afforded by the collinear scattering
singularity \cite{rev,luc,me1,hydro1,schutt,hydro0,mfs,mfss,drag2}
which leads to the existence of two parametrically different length
scales, see Eq.~(\ref{scasep}), and hence of an intermediate momentum
range, ${\ell_{\rm{hydro}}^{-1}\ll{q}\ll\ell_{\rm coll}^{-1}}$. Here a
linear response theory of Ref.~\cite{hydro0} can be used to find the
collective modes. Remarkably, macroscopic equations of this theory
coincide with the linearized hydrodynamic equations \cite{megt} such
that the resulting dispersions should be valid in the hydrodynamic
regime as well and can be compared with the above results.

In doped graphene, the electron system is degenerate and the linear
response theory of Ref.~\cite{hydro0} can be expressed in terms of a
single equation
\begin{equation}
\label{ceqlr}
\frac{\partial\bs{J}}{\partial t} + \frac{v_g^2}{2}\bs{\nabla}\rho
-\nu \Delta\bs{J}
-\frac{v_g^2}{2}\frac{\partial n}{\partial\mu} e^2 \bs{E}
= - \frac{\bs{J}}{\tau_{\rm dis}},
\end{equation}
where $\bs{J}$ is the electric current, see Eq.~(\ref{mc0}), and
$\rho$ denotes the charge density. Taking into account the Vlasov
field (\ref{vlasov}) and continuity equation, one finds the collective
mode with the spectrum
\begin{equation}
\label{flcm}
\omega = 
\sqrt{2e^2\mu q\left(1\!+\!\frac{q}{\varkappa}\right)-\frac{(1\!+\!q^2\ell_G^2)^2}{4\tau^2_{\rm dis}}}
-\frac{i(1\!+\!q^2\ell_G^2)}{2\tau_{\rm dis}},
\end{equation}
where ${D=v_g^2\tau_{\rm dis}/2}$ and
${\sigma=v_g^2(\partial{n}/\partial\mu)\tau_{\rm dis}/2}$ are the
diffusion coefficient and the Drude conductivity.

The spectrum (\ref{flcm}) is exactly the same as the screened sound
mode leading to Eq.~(\ref{cm30}). In the limit
${\tau_{\rm{dis}}\rightarrow\infty}$, one may expand Eq.~(\ref{flcm})
in small ${q\rightarrow0}$. This yields Eq.~(\ref{2Dplasmon}) with the
``wrong'' coefficient, ${\gamma=1/2}$. At the same time, the leading
term (at ${q\ll\varkappa}$) agrees with the Fermi liquid result in the
presence of disorder \cite{zna} (in the absence of viscosity). The
dispersion (\ref{flcm}) is valid for $q\ell_{\rm coll}\ll1$, however,
becomes overdamped already at $q\sim\ell_{\rm hydro}^{-1}$.  For
$q\gg\ell_{\rm coll}^{-1}$, the quasi-equilibrium description leading
to Eq.~(\ref{ceqlr}) breaks down and true plasmons with the dispersion
(\ref{2Dplasmon}) emerge. At these momenta the spectrum (\ref{flcm})
is purely imaginary. Based on this argument, the authors of
Ref.~\cite{megt} argue that the two modes are not connected. Similar
conclusions were reached in Ref.~\cite{ldsp}, where it was argued
that Coulomb interaction precludes the appearance of hydrodynamic
sound in Fermi liquids.

In graphene at charge neutrality, the ``true'' plasmon dispersion was
established in Ref.~\cite{schutt} on the basis of microscopic
theory. The leading behavior of the plasmon dispersion is given by
\begin{equation}
\label{pldp}
\omega = \sqrt{(4\ln2)e^2Tq}.
\end{equation}
This expression can be compared to the results of the linear response
theory in graphene \cite{hydro0,megt}. The linear response theory of
Ref.~\cite{hydro0} is based on the same three-mode approximation as
the hydrodynamics discussed in Sec.~\ref{sec3.2}. Similarly to the
discussion in Sec.~\ref{sec3.3.1}, at charge neutrality the charge
sector decouples from the rest of the theory and can be described by
the equation
\begin{equation}
\label{ceqlrdp}
\frac{\partial\bs{j}}{\partial t} + \frac{v_g^2}{2}\bs{\nabla}n
-\frac{2\ln2}{\pi}e^2T \bs{E}
= - \frac{\bs{j}}{\tau_{\rm dis}} - \frac{\bs{j}}{\tau_{11}},
\end{equation}
where $\tau_{11}$ determines the quantum conductivity (\ref{sq}), see
also Eqs.~(\ref{js}) and (\ref{sigma3}). Combining Eq.~(\ref{ceqlrdp})
with the continuity equation one finds
\begin{subequations}
\label{pddp}
\begin{equation}
\label{pldp0}
\omega^2 +i\omega\left(\frac{1}{\tau_{\rm dis}}+\frac{1}{\tau_{11}}\right)
=
\frac{v_g^2}{2}q^2 + (4\ln2)e^2Tq,
\end{equation}
leading to a plasmon-like spectrum that can be expressed similarly
to Eq.~(\ref{cmdp})
\begin{equation}
\label{pldp1}
\omega = -i \frac{\sigma(\omega)q^2}{e^2\partial n/\partial\mu}
\left[1 + e V_s(q)\frac{\partial n}{\partial\mu}\right],
\end{equation}
where $\sigma(\omega)$ is the optical conductivity \cite{me3} [in
contrast to the static conductivity (\ref{sq}) in Eq.~(\ref{cmdp}]
\begin{equation}
\label{optcon}
\sigma(\omega) = \frac{2e^2T\ln2}{\pi} 
\frac{1}{-i\omega + \tau_{11}^{-1}+\tau_{\rm dis}^{-1}}.
\end{equation}
In the hydrodynamic regime of small frequencies,
${\sigma(\omega\rightarrow0)\rightarrow\sigma_0}$, the mode
(\ref{pldp1}) is purely diffusive recovering Eq.~(\ref{cmdp}).

Resolving Eq.~(\ref{pldp0}) one finds the plasmon dispersion in
the form
\begin{equation}
\label{pd}
\omega = -i\frac{\tau_{\rm dis}\!+\!\tau_{11}}{2\tau_{\rm dis}\tau_{11}}
\!+\!
\sqrt{(4\ln2)e^2Tq  \!+\! \frac{v_g^2}{2}q^2
\!-\! \frac{(\tau_{\rm dis}\!+\!\tau_{11})^2}{4\tau_{\rm dis}^2\tau_{11}^2} }\ .
\end{equation}
\end{subequations}
For ${\omega\gg\tau_{11}^{-1}\gg\tau_{\rm dis}^{-1}}$ and
${q\rightarrow0}$, the leading behavior in Eq.~(\ref{pd}) coincides
with Eq.~(\ref{pldp}). At large momenta the first term in the RHS of
Eq.~(\ref{pldp0}) dominates and the dispersion resembles the
hydrodynamic sound, Eq.~(\ref{cs0}). This contradicts the results of
Ref.~\cite{schutt}: although at large $q$ the true dispersion also
becomes linear, the coefficient (analogous to the speed of sound) is
different (there is no factor of $\sqrt{2}$).

To summarize, the plasmon mode (\ref{pd}) should be contrasted with
the diffusive charge mode (\ref{cmdp}), and not the sound mode
(\ref{csv}). The plasmon and the sound belong largely to different
frequency regimes \cite{fog}, but most importantly, stem from the two
different, decoupled sectors of the theory (the sound mode can also be
obtained from the linear response theory hence one can extend its
region of applicability beyond the hydrodynamic regime). The latter
fact is the reason why the plasmon dispersion is independent of
viscosity, while the sound mode (\ref{csv}) is unaffected by screening
effects (which are essentially responsible for plasmon
excitations). Formally, the two modes coexist but are characterized by
different frequencies that are much higher for the plasmon
mode. Approximately at $q\sim\ell_{\rm coll}^{-1}$, i.e., at the
applicability limit of the linear response theory, the sound mode
becomes overdamped, which does not happen to the plasmon. At that
point the plasmon dispersion is almost linear albeit with the
coefficient that disagrees with the microscopic theory \cite{schutt},
as pointed out above.

An alternative approach to plasmons is to consider the electromagnetic
response of the 2D electron fluid to the high-frequency field
generated by a Hertzian dipole \cite{luskin}. For small enough
frequencies ($\omega\tau_{ee}\ll1$) the electron system responds
hydrodynamically. Coupling the hydrodynamic equations with the 3D
Maxwell's equations one can define a boundary value problem yielding
the full description of the spatial structure of the electromagnetic
field. In particular, the numerical analysis of Ref.~\cite{luskin}
suggests co-existence between the plasmon and diffusive modes in a way
that is somewhat different from the above solution of the purely
hydrodynamic problem (where the electromagnetic field was assumed to
be static). For analytic analysis of edge magnetoplasmons (using the
Wiener-Hopf technique) see Ref.~\cite{gold18}.

\section{Known solutions to hydrodynamic equations in electronic systems}\label{sec3.4}

Once equipped with the hydrodynamic equations and boundary conditions,
one may embark on finding solutions in an attempt to either explain or
predict experimental observations. Since most transport measurements
in solids are performed within linear response, many authors consider
solutions to linearized hydrodynamic equations.

Hydrodynamic charge flow in doped graphene (more generally, in
hydrodynamic Fermi liquids) was considered analytically in
Refs.~\cite{fl2,pol15,pol16,fl0,glazman20,ale,moo,fl1,alex17} and
numerically in Refs.~\cite{sven1,msw2,msw}. Neutral graphene (more
generally, compensated semimetals) was analyzed in
Refs.~\cite{sven2,mr1,mr2,cfl,megt2,mr3,ant20}.

Nonlocal transport properties observed in doped graphene
\cite{geim1,geim3,geim4} were studied in Refs.~\cite{pol15,pol16}
focusing on the appearance of vortices (or ``whirlpools'') in viscous
flows in confined geometries, the effect that is responsible for the
observed negative nonlocal resistance \cite{geim1}. A purely analytic
approach to that problem (albeit in an idealized geometry) was offered
in Refs.~\cite{fl0,fl1}. The authors of Ref.~\cite{fl0} hinted on the
possibility to observe multiple vortices, the effect that was further
explored numerically in Ref.~\cite{sven1} (see Fig.~\ref{fig4:vor}),
where a {\it sign-alternating} nonlocal resistance was suggested as a
consequence. The latter is especially important given that negative
nonlocal resistance is not a unique characteristic of the viscous flow
and can be observed in ballistic systems \cite{geim3,fl18}.
Interestingly enough, complicated patterns of multiple vortices may
arise also in nearly neutral graphene with long-ranged disorder
\cite{ant20}. Further complications with the hydrodynamic
interpretation of the observed nonlocal resistance and the associated
vorticity were discussed in Ref.~\cite{hui20}, where it was argued
that nonlocal (i.e., momentum-dependent) conductivity in disordered
electron systems may mimic the hydrodynamic effects even in the
absence of electron-electron interaction [the idea is to interpret
  Eq.~(\ref{nseqfl}) as the Ohm's law with nonlocal
  conductivity]. However, extracting the viscosity from the nonlocal
conductivity obtained by means of the Kubo formula \cite{read} might
not be straightforward in disordered systems \cite{bur19}. Moreover,
it is unclear why should one use the hydrodynamic ``no-slip'' boundary
conditions [which are needed to obtain Poiseuille-like solutions from
  Eq.~(\ref{nseqfl})] in conventional disordered systems outside of the
hydrodynamic regime.

An alternative measurement providing indirect evidence of hydrodynamic
behavior, namely superballistic transport through a point contact
\cite{geim2} was discussed theoretically in
Refs.~\cite{glazman20,fl2}. Reference~\cite{glazman20} provided a detailed
analysis of the hydrodynamic theory in the slit geometry comparing the
results to those of the ballistic and diffusive (Ohmic) behavior. The
authors of Ref.~\cite{glazman20} concluded that the hydrodynamic
regime represents a relatively narrow intermediate parameter region
between the two more conventional regimes (namely, the diffusive and
ballistic). Further analysis of a viscous flow through a constriction
and the related enhancement of conductivity was reported in
Ref.~\cite{alex21}.

Now, one of the most popular geometries to consider hydrodynamic
effects is the channel (or slab) geometry, see
Figs.~\ref{fig8:diffbal} and \ref{fig18:ch}. The reason for this is
the wide spread of the Hall bar geometry of the experimental samples,
see Figs.~\ref{fig1:hb} and \ref{fig3:nvr}, as well as simplicity of
theoretical solution, since assuming a long channel all physical
quantities depend only on the coordinate along the channel (the
$x$-coordinate in the notations adopted in
Fig.~\ref{fig18:ch}). Assuming the no-slip boundary conditions, one
finds the solution to the Navier-Stokes equation in the form of the
catenary curve, which reduces to the standard Poiseuille flow
\cite{dau6,poi,poise} in the limit of the large Gurzhi length,
$\ell_G\gg W$ (where $W$ is the channel width).

In doped graphene, the electric current is hydrodynamic and is
expected to exhibit this behavior \cite{pol15}, with
\begin{equation}
\label{rj1}
J_x = \sigma_0 E_x \left[1-\frac{\cosh y/\ell_G}{\cosh W/[2\ell_G]}\right],
\end{equation}
where $J_x$ and $E_x$ are the components of the current density and
electric field along the channel and $\sigma_0$ is the Drude
conductivity (due to, e.g., disorder). This effect was later
observed in the imaging experiment of Ref.~\cite{imm}. If the system
is subjected to the magnetic field, then increasing the field
decreases the viscosity, see Eqs.~(\ref{etaflB}), and hence the Gurzhi
length (\ref{lg}) leading to {\it negative} magnetoresistance
(suggested theoretically in Refs.~\cite{ale,moo} and observed
experimentally in Ref.~\cite{geim4}). These effects were also
considered within the two-fluid hydrodynamic model in
Ref.~\cite{mr2}. For an alternative theory of the electronic flows in
narrow channels in magnetic fields describing the interplay of
electron-electron interactions, disorder, and boundary conditions that
goes beyond the hydrodynamic description, see Ref.~\cite{ady2}. For a
detailed discussion of the Hall voltage and more generally the role of
Hall viscosity in 2D Fermi liquids see Ref.~\cite{han20}. The case of
long-range disorder (or general inhomogeneity of the medium) was
considered in Refs.~\cite{alex17,levch17,lucas20}, where a positive
bulk magnetoresistance was found due to the absence of the Hall
voltage \cite{lucas20}. The latter point is reminiscent of the
situation in graphene at charge neutrality (other than the boundary
effects).

In neutral graphene, the picture is more complicated due to decoupling
of the charge and energy flows in the absence of magnetic field. In
that case, the hydrodynamic, Poiseuille-like flow is expected for the
energy current \cite{julia}, while the charge transport exhibits the
usual diffusion with the quantum conductivity (\ref{sq}) due to
electron-electron interaction instead of the standard Drude
conductivity due to disorder. Applying external magnetic field naively
leads to a positive, parabolic magnetoresistance. This is because the
bulk electric current in neutral graphene is accompanied by the
lateral quasiparticle (and energy) current (which in turn leads to the
geometric magnetoresistance). However, due to the compensated Hall
effect and quasiparticle recombination, see Sec.~\ref{sec3.2.1}, there
is a strong boundary effect changing that behavior and leading to
nonsaturating, linear magnetoresistance (at charge neutrality)
\cite{mr1} that is somewhat similar to the edge effects considered in
Ref.~\cite{meg,rashba}. The key point is that the above bulk effect is
incompatible with finite size geometry: assuming that the bulk current
is flowing along the channel, the lateral quasiparticle current must
flow across the channel and hence must vanish at both boundaries. The
resulting inhomogeneity of the individual electron and hole currents
is inconsistent with the standard geometric
magnetoresistance. Moreover, this inhomogeneity is only compatible
with the continuity equation for the total quasiparticle density,
Eq.~(\ref{cei}), if one takes into account recombination. The
resulting quasiparticle density is practically uniform in the bulk
(characterized by the parabolic geometrical magnetoresistance), but is
strongly inhomogeneous in boundary regions of the width of the
recombination length, $\ell_R(B)=\ell_R(B=0)/\sqrt{1\!+\!\mu^2B^2}$
(here $\mu$ stands for carrier mobility). The edge contribution to the
overall resistance is linear in magnetic field \cite{mr1} and can
dominate in classically strong fields. This effect is not specific to
Dirac fermions. Theoretically similar phenomena were considered in
Refs.~\cite{mr2,cfl,mr3}. Experimentally, linear magnetoresistance due
to recombination was studied in bilayer graphene in Ref.~\cite{mrexp}.

\begin{figure}[t]
\centerline{
\includegraphics[width=0.485\columnwidth]{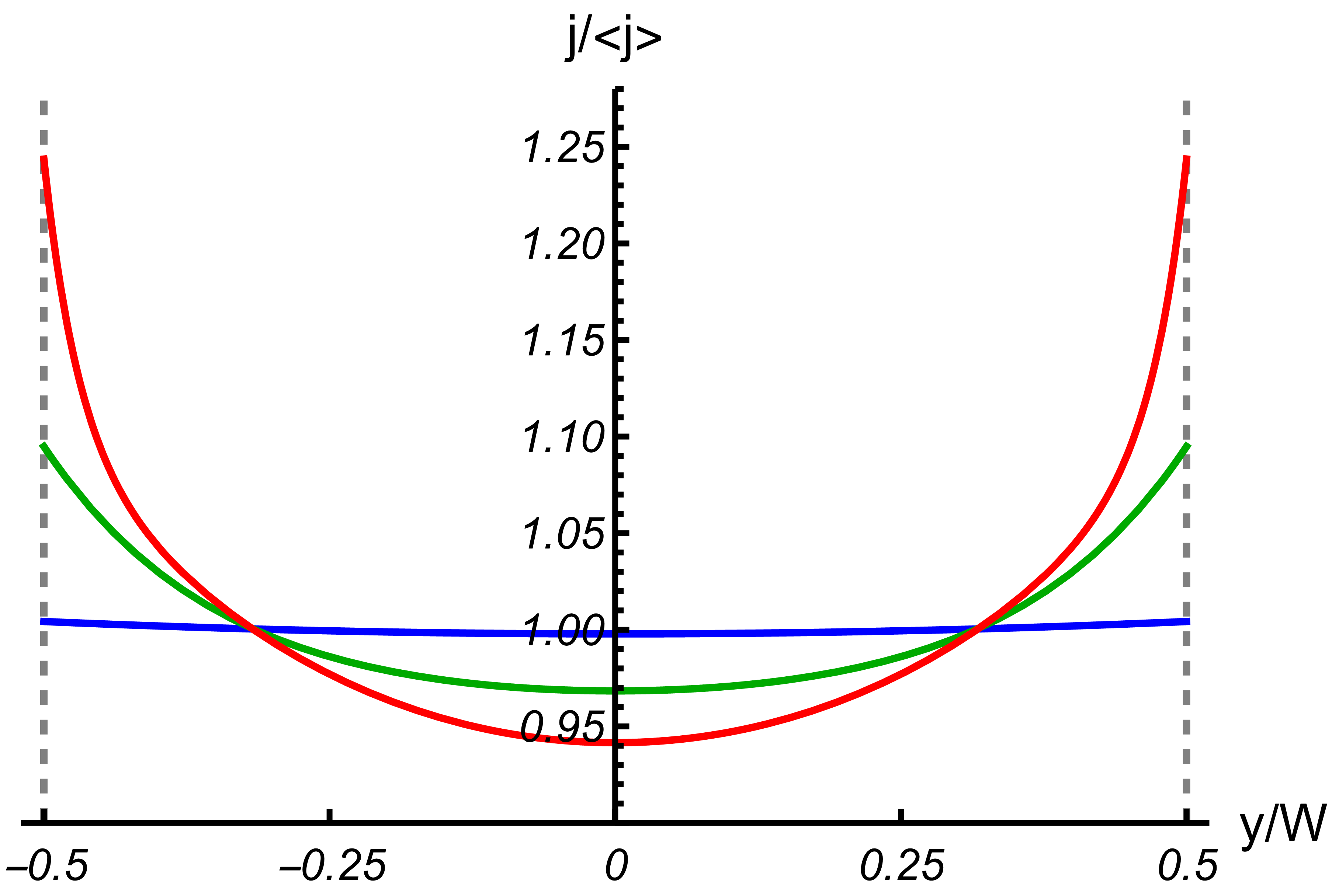}
}
\caption{Anti-Poiseuille flow in narrow channels in graphene in
  perpendicular magnetic field \cite{megt2}. The curves represent the
  inhomogeneous current density in narrow channels of width
  $W=0.1,\,1,\,5\,\mu$m (blue, green, and red curves,
  respectively). Calculations wee performed for typical parameter
  values ${\tau_{\rm dis}\approx0.8}\,$THz \cite{gal},
  ${\alpha_g\approx0.2}$ \cite{gal,sav}, ${\nu\approx0.4}\,$m$^2/$s
  \cite{imh,me2}, ${B=0.1}\,$T, ${T=250}\,$K (Reprinted with
  permission from Ref.~\cite{megt2}. Copyright (2021) by the American
  Physical Society).}
\label{fig:kap}
\end{figure}

Electric current in a neutral graphene channel also becomes
inhomogeneous in magnetic field (where all three modes in the ``three
mode approximation'' discussed in Sec.~\ref{sec3} are
coupled). However, unlike the situation in doped graphene, the current
does not exhibit the Poiseuille-like flow (\ref{rj1}) \cite{megt2},
see Fig.~\ref{fig:kap}. One of the reasons for that is the boundary
conditions: the Poiseuille flow is the solution of the hydrodynamic
equations with the no-slip boundary conditions (which can be
generalized to the Maxwell's boundary conditions with a relatively
small slip length). The electric current in neutral graphene is not
related to any solution of the Navier-Stokes equation and, moreover,
there is no reason to assume that the current vanishes at the channel
boundaries. In fact, for specular boundary conditions the opposite
happens \cite{megt2}: quasiparticle recombination leads to a minimum
of the current density in the center of the channel, while the maximum
value occurs at the boundaries. More general boundary conditions (see
Sec.~\ref{sec3.2.7}) require a numerical solution of the kinetic
equation, which has not yet been carried out in this context.

An alternative geometry to study hydrodynamic flows is offered by the
Corbino disk \cite{corbino}. Here the electric current is
inhomogeneous even in the simplest case of the Ohmic flow in the
absence of magnetic field ($\bs{j}\propto(1/r)\bs{e}_r$, where
$\bs{e}_r$ is the unit vector in the radial direction). Applying an
external magnetic field that is orthogonal to the disk one can induce
an azimuthal, nondissipative Hall current (that is not compensated by
the Hall voltage due to the absence of boundaries). The resulting
inhomogeneous flows represent an excellent opportunity to study
viscous effects \cite{corb}. The Corbino disk with specular boundaries
was analyzed in Ref.~\cite{ady}. Assuming small momentum relaxation,
the authors of Ref.~\cite{ady} concluded that the Hall angle (that can
be determined by the ratio of the azimuthal and radial components of
the current) is directly related to the ratio of the Hall and shear
viscosities such that the resistive Hall angle approaches the viscous
Hall angle. Anomalous thermoelectric response (i.e., violating the
Matthiessen's rule, Wiedemann-Franz law, and Mott relation) exhibited
by hydrodynamic flows in the absence of Galilean invariance was
reported in Ref.~\cite{alex22}.

Recently, the Corbino geometry was used to demonstrate the
``superballistic conduction'' both experimentally \cite{sulp22} and
theoretically \cite{ady22,rai22}. Both theories focused on the
boundary effects. Reference~\cite{rai22} analyzed the radial electric
current (in the absence of magnetic field). In the hydrodynamic
regime, the interface between the lead (assumed to be a perfect
conductor) and the Corbino disk is characterized by the finite Knudsen
layer \cite{fal19} with the boundary conductance that can exceed the
Sharvin conductance \cite{sha}. Ref.~\cite{ady22} came to similar
conclusions arguing that if the number of conducting channels varies
along the current flow (using either a wormhole or Corbino geometries
as examples), the Landauer-Sharvin resistance is detached from the
leads and is spread throughout the bulk of the system. If the length
scale characterizing the spread is larger than $\ell_{ee}$ then the
resistance is reduced leading to superballistic conductance.

More complicated flow patterns can be achieved by considering curved
boundaries or adding artificial obstacles to engineer boundary
conditions \cite{msw2,msw}. In particular, on the basis of numerical
analysis it was shown \cite{msw} that additional barriers on the
channel walls may lead to the effective ``no-slip'' boundary
conditions that are commonly assumed in theoretical calculations.

\section{Nonlinear phenomena in electronic hydrodynamics}\label{sec4}

Nonlinear hydrodynamic effects in electronic systems remain largely
unexplored both theoretically and experimentally. Early numerical work
\cite{men11} suggests that electron flows with the high enough
Reynolds numbers (for samples of the size of $5\,\mu$m and macroscopic
speeds ${u\sim10^5}\,$m/s \cite{mer08}, the authors of
Ref.~\cite{men11} estimate ${{\rm Re}\sim100}$) may exhibit
pre-turbulent phenomena such as vortex shedding.

A representative example of nonlinear phenomena in
graphene –- hot spot relaxation –- was considered in
Ref.~\cite{hydro1}. A hot spot is a particular non-equilibrium state
of the system that is characterized by a locally elevated energy
density. This state can be prepared with the help of a local probe or
focused laser radiation \cite{fei12,chen12}. As expected
\cite{fei12,chen12}, the hot spot loses energy by emitting
plasmon-like waves. At charge neutrality, these are in fact acoustic
energy waves analogous to the long-wavelength oscillations in
interacting systems of relativistic particles [sometimes called the
  “cosmic sound”, see Eq.~(\ref{cs0})]. However, a nonzero excess
energy remains at the hot spot due to compensation between the
thermodynamic pressure and the self-consistent (Vlasov) electric
field. Dissipation tends to destroy the thus achieved
quasi-equilibrium, but the resulting decay is characterized by a
longer time scale as compared to the initial emission of plasmons. At
the same time, the plasmons appear to be damped by viscous effects,
see Sec.~\ref{sec3.3.3}. The plasmon emission can also be expected in
the in the high-frequency regime, where it has been linked to the
Cherenkov effect \cite{che,vav,tam,svin19}.

The above quasi-equilibrium solution \cite{hydro1} may be viewed as an
example of a soliton-like stationary nonlinear wave where charge and
energy fluctuations (otherwise distinct at charge neutrality) are
coupled by nonlinearity of the hydrodynamic theory. Away from charge
neutrality, solitons were considered in the inviscid limit in
Ref.~\cite{svin13} and more generally in Ref.~\cite{zdy19}. In
particular, the authors of Ref.~\cite{zdy19} focused on hydrodynamic
flows in graphene, where the decay of solitonic solutions was
suggested as a possible experimental measure of electronic viscosity.

\begin{figure}[t]
\centerline{
\includegraphics[width=0.3\columnwidth]{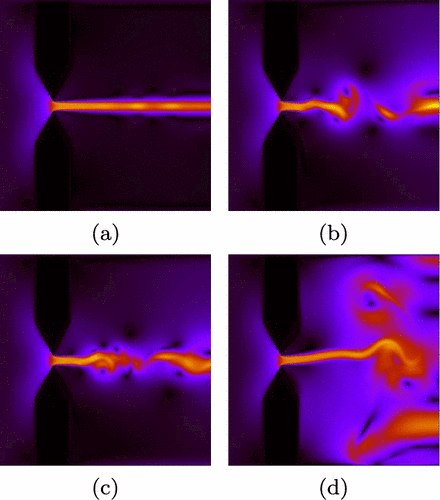}
\qquad
\includegraphics[width=0.3\columnwidth]{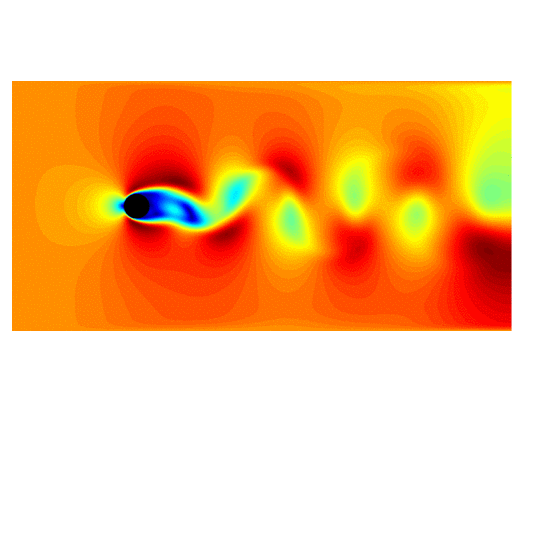}
}
\caption{Preturbulent hydrodynamic phenomena \cite{men11}. Left:
  microscale impurities in graphene can trigger coherent vorticity
  patterns that closely resemble classical 2D turbulence. The color
  represents the magnitude of the velocity. Calculations wee performed
  for ${\rm Re}=25$. Right: Vortex shedding in graphene at ${\rm
    Re}=100$ (Reprinted with permission from
  Ref.~\cite{men11}. Copyright (2011) by the American Physical
  Society).}
\label{fig:preturb}
\end{figure}

One of the most important consequences of nonlinearity of the
hydrodynamic equations -- turbulence \cite{dau6} -- is currently
regarded as unlikely to occur in electronic systems, e.g., in
graphene. In conventional fluids, turbulence can be reached when the
Reynolds number characterizing the flow becomes large, ${\rm
  Re}\gtrsim1000$ \cite{dau6}. In contrast, typical Reynolds numbers
characterizing existing experiments in graphene are rather
small. Indeed, assuming one of the highest reported values of the
drift velocity graphene, $u\sim10^5\,$m/s (based on the ``saturation
velocity measurements'' \cite{vin10}), the experimental estimate for
the kinematic viscosity $\nu\sim0.1\,$m$^2$/s \cite{geim1}, and a
typical sample size $L\sim1\,\mu$m, one can estimate the Reynolds
number as
\[
{\rm Re} = \frac{uL}{\nu} \sim 
\frac{10^5 \frac{\rm m}{\rm s} \times 10^{-6}{\rm m}}{0.1\frac{{\rm m}^2}{\rm s}} = 1.
\]
At such values of the Reynolds number, one may observe
``pre-turbulent'' phenomena, such as vortex shedding, as can be seen
by solving the hydrodynamic equations numerically \cite{men11}
(although at somewhat higher ${\rm Re}$, see
Fig.~\ref{fig:preturb}). For a possibility to achieve turbulence in
electronic systems other than graphene, see Ref.~\cite{ronny20}.

\begin{figure}[t]
\centerline{
\includegraphics[width=0.7\columnwidth]{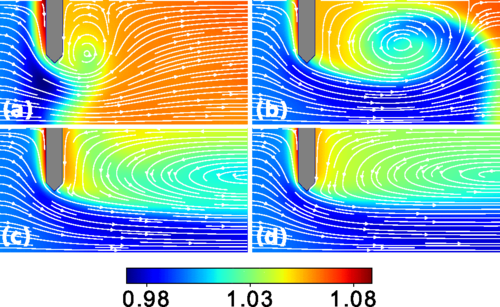}
}
\caption{Kelvin-Helmholz instability in graphene \cite{men17}. The
  color represents density fluctuations relative to the initial
  density. The streamlines show the direction of the hydrodynamic
  velocity. The gray object is the stationary obstacle. The four
  images are respective snapshots of the fluid motion taken at
  different times. Calculations wee performed for ${\rm Re}=53$
  (Reprinted with permission from Ref.~\cite{men17}. Copyright (2017)
  by the American Physical Society).}
\label{fig:kh}
\end{figure}

\begin{figure}[t]
\centerline{
\includegraphics[width=0.5\columnwidth]{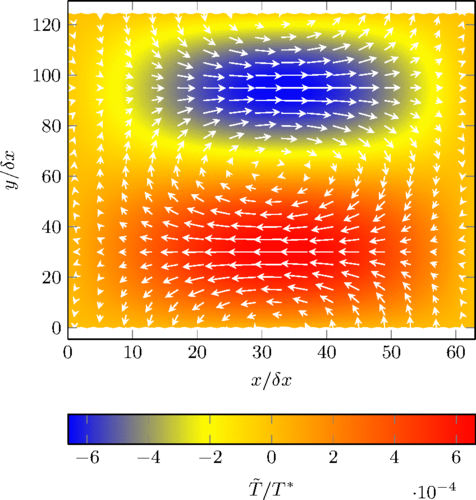}
}
\caption{Rayleigh-B{\'e}nard instability in graphene \cite{men14}. The
  color represents the temperature perturbation field with
  $T^*=100\,$K. The streamlines show the electron velocity. The image
  shows the formation of convection cells and the cosine-shaped
  temperature perturbation vanishing at the thermal contacts.
  (Reprinted with permission from Ref.~\cite{men14}. Copyright (2015)
  by the American Physical Society).}
\label{fig:rb}
\end{figure}

Nonlinearity of the Navier-Stokes equation also leads to a number of
known instabilities, arising in particular in systems with nontrivial
boundary conditions \cite{dau6}. One of these instabilities, the
Kelvin-Helmholz instability \cite{kel,hel}, was studied numerically in
Ref.~\cite{men17}, see Fig.~\ref{fig:kh}. In conventional fluids this
effect (actually visible in the atmosphere as a specific cloud
pattern, the ``fluctus'') occurs in the case of velocity shear within
a continuous fluid or at the interface between two fluids. In an
electronic system this can be achieved by directing a charge flow
through a macroscopic obstacle beyond which one observes vortex
formation \cite{men17} that is reminiscent of the ``whirlpools'' that
have been argued to be at the core of the nonlocal resistance
experiments \cite{geim1,sven1,pol15,pol16,fl0}. Similarly, numerical
simulations demonstrate the Rayleigh-B{\'e}nard instability
\cite{ben,ray}, see Ref.~\cite{men14} and Fig.~\ref{fig:rb}. Note that
the simulations of Ref.~\cite{men17} were performed using a lattice
Boltzmann method for relativistic gases. For more recent work on that
method see Ref.~\cite{baz21}.

In addition to the ``conventional'' instabilities of the hydrodynamic
equations, there is another instability that is predicted to occur in
a ballistic field effect transistor \cite{dsh} or, in other words, in
a gated 2D electron systems. There are two key observations leading to
the appearance of this instability. Firstly, the carrier density in
gated structures is determined by the same electric field (or
voltage), see Eq.~(\ref{v2dg}), that represents the driving term in
the Navier-Stokes equation (\ref{nseq}). In that case, the simplified
Navier-Stokes equation (i.e., in the absence of magnetic field,
neglecting Joule heating and weak disorder scattering) together with
the continuity equation closely resemble the standard hydrodynamic
equations for ``shallow water'' \cite{dau6}. Secondly, one requires
somewhat unusual (but experimentally feasible) boundary conditions: by
connecting the source and drain of the device to a current source and
the gate, while at the same time connecting the source to a voltage
source, one arrives at the setup with a constant value of the voltage
at the source together with the constant value of the current at the
drain. In that case the wave velocities (shallow water waves in
hydrodynamics or plasma waves in the heterostructure) describing
propagation in the opposite directions are different leading to the
instability with respect to plasmon generation. Known as the
``Dyakonov-Shur'' instability, this effect has attracted considerable
attention in literature, including that on hydrodynamic behavior in
graphene \cite{luc18,aiz21}; however, a definitive experimental
observation of the effect is still lacking. For a detailed numerical
analysis of a similar instability in GaAs MESFETs see
Ref.~\cite{Li17}. An alternative suggestion for using viscous
electrons as a source of terahertz radiation was proposed in
Ref.~\cite{luc21}. Dyakonov-Shur instability in the Corbino geometry
was discussed in Ref.~\cite{scaf22}.

\begin{figure}[t]
\centerline{
\includegraphics[width=0.292\columnwidth]{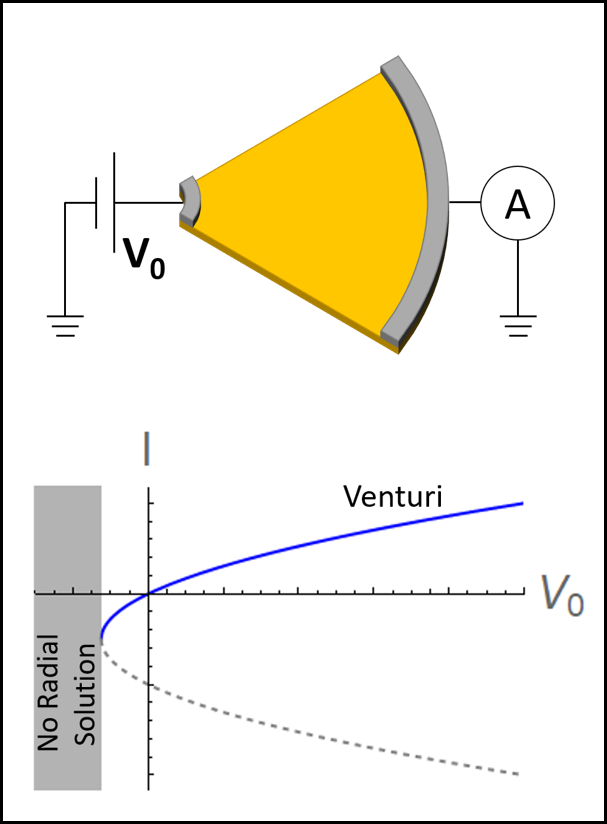}
\quad
\includegraphics[width=0.31\columnwidth]{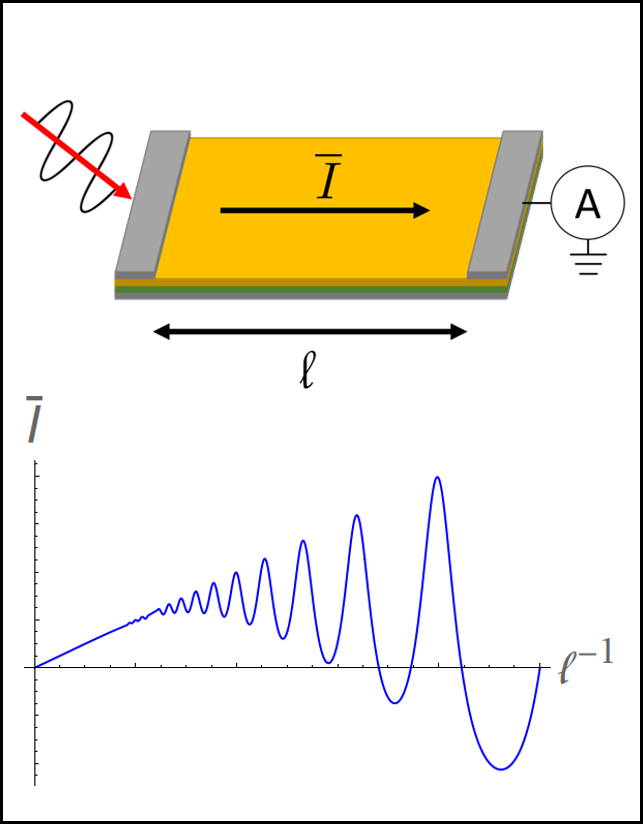}
\quad
\includegraphics[width=0.31\columnwidth]{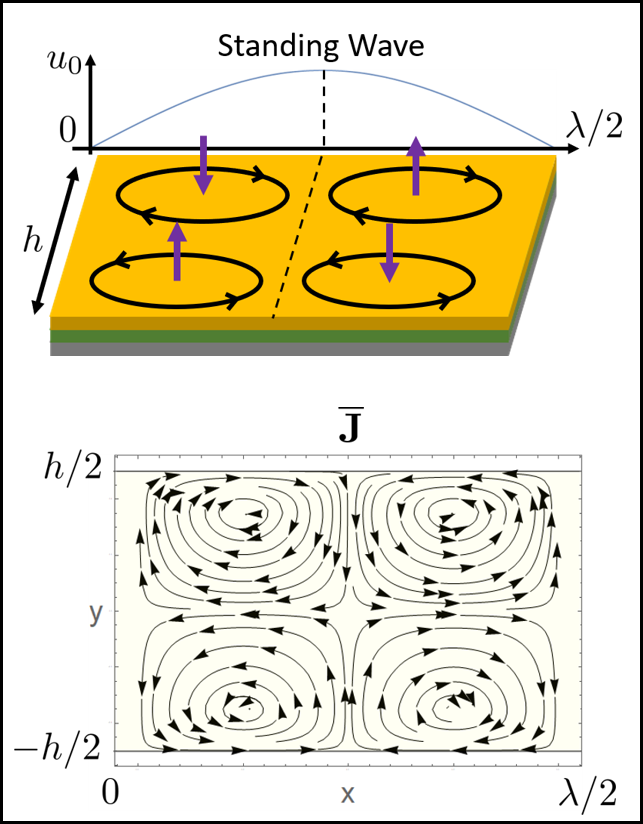}
}
\caption{Nonlinear hydrodynamic phenomena suggested in
  Ref.~\cite{oga21}. Left: the Venturi geometry and the expected
  nonlinear $I-V$ characteristic with $I\sim\sqrt{V}$ (the gray dashed
  line represents and unstable solution, while the gray area
  corresponds to the parameter regime of a possible instability towards
  turbulence). Center: Eckart streaming and the rectification
  effect. Right: Rayleigh streaming. (Reprinted with permission from
  Ref.~\cite{oga21}. Copyright (2021) by the American Physical
  Society).}
\label{fig:venturi}
\end{figure}

Further nonlinear phenomena were discussed in Ref.~\cite{oga21} where
three distinct hydrodynamic effects, namely the Bernoulli effect
\cite{bern}, Eckart streaming \cite{eck}, and Rayleigh streaming
\cite{ray2}, were suggested as possible experiments revealing
nonlinear electron fluid dynamics, see Fig.~\ref{fig:venturi}. The
suggested electronic analog of the Bernoulli effect yields a nonlinear
term in the I-V characteristic ($V\propto I^2$) in the ``Venturi
geometry'' (named after the Venturi tube, the standard device used for
demonstrating the Bernoulli effect), which is essentially a
finite-angle sector of the Corbino disk. The proposed effect is
strongly dependent on sample geometry (e.g., it is expected to vanish
in rectangular samples) and hence the boundary conditions. While the
stationary Bernoulli effect is expected to occur in the ideal
(inviscid) fluid, the dynamic nonlinear phenomena, such as the Eckart
and Rayleigh streaming, are expected to occur in the presence of
dissipation. Applying an oscillatory voltage to one of the sources
while grounding the drain, the authors of Ref.~\cite{oga21} find a dc
current (via the down-conversion). The two effects are distinguished
by whether the dominant dissipation occurs in the bulk (Eckart
streaming) or at the boundaries (Rayleigh streaming).

\section{Theoretical conjectures of hydrodynamic behavior in strongly correlated systems}\label{sec5}

A (relatively) recent discovery of gauge-gravity duality (or AdS/CFT
correspondence) \cite{mal98,hartnoll13,hartnoll18} offers a new
alternative theoretical tool to study strongly correlated systems by
relating strongly coupled quantum field theories to gravity theories
in one additional dimension. The best-known result of this approach is
the conjectured lower bound for the shear viscosity to entropy density
ratio \cite{kov} that has been found to be satisfied in quark-gluon
plasma \cite{Schafer2014}, cold atoms in the unitary limit
\cite{Thomas2009}, and intrinsic graphene \cite{msf}. The same physics
can also be expressed in terms of the diffusivity bound \cite{har15}.
Such bounds reflect not only the interaction strength, but also the
symmetry properties of the system. In particular, in anisotropic
systems the proposed bounds should be modified \cite{julia,ink20}.

In the condensed matter context, the duality has also been applied to
the by now perennial issue of the linear resistivity \cite{bruin} in
“strange metals” \cite{zaa13,zaa14} (cuprates
\cite{proust19,legros19}, iron-based superconductors
\cite{green10,john10,dai,si16}, twisted bilayer graphene at magic
angles \cite{tbg,cao20}, etc). The main premise of this approach is
that excitations in strongly correlated systems are predominantly of
the “collective” nature unlike the quasiparticles in conventional
metals \cite{hartnoll18}. In that case, the system is described by
“hydrodynamic-like” currents, with their relation to the external
fields provided by the standard linear response theory. This way one
can suggest universal bounds on the diffusion coefficient and
conductivity (related by the Einstein relation) of a strange metal, as
well as their scaling with temperature \cite{ink20,har15}. The concept
of diffusion appears through a particular collective mode (the
so-called quasinormal mode \cite{hub00}). Such modes essentially replace
quasiparticles in the qualitative interpretation of the resulting
theory \cite{hartnoll18}. Linear response transport properties can
then be obtained by means of either solving the hydrodynamic equations
or using the memory matrix formalism \cite{forster2018}. The latter
has the advantage of being independent of the concept of
quasiparticles and extending beyond the hydrodynamic regime.

The holographic duality can also be used in the opposite direction,
where solutions of hydrodynamic theories can provide insight into
physical properties of gravitational objects \cite{groz18}.

While they might appear too abstract, the holographic methods can be
put to test by studying the typical condensed matter experiment:
optical pump-probe spectroscopy \cite{bag18}. The idea is to test one
of the characteristic predictions of the bulk (gravity) side of the
duality – instantaneous thermalization \cite{bhat09}. This feature
(impossible in the usual semiclassical description of transport) is
the natural consequence of causality and is related to the “eigenstate
thermalization hypothesis” \cite{deutsch,sred}. As a result, measuring
the optical conductivity in a strange metal excited by a short,
intense laser pulse that does not contain a zero-frequency component
one should obtain the exact same results as in the same system at
equilibrium (characterized by the final temperature) at all times
after the pulse.

The linear resistivity has also been interpreted as a signature of
``Planckian dissipation'' \cite{zaa19,lucas19} (which is also related
to the above proposed bounds). The idea comes from the fact that the
observed optical conductivity in strange metals often allows for a
good fit with the standard Drude expression
\cite{bruin,legros19,zaa03,coop09} which is described by a timescale
typically referred to as the ``transport scattering time'',
$\tau_{\rm{tr}}$, \cite{ziman}. The linear temperature dependence of
the resistivity thus translates into the $\tau_{\rm tr}$ being inverse
proportional to temperature or, in other words, proportional to the
``Planckian'' timescale
\begin{equation}
\label{planck}
\tau_{\rm tr} \propto \tau_{P} = \frac{\hbar}{k_B T},
\end{equation}
where the Planck's and Boltzmann's constants ($\hbar$ and $k_B$) are
restored for clarity. While completely natural in neutral graphene,
see Eqs.~(\ref{sigma3}), where the temperature dependence
(\ref{planck}) follows already from dimensional analysis (in graphene
at charge neutrality, $T$ is the only energy scale), application of
the concept of the scattering time to strongly correlated systems is
more problematic. One possibility is that one can trace the decay of
correlation functions (which can be characterized through a
``transport'' time scale) to the decay of local operators, as
suggested in Ref.~\cite{lucas19}.

The hypothesis of the near-hydrodynamic behavior in strange metals (at
least, at low temperatures where the measured optical conductivity has
a Drude form) might sound attractive, but it certainly does not solve
all the problems \cite{zaa19}. At higher temperatures, there appears
the state of a “bad metal”, where the optical conductivity is no
longer of the Drude form \cite{del17}, while the temperature
dependence of the resistivity is still linear. Quantum Monte Carlo
simulations \cite{huang17} suggest that this state is accompanied by
hints of spin-stripe correlations \cite{and18}. While there might be a
way to include that physics into holographic modeling \cite{amo18},
the role of electron-phonon coupling, quantum criticality, and their
relation to the seemingly “universal” linear resistivity across
several distinct families of materials remains to be understood.

The above ideas on applying holographic methods to strange metals (in
particular in cuprates) remain controversial. For a recent critique of
this approach, see Ref.~\cite{khvesh21}.

A detailed discussion of relativistic hydrodynamics on the basis of
the AdS/CFT correspondence was offered in Ref.~\cite{erd18}. In the 2D
wire (channel) geometry with no-slip boundary conditions, this theory
yields the Poiseuille behavior (see Sec.~\ref{sec2}) for all
velocities up to the ultrarelativistic limit $u\rightarrow{v}_g$. In
the latter case, however, the differential resistance of the channel
vanishes as a consequence of the kinematics of special relativity. The
theory of Ref.~\cite{erd18} also offers further insights into the
importance of the shear viscosity to entropy density ratio,
$\eta/s$. Firstly, the channel resistance strongly scales with
$\eta/s$, such that ``holographic strongly coupled fluids'' (either at
or near the proposed bound $\eta/s\simeq1/(4\pi)$ \cite{kov}) are
characterized by smaller resistance in comparison to conventional
fluids. Secondly, the boundary relaxation time (i.e., the timescale
describing the rate of the loss of momentum at the channel boundaries
with no-slip boundary conditions) is inverse proportional to $\eta/s$.

\section{Open questions and perspectives}\label{sec6}

The scope of this review was mostly limited to observable effects that
can be interpreted as evidence of electronic hydrodynamics in graphene
and other 2D materials as well as theoretical work exploring
hydrodynamic phenomena in electronic systems. Several important topics
were purposefully left out, most notably the hydrodynamic behavior of
non-electronic excitations in solids, topological hydrodynamics, and
generalized hydrodynamics in 1D systems.

The initial argument for electronic hydrodynamics requiring the
electron-electron interaction to be the dominant scattering mechanism
implies the existence of scale separation between electronic
thermalization and energy relaxation due to, e.g., electron-phonon
interaction. The latter typically assumes that the phonons are in
thermal equilibrium. However, the current-carrying distribution of
electrons is generally nonequilibrium and hence electron-phonon
coupling can drive the phonons out of equilibrium as well
\cite{peierls}. The resulting phenomenon of phonon drag is
well studied \cite{lgu1,lgu2} and in particular allows for a
hydrodynamic description \cite{gum}. Recently, evidence of the coupled
electron-phonon fluid was reported in the Dirac semimetal PtSn$_4$
\cite{gooar} (for the theory see Ref.~\cite{lucas21}), the material
characterized by very low resistivity as well as showing a pronounced
phonon drag peak \cite{goo18} at low temperatures. Moreover, it was
argued \cite{lev20} that near-hydrodynamic behavior of electronic
transport in the delafossite metals PdCoO$_2$ and PtCoO$_2$
\cite{mac,nandi18} should be understood in the context of phonon drag.

Another aspect of the strong coupling between the electronic system
and the crystal lattice is the interplay between electronic viscosity
and elasticity of the crystal \cite{read,brad,julia1,cook,julia2}.
Moreover, static deformations in graphene are known to lead to the
appearance of giant pseudomagnetic fields \cite{levy10}. From a
general perspective, elasticity and hydrodynamics belong to a broader
class of tensor-field theories that also includes gravitation theories
and the theory of critical phenomena in spaces with nontrivial metrics
\cite{mnas}.

Observations of viscous hydrodynamics in electronic transport raised
the question of whether other excitations in solids might behave
hydrodynamically as well. In particular, the classic proposal for the
hydrodynamic behavior of spin waves \cite{hh69} recently came under
intense scrutiny both experimentally \cite{prasai} and theoretically
\cite{duine19,demler18}. Emergent hydrodynamics in a strongly
interacting dipolar spin ensemble (consisting of substitutional
nitrogen defects -- P1 centers -- and nitrogen-vacancy centers in
diamonds) was studied experimentally in Ref.~\cite{moo21}.

Generally speaking, hydrodynamic flows represent a macroscopic, long
wavelength motion governed by global conservation laws \cite{chai}. In
conventional fluids, these include the particle number, energy, and
momentum conservation allowing for a statistical description of the
system based on traditional Gibbs approach \cite{dau10}. In two- (and
three-) dimensional electronic systems energy and momentum are
conserved only approximately, which limits the applicability of the
hydrodynamic approach to a relatively narrow temperature interval
\cite{rev,luc,megt} as well as leads to unconventional behavior
\cite{mr1,mr2,sven2}. In one spatial dimension, in particular in the
context of integrable (exactly solvable) models, the situation is
different: here the system is characterized by a large number of
integrals of motion leading to the concept of {\it generalized} Gibbs
ensembles \cite{fabs15,langen15}. Applying the hydrodynamic approach
to the generalized Gibbs statistics yields generalized hydrodynamics
offering new possibilities in describing quantum transport in systems
with predominantly ballistic behavior (due to the large number of
conservation laws). This approach was introduced in the context of
integrable field theories \cite{doyon16} and quantum spin chains
\cite{nardis16} and was successfully applied to a number of other
integrable systems \cite{bert20}. The resulting framework was used to
describe one-dimensional cold atomic gases at large wavelengths
\cite{konik19} and has been observed experimentally
\cite{doyon19}. Generalized hydrodynamics in nonintegrable systems was
studied in Ref.~\cite{vass21}.

Another topic outside of the scope of this review is topological
hydrodynamics, see Ref.~\cite{tser18} and references
therein. Recently, an optical topological invariant (measurable via
the evanescent magneto-optic Kerr effect \cite{kerr}) was proposed to
describe properties of the viscous Hall fluid \cite{jack21} suggesting
that graphene with the ``repulsive'' Hall viscosity (i.e.,
$\omega_c\nu_H>0$) may be used to create a topological
electromagnetic phase of matter. Especially interesting in this
context is the interplay of topological band structure and
electron-electron interactions (responsible for establishing the local
equilibrium underlying the usual hydrodynamic theory). A related issue
is quantum hydrodynamics of vorticity \cite{tser19} describing
vortex-antivortex dynamics in 2D bosonic lattices pertaining to the
superfluid-insulator transition.

While some experimental work on hydrodynamics in topological materials
was addressed in Sec.~\ref{sec2.5.2}, the theoretical discussion of
Sec.~\ref{sec3} focused on the well-studied cases of the 2D Dirac and
Fermi liquids. In contrast, a hydrodynamic theory of topological
materials (including Weyl semimetals \cite{lucnmr,gorbar} and
conducting surface states of topological insulators) has not been
hammered out yet. For recent literature on this subject see
Refs.~\cite{spivak13,gal18,yin19,cop19,kawa20,has21}. The effects of
band topology on the shear viscosity were considered in
Ref.~\cite{sur20}.

Despite the impressive amount of recent work on the subject,
electronic hydrodynamics remains a young field with many unanswered
questions. So far, the main focus of the community was on
Fermi-liquid-type materials (including doped graphene), where the
hydrodynamic equations are basically equivalent to the standard
Navier-Stokes equation and the velocity field completely determines
the electric current. Even in this simplest setting, the question of
boundary conditions remains largely unresolved, especially in view of
the experiment of Ref.~\cite{zel}. Furthermore, practical applications
of hydrodynamic equations require reliable tools for their numerical
solution. Although there exists a massive amount of literature devoted
to solution of differential equations (as well as commercial and open
source software packages dealing with their numerical solutions),
equations of electronic hydrodynamics have to be solved together with
the equations describing the electrostatic environment of the system,
the electronic circuit into which the system is integrated, and in the
case of spintronics applications – the magnetic environment. Combining
all these aspects of the problem with the realistic boundary
conditions and specific symmetries represents a formidable
computational problem that is rather difficult to solve using the
available “canned” solvers \cite{che1,che2}.

Transport properties addressed with the hydrodynamic approach so far
remain at the semiclassical ``Drude'' level (which is not surprising
given that the hydrodynamic equations, see Sec.~\ref{sec3.2.6}, were
derived from the semiclassical Boltzmann equation, see
Sec.~\ref{sec3}. In contrast, the traditional transport theory
considers also ``higher order'' processes leading to the so-called
``quantum corrections'' \cite{aar,zna,aag}. While typically discussed
using field-theoretic methods, these results can also be obtained
within the kinetic approach (for the corresponding ``quantum kinetic
equation'' see Ref.~\cite{zna}). It remains to be seen, whether this
physics can be included in a macroscopic, hydrodynamic-type
description. Moreover, it is unclear whether one can establish any
relation between the well-known hydrodynamic fluctuations \cite{dau6}
and mesoscopic fluctuations in conventional diffusive conductors
\cite{AltLeeWeb1991,df1,df2}.

One of the most intriguing promises of the hydrodynamic approach is
its supposed ability to describe properties of more complicated
systems, including the ``strange metals'' \cite{zaa13}. This direction
of research is still in its infancy. Many novel materials (including
van der Waals heterostructures \cite{tbg,geimgrig}, conducting surface
states of topological insulators, and Weyl semimetals) are
characterized by strong spin-orbit coupling. Up until now, a coherent
kinetic theory for electrons with spin-orbit interaction has not been
established, see Refs.~\cite{aka,mish03,sin07,sin08}. An advance in
this direction could provide a substantial contribution to the
application-oriented field of spintronics \cite{fabian04,sinova15},
which has been under active development in the last two decades.

The conjecture of Planckian dissipation does not explain why does the
observed resistivity remains linear (i.e., does not vanish faster) in
different materials where different scattering mechanisms are presumed
to be relevant in different temperature regimes \cite{bruin}. This
could indicate an existence of a universal principle limiting the
decay rate of the longest lived modes in strongly correlated systems
(similarly to the phase space limitations on the quasiparticle
properties in Fermi liquids). Such principle has not been
established yet.

Finally, the issues raised in the course of the rapid development of
electronic hydrodynamics are of fundamental importance for the physics
of novel electronic systems necessary for the future development of
functional materials. Future advances in this field will have
far-reaching implications beyond the scope of particular systems
considered in this review. They will substantially improve our
understanding of interrelation of macroscopic transport properties (of
charge, spin, and heat) and microscopic structure (symmetry
properties, band structure, electronic correlations) of materials
allowing for material engineering and functionalization.

\begin{acknowledgements}
The author is grateful to I. Aleiner, P. Alekseev, U. Briskot,
I. Burmistrov, A. Dmitriev, I. Gornyi, V. Kachorovskii, E. Kiselev,
A. Mirlin, J. Schmalian, M. Sch\"utt, A. Shnirman, and M. Titov for
fruitful discussions. This work was supported by the German Research
Foundation DFG project NA 1114/5-1, by the German Research Foundation
DFG within FLAG-ERA Joint Transnational Call (Project GRANSPORT), by
the European Commission under the EU Horizon 2020 MSCA-RISE-2019
program (Project 873028 HYDROTRONICS).
\end{acknowledgements}

%
%

\bibliographystyle{spphys}       
\bibliography{refs-books,hydro-refs}   

\begin{thebibliography}{100}
\providecommand{\url}[1]{{#1}}
\providecommand{\urlprefix}{URL }
\expandafter\ifx\csname urlstyle\endcsname\relax
  \providecommand{\doi}[1]{DOI \discretionary{}{}{}#1}\else
  \providecommand{\doi}{DOI \discretionary{}{}{}\begingroup
  \urlstyle{rm}\Url}\fi

\bibitem{chai}
P.M. Chaikin, T.C. Lubensky, \emph{Principles of Condensed Matter Physics}
  (Cambridge University Press, 1995).
\newblock \leavevmode\\ \doi{10.1017/CBO9780511813467}

\bibitem{blo}
N.~Bloembergen, Physica \textbf{15}, 386 (1949).
\newblock \leavevmode\\ \doi{10.1016/0031-8914(49)90114-7}

\bibitem{kad}
L.P. Kadanoff, P.C. Martin, Ann. Phys. (NY) \textbf{24}, 419 (1963).
\newblock \leavevmode\\ \doi{10.1016/0003-4916(63)90078-2}

\bibitem{tuc}
A.~Tucciarone, J.M. Hastings, L.M. Corliss, Phys. Rev. B \textbf{8}, 1103
  (1973).
\newblock \leavevmode\\ \doi{10.1103/PhysRevB.8.1103}

\bibitem{sdme}
B.N. Narozhny, Phys. Rev. B \textbf{54}, 3311 (1996).
\newblock \leavevmode\\ \doi{10.1103/PhysRevB.54.3311}

\bibitem{xxz}
B.N. Narozhny, A.J. Millis, N.~Andrei, Phys. Rev. B \textbf{58}, R2921 (1998).
\newblock \leavevmode\\ \doi{10.1103/PhysRevB.58.R2921}

\bibitem{sdnat}
A.~Rosch, N.~Andrei, Phys. Rev. Lett. \textbf{85}, 1092 (2000).
\newblock \leavevmode\\ \doi{10.1103/PhysRevLett.85.1092}

\bibitem{sdaff}
J.~Sirker, R.G. Pereira, I.~Affleck, Phys. Rev. Lett. \textbf{103}, 216602
  (2009).
\newblock \leavevmode\\ \doi{10.1103/PhysRevLett.103.216602}

\bibitem{aar}
B.L. Altshuler, A.G. Aronov, in \emph{Electron-Electron Interactions in
  Disordered Systems}, ed. by A.L. Efros, M.~Pollak (North-Holland, Amsterdam,
  1985).
\newblock \leavevmode\\ \doi{10.1016/B978-0-444-86916-6.50007-7}

\bibitem{ziman}
J.M. Ziman, \emph{Principles of the Theory of Solids} (Cambridge University
  Press, Cambridge, 1965).
\newblock \leavevmode\\ \doi{10.1017/CBO9781139644075}

\bibitem{bee}
C.~Beenakker, H.~{van Houten}, in \emph{Semiconductor Heterostructures and
  Nanostructures}, \emph{Solid State Physics}, vol.~44, ed. by H.~Ehrenreich,
  D.~Turnbull (Academic Press, 1991), pp. 1--228.
\newblock \leavevmode\\ \doi{10.1016/S0081-1947(08)60091-0}

\bibitem{zna}
G.~Zala, B.N. Narozhny, I.L. Aleiner, Phys. Rev. B \textbf{64}, 214204 (2001).
\newblock \leavevmode\\ \doi{10.1103/PhysRevB.64.214204}

\bibitem{dau6}
L.D. Landau, E.M. Lifshitz, \emph{Fluid Mechanics} (Pergamon Press, London,
  1987).
\newblock \leavevmode\\ \doi{10.1016/C2013-0-03799-1}

\bibitem{lamb}
H.~Lamb, \emph{Hydrodynamics} (Dover, New York, 1945)

\bibitem{wolf}
D.~Vollhardt, P.~W\"olfle, \emph{The Superfluid Phases of Helium 3} (Taylor and
  Francis, London, 1990).
\newblock \leavevmode\\ \doi{10.1201/b12808}

\bibitem{dau10}
E.M. Lifshitz, L.P. Pitaevskii, \emph{Physical Kinetics} (Pergamon Press,
  London, 1981).
\newblock \leavevmode\\ \doi{10.1016/C2009-0-25523-1}

\bibitem{gurzhi}
R.N. Gurzhi, Soviet Physics Uspekhi \textbf{11}(2), 255 (1968).
\newblock [Usp. Fiz. Nauk {\bf 94}, 689 (1968)].
\newblock \leavevmode\\ \doi{10.1070/PU1968v011n02ABEH003815}

\bibitem{beh}
Y.~Machida, N.~Matsumoto, T.~Isono, K.~Behnia, Science \textbf{367}(6475), 309
  (2020).
\newblock \leavevmode\\ \doi{10.1126/science.aaz8043}

\bibitem{hh69}
B.I. Halperin, P.C. Hohenberg, Phys. Rev. \textbf{188}, 898 (1969).
\newblock \leavevmode\\ \doi{10.1103/PhysRev.188.898}

\bibitem{gurzhi1}
R.N. Gurzhi, Zh. Eksp. Teor. Fiz. \textbf{44}, 771 (1963).
\newblock [Sov. Phys. JETP {\bf 17}, 521 (1963)]

\bibitem{gurzhi2}
R.N. Gurzhi, Zh. Eksp. Teor. Fiz. \textbf{47}, 1415 (1965).
\newblock [Sov. Phys. JETP {\bf 20}, 953 (1965)]

\bibitem{dsh}
M.~Dyakonov, M.~Shur, Phys. Rev. Lett. \textbf{71}, 2465 (1993).
\newblock \leavevmode\\ \doi{10.1103/PhysRevLett.71.2465}

\bibitem{rev}
B.N. Narozhny, I.V. Gornyi, A.D. Mirlin, J.~Schmalian, Annalen der Physik
  \textbf{529}(11), 1700043 (2017).
\newblock \leavevmode\\ \doi{10.1002/andp.201700043}

\bibitem{luc}
A.~Lucas, K.C. Fong, J. Phys: Condens. Matter \textbf{30}(5), 053001 (2018).
\newblock \leavevmode\\ \doi{10.1088/1361-648x/aaa274}

\bibitem{pg}
M.~Polini, A.K. Geim, Phys. Today \textbf{73}(6), 28 (2020).
\newblock \leavevmode\\ \doi{10.1063/PT.3.4497}

\bibitem{akha}
A.A. Abrikosov, I.M. Khalatnikov, Rep. Prog. Phys. \textbf{22}(1), 329 (1959).
\newblock \leavevmode\\ \doi{10.1088/0034-4885/22/1/310}

\bibitem{geim1}
D.A. Bandurin, I.~Torre, R.~{Krishna Kumar}, M.~Ben~Shalom, A.~Tomadin,
  A.~Principi, G.H. Auton, E.~Khestanova, K.S. Novoselov, I.V. Grigorieva, L.A.
  Ponomarenko, A.K. Geim, M.~Polini, Science \textbf{351}, 1055 (2016).
\newblock \leavevmode\\ \doi{10.1126/science.aad0201}

\bibitem{kim1}
J.~Crossno, J.K. Shi, K.~Wang, X.~Liu, A.~Harzheim, A.~Lucas, S.~Sachdev,
  P.~Kim, T.~Taniguchi, K.~Watanabe, T.A. Ohki, K.C. Fong, Science
  \textbf{351}(6277), 1058 (2016).
\newblock \leavevmode\\ \doi{10.1126/science.aad0343}

\bibitem{mac}
P.J.W. Moll, P.~Kushwaha, N.~Nandi, B.~Schmidt, A.P. Mackenzie, Science
  \textbf{351}(6277), 1061 (2016).
\newblock \leavevmode\\ \doi{10.1126/science.aac8385}

\bibitem{ihn}
B.A. Braem, F.M.D. Pellegrino, A.~Principi, M.~R\"o\"osli, C.~Gold, S.~Hennel,
  J.V. Koski, M.~Berl, W.~Dietsche, W.~Wegscheider, M.~Polini, T.~Ihn,
  K.~Ensslin, Phys. Rev. B \textbf{98}, 241304(R) (2018).
\newblock \leavevmode\\ \doi{10.1103/PhysRevB.98.241304}

\bibitem{goo}
A.~Jaoui, B.~Fauqu\'e, C.W. Rischau, A.~Subedi, C.~Fu, J.~Gooth, N.~Kumar,
  V.~S\"u{\ss}, D.L. Maslov, C.~Felser, K.~Behnia, npj Quant. Mater.
  \textbf{3}, 64 (2018).
\newblock \leavevmode\\ \doi{10.1038/s41535-018-0136-x}

\bibitem{gus20}
G.M. Gusev, A.S. Jaroshevich, A.D. Levin, Z.D. Kvon, A.K. Bakarov, Sci. Rep.
  \textbf{10}(1), 7860 (2020).
\newblock \leavevmode\\ \doi{10.1038/s41598-020-64807-6}

\bibitem{var20}
G.~Varnavides, A.S. Jermyn, P.~Anikeeva, C.~Felser, P.~Narang, Nat. Commun.
  \textbf{11}(1), 4710 (2020).
\newblock \leavevmode\\ \doi{10.1038/s41467-020-18553-y}

\bibitem{vool}
U.~Vool, A.~Hamo, G.~Varnavides, Y.~Wang, T.X. Zhou, N.~Kumar, Y.~Dovzhenko,
  Z.~Qiu, C.A.C. Garcia, A.T. Pierce, J.~Gooth, P.~Anikeeva, C.~Felser,
  P.~Narang, A.~Yacoby, Nat. Phys.  (2021).
\newblock \leavevmode\\ \doi{10.1038/s41567-021-01341-w}

\bibitem{jao}
A.~Jaoui, B.~Fauqu{\'e}, K.~Behnia, Nat. Commun. \textbf{12}(1), 195 (2021).
\newblock \leavevmode\\ \doi{10.1038/s41467-020-20420-9}

\bibitem{gupta}
A.~Gupta, J.J. Heremans, G.~Kataria, M.~Chandra, S.~Fallahi, G.C. Gardner, M.J.
  Manfra, Phys. Rev. Lett. \textbf{126}, 076803 (2021).
\newblock \leavevmode\\ \doi{10.1103/PhysRevLett.126.076803}

\bibitem{glazman20}
S.S. Pershoguba, A.F. Young, L.I. Glazman, Phys. Rev. B \textbf{102}, 125404
  (2020).
\newblock \leavevmode\\ \doi{10.1103/PhysRevB.102.125404}

\bibitem{adam}
D.Y.H. Ho, I.~Yudhistira, N.~Chakraborty, S.~Adam, Phys. Rev. B \textbf{97},
  121404(R) (2018).
\newblock \leavevmode\\ \doi{10.1103/PhysRevB.97.121404}

\bibitem{geim2}
R.~{Krishna Kumar}, D.A. Bandurin, F.M.D. Pellegrino, Y.~Cao, A.~Principi,
  H.~Guo, G.H. Auton, M.~{Ben Shalom}, L.A. Ponomarenko, G.~Falkovich,
  K.~Watanabe, T.~Taniguchi, I.V. Grigorieva, L.S. Levitov, M.~Polini, A.K.
  Geim, Nat. Phys. \textbf{13}(12), 1182 (2017).
\newblock \leavevmode\\ \doi{10.1038/nphys4240}

\bibitem{kim2}
F.~Ghahari, H.Y. Xie, T.~Taniguchi, K.~Watanabe, M.S. Foster, P.~Kim, Phys.
  Rev. Lett. \textbf{116}, 136802 (2016).
\newblock \leavevmode\\ \doi{10.1103/PhysRevLett.116.136802}

\bibitem{geim3}
D.A. Bandurin, A.V. Shytov, L.S. Levitov, R.~{Krishna Kumar}, A.I. Berdyugin,
  M.~{Ben Shalom}, I.V. Grigorieva, A.K. Geim, G.~Falkovich, Nat. Commun.
  \textbf{9}(1), 4533 (2018).
\newblock \leavevmode\\ \doi{10.1038/s41467-018-07004-4}

\bibitem{geim4}
A.I. Berdyugin, S.G. Xu, F.M.D. Pellegrino, R.~{Krishna Kumar}, A.~Principi,
  I.~Torre, M.B. Shalom, T.~Taniguchi, K.~Watanabe, I.V. Grigorieva, M.~Polini,
  A.K. Geim, D.A. Bandurin, Science \textbf{364}(6436), 162 (2019).
\newblock \leavevmode\\ \doi{10.1126/science.aau0685}

\bibitem{gal}
P.~Gallagher, C.S. Yang, T.~Lyu, F.~Tian, R.~Kou, H.~Zhang, K.~Watanabe,
  T.~Taniguchi, F.~Wang, Science \textbf{364}(6436), 158 (2019).
\newblock \leavevmode\\ \doi{10.1126/science.aat8687}

\bibitem{young}
A.~Jenkins, S.~Baumann, H.~Zhou, S.A. Meynell, D.~Yang, T.T. K.~Watanabe,
  A.~Lucas, A.F. Young, A.C. {Bleszynski Jayich} (2020).
\newblock \leavevmode\\ \doi{10.48550/arXiv.2002.05065}

\bibitem{brar}
Z.J. Krebs, W.A. Behn, S.~Li, K.J. Smith, K.~Watanabe, T.~Taniguchi,
  A.~Levchenko, V.W. Brar (2021).
\newblock \leavevmode\\ \doi{10.48550/arXiv.2106.07212}

\bibitem{fink}
A.~Finkler, Y.~Segev, Y.~Myasoedov, M.L. Rappaport, L.~Ne'eman, D.~Vasyukov,
  E.~Zeldov, M.E. Huber, J.~Martin, A.~Yacoby, Nano Lett. \textbf{10}(3), 1046
  (2010).
\newblock \leavevmode\\ \doi{10.1021/nl100009r}

\bibitem{halb}
D.~Halbertal, J.~Cuppens, M.~{Ben Shalom}, L.~Embon, N.~Shadmi, Y.~Anahory,
  H.R. Naren, J.~Sarkar, A.~Uri, Y.~Ronen, Y.~Myasoedov, L.S. Levitov,
  E.~Joselevich, A.K. Geim, E.~Zeldov, Nature \textbf{539}(7629), 407 (2016).
\newblock \leavevmode\\ \doi{10.1038/nature19843}

\bibitem{zel19}
A.~Marguerite, J.~Birkbeck, A.~Aharon-Steinberg, D.~Halbertal, K.~Bagani,
  I.~Marcus, Y.~Myasoedov, A.K. Geim, D.J. Perello, E.~Zeldov, Nature
  \textbf{575}(7784), 628 (2019).
\newblock \leavevmode\\ \doi{10.1038/s41586-019-1704-3}

\bibitem{uri}
A.~Uri, Y.~Kim, K.~Bagani, C.K. Lewandowski, S.~Grover, N.~Auerbach, E.O.
  Lachman, Y.~Myasoedov, T.~Taniguchi, K.~Watanabe, J.~Smet, E.~Zeldov, Nat.
  Phys. \textbf{16}(2), 164 (2020).
\newblock \leavevmode\\ \doi{10.1038/s41567-019-0713-3}

\bibitem{imh}
M.J.H. Ku, T.X. Zhou, Q.~Li, Y.J. Shin, J.K. Shi, C.~Burch, L.E. Anderson, A.T.
  Pierce, Y.~Xie, A.~Hamo, U.~Vool, H.~Zhang, F.~Casola, T.~Taniguchi,
  K.~Watanabe, M.M. Fogler, P.~Kim, A.~Yacoby, R.L. Walsworth, Nature
  \textbf{583}, 537 (2020).
\newblock \leavevmode\\ \doi{10.1038/s41586-020-2507-2}

\bibitem{imm}
J.A. Sulpizio, L.~Ella, A.~Rozen, J.~Birkbeck, D.J. Perello, D.~Dutta,
  M.~Ben-Shalom, T.~Taniguchi, K.~Watanabe, T.~Holder, R.~Queiroz, A.~Stern,
  T.~Scaffidi, A.K. Geim, S.~Ilani, Nature \textbf{576}, 75 (2019).
\newblock \leavevmode\\ \doi{10.1038/s41586-019-1788-9}

\bibitem{sulp}
L.~Ella, A.~Rozen, J.~Birkbeck, M.~Ben-Shalom, D.~Perello, J.~Zultak,
  T.~Taniguchi, K.~Watanabe, A.K. Geim, S.~Ilani, J.A. Sulpizio, Nat.
  Nanotechnol. \textbf{14}, 480 (2019).
\newblock \leavevmode\\ \doi{10.1038/s41565-019-0398-x}

\bibitem{zel}
A.~Aharon-Steinberg, A.~Marguerite, D.J. Perello, K.~Bagani, T.~Holder,
  Y.~Myasoedov, L.S. Levitov, A.K. Geim, E.~Zeldov, Nature \textbf{593}, 528
  (2021).
\newblock \leavevmode\\ \doi{10.1038/s41586-021-03501-7}

\bibitem{sulp22}
C.~Kumar, J.~Birkbeck, J.A. Sulpizio, D.J. Perello, T.~Taniguchi, K.~Watanabe,
  O.~Reuven, T.~Scaffidi, A.~Stern, A.K. Geim, S.~Ilani (2021).
\newblock \leavevmode\\ \doi{10.48550/arXiv.2111.06412}

\bibitem{gus18}
G.M. Gusev, A.D. Levin, E.V. Levinson, A.K. Bakarov, Phys. Rev. B \textbf{98},
  161303(R) (2018).
\newblock \leavevmode\\ \doi{10.1103/PhysRevB.98.161303}

\bibitem{rai20}
O.E. Raichev, G.M. Gusev, A.D. Levin, A.K. Bakarov, Phys. Rev. B \textbf{101},
  235314 (2020).
\newblock \leavevmode\\ \doi{10.1103/PhysRevB.101.235314}

\bibitem{gus21}
G.M. Gusev, A.S. Jaroshevich, A.D. Levin, Z.D. Kvon, A.K. Bakarov, Phys. Rev. B
  \textbf{103}, 075303 (2021).
\newblock \leavevmode\\ \doi{10.1103/PhysRevB.103.075303}

\bibitem{mol95}
M.J.M. de~Jong, L.W. Molenkamp, Phys. Rev. B \textbf{51}, 13389 (1995).
\newblock \leavevmode\\ \doi{10.1103/PhysRevB.51.13389}

\bibitem{lucnmr}
A.~Lucas, R.A. Davison, S.~Sachdev, Proc. Natl. Acad. Sci. \textbf{113}(34),
  9463 (2016).
\newblock \leavevmode\\ \doi{10.1073/pnas.1608881113}

\bibitem{gorbar}
E.V. Gorbar, V.A. Miransky, I.A. Shovkovy, P.O. Sukhachov, Phys. Rev. B
  \textbf{97}, 205119 (2018).
\newblock \leavevmode\\ \doi{10.1103/PhysRevB.97.205119}

\bibitem{jackiw69}
J.S. Bell, R.~Jackiw, Nuovo Cimento A \textbf{60}, 47 (1969).
\newblock \leavevmode\\ \doi{10.1007/BF02823296}

\bibitem{adler}
S.L. Adler, Phys. Rev. \textbf{177}, 2426 (1969).
\newblock \leavevmode\\ \doi{10.1103/PhysRev.177.2426}

\bibitem{jackiw2010}
R.~Jackiw, Int. J. Mod. Phys. A \textbf{25}(04), 659 (2010).
\newblock \leavevmode\\ \doi{10.1142/S0217751X10048391}

\bibitem{fel16}
C.~Felser, B.~Yan, Nat. Mater. \textbf{15}(11), 1149 (2016).
\newblock \leavevmode\\ \doi{10.1038/nmat4741}

\bibitem{ong16}
M.~Hirschberger, S.~Kushwaha, Z.~Wang, Q.~Gibson, S.~Liang, C.A. Belvin, B.A.
  Bernevig, R.J. Cava, N.P. Ong, Nat. Mater. \textbf{15}(11), 1161 (2016).
\newblock \leavevmode\\ \doi{10.1038/nmat4684}

\bibitem{spivak13}
D.T. Son, B.Z. Spivak, Phys. Rev. B \textbf{88}, 104412 (2013).
\newblock \leavevmode\\ \doi{10.1103/PhysRevB.88.104412}

\bibitem{fel162}
F.~Arnold, C.~Shekhar, S.C. Wu, Y.~Sun, R.D. dos Reis, N.~Kumar, M.~Naumann,
  M.O. Ajeesh, M.~Schmidt, A.G. Grushin, J.H. Bardarson, M.~Baenitz,
  D.~Sokolov, H.~Borrmann, M.~Nicklas, C.~Felser, E.~Hassinger, B.~Yan, Nat.
  Commun. \textbf{7}(1), 11615 (2016).
\newblock \leavevmode\\ \doi{10.1038/ncomms11615}

\bibitem{knu}
M.~Knudsen, Annalen der Physik \textbf{333}(1), 75 (1909).
\newblock [Ann. Phys. (Leipzig) {\bf 28}, 75 (1909)].
\newblock \leavevmode\\ \doi{10.1002/andp.19093330106}

\bibitem{pauw}
L.J. {van der Pauw}, Philips Tech. Rev. \textbf{20}, 223 (1958)

\bibitem{nlr}
D.A. Abanin, S.V. Morozov, L.A. Ponomarenko, R.V. Gorbachev, A.S. Mayorov, M.I.
  Katsnelson, K.~Watanabe, T.~Taniguchi, K.S. Novoselov, L.S. Levitov, A.K.
  Geim, Science \textbf{332}, 328 (2011).
\newblock \leavevmode\\ \doi{10.1126/science.1199595}

\bibitem{dru}
P.~Drude, Annalen der Physik \textbf{306}(1), 566 (1900).
\newblock [Ann. Phys. (Leipzig) {\bf 1}, 566 (1900)].
\newblock \leavevmode\\ \doi{10.1002/andp.19003060312}

\bibitem{skocpol}
W.J. Skocpol, P.M. Mankiewich, R.E. Howard, L.D. Jackel, D.M. Tennant, A.D.
  Stone, Phys. Rev. Lett. \textbf{58}, 2347 (1987).
\newblock \leavevmode\\ \doi{10.1103/PhysRevLett.58.2347}

\bibitem{vwees1}
H.~{van Houten}, C.W.J. Beenakker, J.G. Williamson, M.E.I. Broekaart, P.H.M.
  {van Loosdrecht}, B.J. {van Wees}, J.E. Mooij, C.T. Foxon, J.J. Harris, Phys.
  Rev. B \textbf{39}, 8556 (1989).
\newblock \leavevmode\\ \doi{10.1103/PhysRevB.39.8556}

\bibitem{geimnl1}
A.K. Geim, P.C. Main, P.H. Beton, P.~Steda, L.~Eaves, C.D.W. Wilkinson, S.P.
  Beaumont, Phys. Rev. Lett. \textbf{67}, 3014 (1991).
\newblock \leavevmode\\ \doi{10.1103/PhysRevLett.67.3014}

\bibitem{roukes}
K.L. Shepard, M.L. Roukes, B.P. {van der Gaag}, Phys. Rev. Lett. \textbf{68},
  2660 (1992).
\newblock \leavevmode\\ \doi{10.1103/PhysRevLett.68.2660}

\bibitem{vonklit}
Y.~Hirayama, A.D. Wieck, T.~Bever, K.~von Klitzing, K.~Ploog, Phys. Rev. B
  \textbf{46}, 4035 (1992).
\newblock \leavevmode\\ \doi{10.1103/PhysRevB.46.4035}

\bibitem{nlgold}
G.~Mihajlovi\ifmmode~\acute{c}\else \'{c}\fi{}, J.E. Pearson, M.A. Garcia, S.D.
  Bader, A.~Hoffmann, Phys. Rev. Lett. \textbf{103}, 166601 (2009).
\newblock \leavevmode\\ \doi{10.1103/PhysRevLett.103.166601}

\bibitem{geimnl2}
R.V. Gorbachev, J.C.W. Song, G.L. Yu, A.V. Kretinin, F.~Withers, Y.~Cao,
  A.~Mishchenko, I.V. Grigorieva, K.S. Novoselov, L.S. Levitov, A.K. Geim,
  Science \textbf{346}(6208), 448 (2014).
\newblock \leavevmode\\ \doi{10.1126/science.1254966}

\bibitem{mceu}
P.L. McEuen, A.~Szafer, C.A. Richter, B.W. Alphenaar, J.K. Jain, A.D. Stone,
  R.G. Wheeler, R.N. Sacks, Phys. Rev. Lett. \textbf{64}, 2062 (1990).
\newblock \leavevmode\\ \doi{10.1103/PhysRevLett.64.2062}

\bibitem{goldman}
J.K. Wang, V.J. Goldman, Phys. Rev. B \textbf{45}, 13479 (1992).
\newblock \leavevmode\\ \doi{10.1103/PhysRevB.45.13479}

\bibitem{roth}
A.~Roth, C.~Brune, H.~Buhmann, L.W. Molenkamp, J.~Maciejko, X.L. Qi, S.C.
  Zhang, Science \textbf{325}(5938), 294 (2009).
\newblock \leavevmode\\ \doi{10.1126/science.1174736}

\bibitem{caza}
X.P. Zhang, C.~Huang, M.A. Cazalilla, 2D Materials \textbf{4}(2), 024007
  (2017).
\newblock \leavevmode\\ \doi{10.1088/2053-1583/aa5e9b}

\bibitem{koma}
K.~Komatsu, Y.~Morita, E.~Watanabe, D.~Tsuya, K.~Watanabe, T.~Taniguchi,
  S.~Moriyama, Science Advances \textbf{4}(5), eaaq0194 (2018).
\newblock \leavevmode\\ \doi{10.1126/sciadv.aaq0194}

\bibitem{chiap}
F.~Chiappini, S.~Wiedmann, M.~Titov, A.K. Geim, R.V. Gorbachev, E.~Khestanova,
  A.~Mishchenko, K.S. Novoselov, J.C. Maan, U.~Zeitler, Phys. Rev. B
  \textbf{94}, 085302 (2016).
\newblock \leavevmode\\ \doi{10.1103/PhysRevB.94.085302}

\bibitem{sven2}
S.~Danz, M.~Titov, B.N. Narozhny, Phys. Rev. B \textbf{102}, 081114(R) (2020).
\newblock \leavevmode\\ \doi{10.1103/PhysRevB.102.081114}

\bibitem{meg}
M.~Titov, R.V. Gorbachev, B.N. Narozhny, T.~Tudorovskiy, M.~Sch\"utt, P.M.
  Ostrovsky, I.V. Gornyi, A.D. Mirlin, M.I. Katsnelson, K.S. Novoselov, A.K.
  Geim, L.A. Ponomarenko, Phys. Rev. Lett. \textbf{111}, 166601 (2013).
\newblock \leavevmode\\ \doi{10.1103/PhysRevLett.111.166601}

\bibitem{mr1}
P.S. Alekseev, A.P. Dmitriev, I.V. Gornyi, V.Y. Kachorovskii, B.N. Narozhny,
  M.~Sch\"utt, M.~Titov, Phys. Rev. Lett. \textbf{114}, 156601 (2015).
\newblock \leavevmode\\ \doi{10.1103/PhysRevLett.114.156601}

\bibitem{mrexp}
G.Y. Vasileva, D.~Smirnov, Y.L. Ivanov, Y.B. Vasilyev, P.S. Alekseev, A.P.
  Dmitriev, I.V. Gornyi, V.Y. Kachorovskii, M.~Titov, B.N. Narozhny, R.J. Haug,
  Phys. Rev. B \textbf{93}, 195430 (2016).
\newblock \leavevmode\\ \doi{10.1103/PhysRevB.93.195430}

\bibitem{alf}
M.S. Foster, I.L. Aleiner, Phys. Rev. B \textbf{79}, 085415 (2009).
\newblock \leavevmode\\ \doi{10.1103/PhysRevB.79.085415}

\bibitem{navier}
C.L. Navier, in \emph{M{\'e}moires de l'Acad{\'e}mie des sciences de l'Institut
  de France - Ann{\'e}e 1823} (Gauthier-Villars, Paris, 1827), pp. 389--440

\bibitem{stokes}
G.G. Stokes, Transactions of the Cambridge Philosophical Society \textbf{8},
  287 (1845)

\bibitem{stokes51}
G.G. Stokes, Transactions of the Cambridge Philosophical Society \textbf{9}, 8
  (1851)

\bibitem{sven1}
S.~Danz, B.N. Narozhny, 2D Materials \textbf{7}(3), 035001 (2020).
\newblock \leavevmode\\ \doi{10.1088/2053-1583/ab7bfa}

\bibitem{pol15}
I.~Torre, A.~Tomadin, A.K. Geim, M.~Polini, Phys. Rev. B \textbf{92}, 165433
  (2015).
\newblock \leavevmode\\ \doi{10.1103/PhysRevB.92.165433}

\bibitem{pol16}
F.M.D. Pellegrino, I.~Torre, A.K. Geim, M.~Polini, Phys. Rev. B \textbf{94},
  155414 (2016).
\newblock \leavevmode\\ \doi{10.1103/PhysRevB.94.155414}

\bibitem{fl0}
L.S. Levitov, G.~Falkovich, Nat. Phys. \textbf{12}(7), 672 (2016).
\newblock \leavevmode\\ \doi{10.1038/nphys3667}

\bibitem{msw2}
R.~Moessner, P.~Sur\'owka, P.~Witkowski, Phys. Rev. B \textbf{97}, 161112
  (2018).
\newblock \leavevmode\\ \doi{10.1103/PhysRevB.97.161112}

\bibitem{msw}
R.~Moessner, N.~Morales-Dur\'an, P.~Sur\'owka, P.~Witkowski, Phys. Rev. B
  \textbf{100}, 155115 (2019).
\newblock \leavevmode\\ \doi{10.1103/PhysRevB.100.155115}

\bibitem{stein19}
J.~Mayzel, V.~Steinberg, A.~Varshney, Nat. Commun. \textbf{10}, 937 (2019).
\newblock \leavevmode\\ \doi{10.1038/s41467-019-08916-5}

\bibitem{fl18}
A.V. Shytov, J.F. Kong, G.~Falkovich, L.S. Levitov, Phys. Rev. Lett.
  \textbf{121}, 176805 (2018).
\newblock \leavevmode\\ \doi{10.1103/PhysRevLett.121.176805}

\bibitem{fl2}
H.~Guo, E.~Ilseven, G.~Falkovich, L.S. Levitov, Proc. Natl. Acad. Sci.
  \textbf{114}(12), 3068 (2017).
\newblock \leavevmode\\ \doi{10.1073/pnas.1612181114}

\bibitem{qpc1}
B.J. van Wees, H.~van Houten, C.W.J. Beenakker, J.G. Williamson, L.P.
  Kouwenhoven, D.~van~der Marel, C.T. Foxon, Phys. Rev. Lett. \textbf{60}, 848
  (1988).
\newblock \leavevmode\\ \doi{10.1103/PhysRevLett.60.848}

\bibitem{qpc2}
D.A. Wharam, T.J. Thornton, R.~Newbury, M.~Pepper, H.~Ahmed, J.E.F. Frost, D.G.
  Hasko, D.C. Peacock, D.A. Ritchie, G.A.C. Jones, Journal of Physics C
  \textbf{21}(8), L209 (1988).
\newblock \leavevmode\\ \doi{10.1088/0022-3719/21/8/002}

\bibitem{sha}
Y.V. Sharvin, Zh. Eksp. Teor. Fiz. \textbf{48}, 984 (1965).
\newblock [Sov. Phys. JETP {\bf 21}, 655 (1965)]

\bibitem{gor21}
G.~Zhang, V.~Kachorovskii, K.~Tikhonov, I.~Gornyi, Phys. Rev. B \textbf{104},
  075417 (2021).
\newblock \leavevmode\\ \doi{10.1103/PhysRevB.104.075417}

\bibitem{lucas17}
A.~Lucas, Phys. Rev. B \textbf{95}, 115425 (2017).
\newblock \leavevmode\\ \doi{10.1103/PhysRevB.95.115425}

\bibitem{guo17}
H.~Guo, E.~Ilseven, G.~Falkovich, L.~Levitov.
\newblock {Stokes Paradox, Back Reflections and Interaction-Enhanced
  Conduction} (2017).
\newblock \leavevmode\\ \doi{10.48550/arXiv.1612.09239}

\bibitem{oseen}
C.W. Oseen, Ark. f. Mat., Astr. och Fysik (Stockholm) \textbf{6}, 29 (1910)

\bibitem{ale}
P.S. Alekseev, Phys. Rev. Lett. \textbf{117}, 166601 (2016).
\newblock \leavevmode\\ \doi{10.1103/PhysRevLett.117.166601}

\bibitem{moo}
T.~Scaffidi, N.~Nandi, B.~Schmidt, A.P. Mackenzie, J.E. Moore, Phys. Rev. Lett.
  \textbf{118}, 226601 (2017).
\newblock \leavevmode\\ \doi{10.1103/PhysRevLett.118.226601}

\bibitem{tar}
S.~Tarucha, T.~Saku, Y.~Tokura, Y.~Hirayama, Phys. Rev. B \textbf{47}, 4064
  (1993).
\newblock \leavevmode\\ \doi{10.1103/PhysRevB.47.4064}

\bibitem{dat}
S.~Datta, \emph{{Electron Transport in Mesoscopic Systems}} (Cambridge
  University Press, 1997).
\newblock \leavevmode\\ \doi{10.1017/CBO9780511805776}

\bibitem{pol17}
F.M.D. Pellegrino, I.~Torre, M.~Polini, Phys. Rev. B \textbf{96}, 195401
  (2017).
\newblock \leavevmode\\ \doi{10.1103/PhysRevB.96.195401}

\bibitem{mr2}
P.S. Alekseev, A.P. Dmitriev, I.V. Gornyi, V.Y. Kachorovskii, B.N. Narozhny,
  M.~Titov, Phys. Rev. B \textbf{97}, 085109 (2018).
\newblock \leavevmode\\ \doi{10.1103/PhysRevB.97.085109}

\bibitem{cfl}
P.S. Alekseev, A.P. Dmitriev, I.V. Gornyi, V.Y. Kachorovskii, B.N. Narozhny,
  M.~Titov, Phys. Rev. B \textbf{98}, 125111 (2018).
\newblock \leavevmode\\ \doi{10.1103/PhysRevB.98.125111}

\bibitem{poi}
J.L.M. Poiseuille, C. R. Acad. Sci. \textbf{11}, 961 (1840)

\bibitem{ks19}
E.I. Kiselev, J.~Schmalian, Phys. Rev. B \textbf{99}, 035430 (2019).
\newblock \leavevmode\\ \doi{10.1103/PhysRevB.99.035430}

\bibitem{luk}
J.R. Maze, P.L. Stanwix, J.S. Hodges, S.~Hong, J.M. Taylor, P.~Cappellaro,
  L.~Jiang, M.V. {Gurudev Dutt}, E.~Togan, A.S. Zibrov, A.~Yacoby, R.L.
  Walsworth, M.D. Lukin, Nature \textbf{455}, 644 (2008).
\newblock \leavevmode\\ \doi{10.1038/nature07279}

\bibitem{me2}
B.N. Narozhny, M.~Sch\"utt, Phys. Rev. B \textbf{100}, 035125 (2019).
\newblock \leavevmode\\ \doi{10.1103/PhysRevB.100.035125}

\bibitem{me1}
B.N. Narozhny, Ann. Phys. \textbf{411}, 167979 (2019).
\newblock \leavevmode\\ \doi{10.1016/j.aop.2019.167979}

\bibitem{hydro1}
U.~Briskot, M.~Sch\"utt, I.V. Gornyi, M.~Titov, B.N. Narozhny, A.D. Mirlin,
  Phys. Rev. B \textbf{92}, 115426 (2015).
\newblock \leavevmode\\ \doi{10.1103/PhysRevB.92.115426}

\bibitem{schutt}
M.~Sch\"utt, P.M. Ostrovsky, I.V. Gornyi, A.D. Mirlin, Phys. Rev. B
  \textbf{83}, 155441 (2011).
\newblock \leavevmode\\ \doi{10.1103/PhysRevB.83.155441}

\bibitem{mus}
M.~M\"uller, S.~Sachdev, Phys. Rev. B \textbf{78}, 115419 (2008).
\newblock \leavevmode\\ \doi{10.1103/PhysRevB.78.115419}

\bibitem{hydro0}
B.N. Narozhny, I.V. Gornyi, M.~Titov, M.~Sch{\"u}tt, A.D. Mirlin, Phys. Rev. B
  \textbf{91}, 035414 (2015).
\newblock \leavevmode\\ \doi{10.1103/PhysRevB.91.035414}

\bibitem{megt2}
B.N. Narozhny, I.V. Gornyi, M.~Titov, Phys. Rev. B \textbf{104}, 075443 (2021).
\newblock \leavevmode\\ \doi{10.1103/PhysRevB.104.075443}

\bibitem{eis}
J.P. Eisenstein, in \emph{Perspectives in Quantum Hall Effects}, ed. by S.D.
  Sarma, A.~Pinczuk (Wiley, New York, 1997).
\newblock \leavevmode\\ \doi{10.1002/9783527617258.ch2}

\bibitem{gir}
S.M. Girvin, A.H. MacDonald, in \emph{Perspectives in Quantum Hall Effects},
  ed. by S.D. Sarma, A.~Pinczuk (Wiley, New York, 1997).
\newblock \leavevmode\\ \doi{10.1002/9783527617258.ch5}

\bibitem{ber}
B.A. Bernevig, T.L. Hughes, \emph{{Topological Insulators and Topological
  Superconductors}} (Princeton University Press, 2013)

\bibitem{dgg16}
Y.T. Cui, B.~Wen, E.Y. Ma, G.~Diankov, Z.~Han, F.~Amet, T.~Taniguchi,
  K.~Watanabe, D.~Goldhaber-Gordon, C.R. Dean, Z.X. Shen, Phys. Rev. Lett.
  \textbf{117}, 186601 (2016).
\newblock \leavevmode\\ \doi{10.1103/PhysRevLett.117.186601}

\bibitem{zhit}
J.~Chae, S.~Jung, S.~Woo, H.~Baek, J.~Ha, Y.J. Song, Y.W. Son, N.B. Zhitenev,
  J.A. Stroscio, Y.~Kuk, Nano Letters \textbf{12}(4), 1839 (2012).
\newblock \leavevmode\\ \doi{10.1021/nl2041222}

\bibitem{nic18}
J.M. Marmolejo-Tejada, J.H. Garc{\'{\i}}a, M.D. Petrovi{\'{c}}, P.H. Chang,
  X.L. Sheng, A.~Cresti, P.~Plech{\'{a}}{\v{c}}, S.~Roche, B.K. Nikoli{\'{c}},
  Journal of Physics: Materials \textbf{1}(1), 015006 (2018).
\newblock \leavevmode\\ \doi{10.1088/2515-7639/aad585}

\bibitem{yac16}
M.T. Allen, O.~Shtanko, I.C. Fulga, A.R. Akhmerov, K.~Watanabe, T.~Taniguchi,
  P.~Jarillo-Herrero, L.S. Levitov, A.~Yacoby, Nat. Phys. \textbf{12}, 128
  (2016).
\newblock \leavevmode\\ \doi{10.1038/nphys3534}

\bibitem{sze}
S.M. Sze, K.K. Ng, \emph{{Physics of Semiconductor Devices}} (Wiley, New York,
  2006).
\newblock \leavevmode\\ \doi{10.1002/0470068329}

\bibitem{schottky}
W.~Schottky, Naturwissenschaften \textbf{26}, 843 (1938).
\newblock \leavevmode\\ \doi{10.1007/BF01774216}

\bibitem{geimedge}
M.J. Zhu, A.V. Kretinin, M.D. Thompson, D.A. Bandurin, S.~Hu, G.L. Yu,
  J.~Birkbeck, A.~Mishchenko, I.J. Vera-Marun, K.~Watanabe, T.~Taniguchi,
  M.~Polini, J.R. Prance, K.S. Novoselov, A.K. Geim, M.~{Ben Shalom}, Nat.
  Commun. \textbf{8}, 14552 (2017).
\newblock \leavevmode\\ \doi{10.1038/ncomms14552}

\bibitem{halb17}
D.~Halbertal, M.~{Ben Shalom}, A.~Uri, K.~Bagani, A.Y. Meltzer, I.~Markus,
  Y.~Myasoedov, J.~Birkbeck, L.S. Levitov, A.K. Geim, E.~Zeldov, Science
  \textbf{358}, 1303 (2017).
\newblock \leavevmode\\ \doi{10.1126/science.aan0877}

\bibitem{bra18}
B.A. Braem, C.~Gold, S.~Hennel, M.~R{\"o}{\"o}sli, M.~Berl, W.~Dietsche,
  W.~Wegscheider, K.~Ensslin, T.~Ihn, New Journal of Physics \textbf{20}(7),
  073015 (2018).
\newblock \leavevmode\\ \doi{10.1088/1367-2630/aad068}

\bibitem{fal19}
M.~Shavit, A.V. Shytov, G.~Falkovich, Phys. Rev. Lett. \textbf{123}, 026801
  (2019).
\newblock \leavevmode\\ \doi{10.1103/PhysRevLett.123.026801}

\bibitem{corbino}
{O. M. Corbino}, Nuovo Cimento \textbf{1}, 397 (1911).
\newblock \leavevmode\\ \doi{10.1007/BF02958241}

\bibitem{corb}
A.~Tomadin, G.~Vignale, M.~Polini, Phys. Rev. Lett. \textbf{113}, 235901
  (2014).
\newblock \leavevmode\\ \doi{10.1103/PhysRevLett.113.235901}

\bibitem{wfl}
R.~Franz, G.~Wiedemann, Annalen der Physik \textbf{165}(8), 497 (1853).
\newblock \leavevmode\\ \doi{10.1002/andp.18531650802}

\bibitem{ll}
L.~Lorenz, Annalen der Physik \textbf{223}(11), 429 (1872).
\newblock \leavevmode\\ \doi{10.1002/andp.18722231107}

\bibitem{hust}
J.G. Hust, A.B. Lankford, \emph{{Thermal conductivity of aluminum, copper, iron
  and tungsten for temperatures from 1 K to the melting point}} (1984).
\newblock National Bureau of Standards, Boulder, Colorado; NBSIR 84-3007

\bibitem{kim0}
Y.M. Zuev, W.~Chang, P.~Kim, Phys. Rev. Lett. \textbf{102}, 096807 (2009).
\newblock \leavevmode\\ \doi{10.1103/PhysRevLett.102.096807}

\bibitem{ong}
J.G. Checkelsky, N.P. Ong, Phys. Rev. B \textbf{80}, 081413(R) (2009).
\newblock \leavevmode\\ \doi{10.1103/PhysRevB.80.081413}

\bibitem{kopp21}
A.~Block, A.~Principi, N.C.H. Hesp, A.W. Cummings, M.~Liebel, K.~Watanabe,
  T.~Taniguchi, S.~Roche, F.H.L. Koppens, N.F. {van Hulst}, K.J. Tielrooij,
  Nature Nanotechnology \textbf{16}, 1195 (2021).
\newblock \leavevmode\\ \doi{10.1038/s41565-021-00957-6}

\bibitem{har}
S.A. Hartnoll, P.K. Kovtun, M.~M\"uller, S.~Sachdev, Phys. Rev. B \textbf{76},
  144502 (2007).
\newblock \leavevmode\\ \doi{10.1103/PhysRevB.76.144502}

\bibitem{wal}
P.R. Wallace, Phys. Rev. \textbf{71}, 622 (1947).
\newblock \leavevmode\\ \doi{10.1103/PhysRev.71.622}

\bibitem{sem}
G.W. Semenoff, Phys. Rev. Lett. \textbf{53}, 2449 (1984).
\newblock \leavevmode\\ \doi{10.1103/PhysRevLett.53.2449}

\bibitem{dvm}
D.P. DiVincenzo, E.J. Mele, Phys. Rev. B \textbf{29}, 1685 (1984).
\newblock \leavevmode\\ \doi{10.1103/PhysRevB.29.1685}

\bibitem{geimqhe}
K.S. Novoselov, A.K. Geim, S.V. Morozov, D.~Jiang, M.I. Katsnelson, I.V.
  Grigorieva, S.V. Dubonos, A.A. Firsov, Nature \textbf{438}, 197 (2005).
\newblock \leavevmode\\ \doi{10.1038/nature04233}

\bibitem{mfs}
M.~M\"uller, L.~Fritz, S.~Sachdev, Phys. Rev. B \textbf{78}, 115406 (2008).
\newblock \leavevmode\\ \doi{10.1103/PhysRevB.78.115406}

\bibitem{mfss}
L.~Fritz, J.~Schmalian, M.~M\"uller, S.~Sachdev, Phys. Rev. B \textbf{78},
  085416 (2008).
\newblock \leavevmode\\ \doi{10.1103/PhysRevB.78.085416}

\bibitem{mish1}
E.G. Mishchenko, Phys. Rev. Lett. \textbf{98}, 216801 (2007).
\newblock \leavevmode\\ \doi{10.1103/PhysRevLett.98.216801}

\bibitem{mish}
E.G. Mishchenko, {EPL} (Europhysics Letters) \textbf{83}(1), 17005 (2008).
\newblock \leavevmode\\ \doi{10.1209/0295-5075/83/17005}

\bibitem{goo18}
J.~Gooth, F.~Menges, N.~Kumar, V.~S{\"{u}}$\beta$, C.~Shekhar, Y.~Sun,
  U.~Drechsler, R.~Zierold, C.~Felser, B.~Gotsmann, Nat. Commun. \textbf{9},
  4093 (2018).
\newblock \leavevmode\\ \doi{10.1038/s41467-018-06688-y}

\bibitem{Kumar19}
N.~Kumar, Y.~Sun, M.~Nicklas, S.J. Watzman, O.~Young, I.~Leermakers,
  J.~Hornung, J.~Klotz, J.~Gooth, K.~Manna, V.~S{\"{u}}{\ss}, S.N. Guin,
  T.~F{\"{o}}rster, M.~Schmidt, L.~Muechler, B.~Yan, P.~Werner, W.~Schnelle,
  U.~Zeitler, J.~Wosnitza, S.~Parkin, C.~Felser, C.~Shekhar, Nat. Commun.
  \textbf{10}(1), 2475 (2019).
\newblock \leavevmode\\ \doi{10.1038/s41467-019-10126-y}

\bibitem{fos16}
H.Y. Xie, M.S. Foster, Phys. Rev. B \textbf{93}, 195103 (2016).
\newblock \leavevmode\\ \doi{10.1103/PhysRevB.93.195103}

\bibitem{vig15}
A.~Principi, G.~Vignale, Phys. Rev. Lett. \textbf{115}, 056603 (2015).
\newblock \leavevmode\\ \doi{10.1103/PhysRevLett.115.056603}

\bibitem{lds2}
A.~Lucas, S.~Das~Sarma, Phys. Rev. B \textbf{97}, 245128 (2018).
\newblock \leavevmode\\ \doi{10.1103/PhysRevB.97.245128}

\bibitem{moll21}
M.R. {van Delft}, Y.~Wang, C.~Putzke, J.~Oswald, G.~Varnavides, C.A.C. Garcia,
  C.~Guo, H.~Schmid, V.~S{\"u}ss, H.~Borrmann, J.~Diaz, Y.~Sun, C.~Felser,
  B.~Gotsmann, P.~Narang, P.J.W. Moll, Nat. Commun. \textbf{12}, 4799 (2021).
\newblock \leavevmode\\ \doi{10.1038/s41467-021-25037-0}

\bibitem{sond}
E.H. Sondheimer, Phys. Rev. \textbf{80}, 401 (1950).
\newblock \leavevmode\\ \doi{10.1103/PhysRev.80.401}

\bibitem{shu}
E.~Shuryak, {Prog. Part. Nucl. Phys.} \textbf{62}(1), 48 (2009).
\newblock \leavevmode\\ \doi{10.1016/j.ppnp.2008.09.001}

\bibitem{sho}
S.N. Shore, \emph{{Astrophysical Hydrodynamics: An Introduction}} (Wiley,
  Weinheim, 2007).
\newblock \leavevmode\\ \doi{10.1002/9783527619054}

\bibitem{stein}
M.S. Steinberg, Phys. Rev. \textbf{109}, 1486 (1958).
\newblock \leavevmode\\ \doi{10.1103/PhysRev.109.1486}

\bibitem{kats}
M.I. Katsnelson, \emph{Graphene} (Cambridge University Press, 2012).
\newblock \leavevmode\\ \doi{10.1017/CBO9781139031080}

\bibitem{job}
M.J. Bhaseen, A.G. Green, S.L. Sondhi, Phys. Rev. B \textbf{79}, 094502 (2009).
\newblock \leavevmode\\ \doi{10.1103/PhysRevB.79.094502}

\bibitem{euler}
L.~Euler, M{\'e}moires de l'acad{\'e}mie des sciences de Berlin \textbf{11},
  274 (1757)

\bibitem{cosmic}
T.V. Phan, J.C.W. Song, L.S. Levitov (2013).
\newblock \leavevmode\\ \doi{10.48550/arXiv.1306.4972}

\bibitem{lev13}
L.S. Levitov, A.V. Shtyk, M.V. Feigelman, Phys. Rev. B \textbf{88}, 235403
  (2013).
\newblock \leavevmode\\ \doi{10.1103/PhysRevB.88.235403}

\bibitem{fog}
Z.~Sun, D.N. Basov, M.M. Fogler, Proc. Natl. Acad. Sci. \textbf{115}(13), 3285
  (2018).
\newblock \leavevmode\\ \doi{10.1073/pnas.1717010115}

\bibitem{svin}
D.~Svintsov, Phys. Rev. B \textbf{97}, 121405(R) (2018).
\newblock \leavevmode\\ \doi{10.1103/PhysRevB.97.121405}

\bibitem{ldsp}
A.~Lucas, S.~Das~Sarma, Phys. Rev. B \textbf{97}, 115449 (2018).
\newblock \leavevmode\\ \doi{10.1103/PhysRevB.97.115449}

\bibitem{ks20}
E.I. Kiselev, J.~Schmalian, Phys. Rev. B \textbf{102}, 245434 (2020).
\newblock \leavevmode\\ \doi{10.1103/PhysRevB.102.245434}

\bibitem{fat}
D.V. Fateev, V.V. Popov, Semiconductors \textbf{54}, 941 (2020).
\newblock \leavevmode\\ \doi{10.1134/S1063782620080084}

\bibitem{megt}
B.N. Narozhny, I.V. Gornyi, M.~Titov, Phys. Rev. B \textbf{103}, 115402 (2021).
\newblock \leavevmode\\ \doi{10.1103/PhysRevB.103.115402}

\bibitem{aka}
M.~Auslender, M.I. Katsnelson, Phys. Rev. B \textbf{76}, 235425 (2007).
\newblock \leavevmode\\ \doi{10.1103/PhysRevB.76.235425}

\bibitem{sav}
A.A. Kozikov, A.K. Savchenko, B.N. Narozhny, A.V. Shytov, Phys. Rev. B
  \textbf{82}, 075424 (2010).
\newblock \leavevmode\\ \doi{10.1103/PhysRevB.82.075424}

\bibitem{shsch}
D.E. Sheehy, J.~Schmalian, Phys. Rev. Lett. \textbf{99}, 226803 (2007).
\newblock \leavevmode\\ \doi{10.1103/PhysRevLett.99.226803}

\bibitem{msf}
M.~M\"uller, J.~Schmalian, L.~Fritz, Phys. Rev. Lett. \textbf{103}, 025301
  (2009).
\newblock \leavevmode\\ \doi{10.1103/PhysRevLett.103.025301}

\bibitem{paconpb}
J.~Gonzalez, F.~Guinea, M.A.H. Vozmediano, Nuclear Physics B \textbf{424}(3),
  595  (1994).
\newblock \leavevmode\\ \doi{10.1016/0550-3213(94)90410-3}

\bibitem{paco99}
J.~Gonz\'alez, F.~Guinea, M.A.H. Vozmediano, Phys. Rev. B \textbf{59}, R2474
  (1999).
\newblock \leavevmode\\ \doi{10.1103/PhysRevB.59.R2474}

\bibitem{julia}
J.M. Link, B.N. Narozhny, E.I. Kiselev, J.~Schmalian, Phys. Rev. Lett.
  \textbf{120}, 196801 (2018).
\newblock \leavevmode\\ \doi{10.1103/PhysRevLett.120.196801}

\bibitem{aa17}
I.L. Aleiner, O.~Agam, Ann. Phys. \textbf{385}, 716 (2017).
\newblock \leavevmode\\ \doi{10.1016/j.aop.2017.08.017}

\bibitem{can}
A.H. Castro~Neto, F.~Guinea, N.M.R. Peres, K.S. Novoselov, A.K. Geim, Rev. Mod.
  Phys. \textbf{81}, 109 (2009).
\newblock \leavevmode\\ \doi{10.1103/RevModPhys.81.109}

\bibitem{sim20}
G.~Wagner, D.X. Nguyen, S.H. Simon, Phys. Rev. B \textbf{101}, 245438 (2020).
\newblock \leavevmode\\ \doi{10.1103/PhysRevB.101.245438}

\bibitem{adam18}
H.K. Tang, J.N. Leaw, J.N.B. Rodrigues, I.F. Herbut, P.~Sengupta, F.F. Assaad,
  S.~Adam, Science \textbf{361}(6402), 570 (2018).
\newblock \leavevmode\\ \doi{10.1126/science.aao2934}

\bibitem{vgeim1}
D.C. Elias, R.V. Gorbachev, A.S. Mayorov, S.V. Morozov, A.A. Zhukov, P.~Blake,
  L.A. Ponomarenko, I.V. Grigorieva, K.S. Novoselov, F.~Guinea, A.K. Geim, Nat.
  Phys. \textbf{7}, 701 (2011).
\newblock \leavevmode\\ \doi{10.1038/NPHYS2049}

\bibitem{svin18}
G.~Alymov, V.~Vyurkov, V.~Ryzhii, A.~Satou, D.~Svintsov, Phys. Rev. B
  \textbf{97}, 205411 (2018).
\newblock \leavevmode\\ \doi{10.1103/PhysRevB.97.205411}

\bibitem{bri}
A.~Tomadin, D.~Brida, G.~Cerullo, A.C. Ferrari, M.~Polini, Phys. Rev. B
  \textbf{88}, 035430 (2013).
\newblock \leavevmode\\ \doi{10.1103/PhysRevB.88.035430}

\bibitem{kash}
A.B. Kashuba, Phys. Rev. B \textbf{78}, 085415 (2008).
\newblock \leavevmode\\ \doi{10.1103/PhysRevB.78.085415}

\bibitem{drag2}
M.~Sch\"utt, P.M. Ostrovsky, M.~Titov, I.V. Gornyi, B.N. Narozhny, A.D. Mirlin,
  Phys. Rev. Lett. \textbf{110}, 026601 (2013).
\newblock \leavevmode\\ \doi{10.1103/PhysRevLett.110.026601}

\bibitem{srl}
J.C.W. Song, M.Y. Reizer, L.S. Levitov, Phys. Rev. Lett. \textbf{109}, 106602
  (2012).
\newblock \leavevmode\\ \doi{10.1103/PhysRevLett.109.106602}

\bibitem{ralph13}
M.W. Graham, S.F. Shi, D.C. Ralph, J.~Park, P.L. McEuen, Nat. Phys. \textbf{9},
  103 (2013).
\newblock \leavevmode\\ \doi{10.1038/nphys2493}

\bibitem{betz}
A.C. Betz, S.H. Jhang, E.~Pallecchi, R.~Ferreira, G.~F{\`e}ve, J.M. Berroir,
  B.~Pla{\c{c}}ais, Nat. Phys. \textbf{9}, 109 (2013).
\newblock \leavevmode\\ \doi{10.1038/nphys2494}

\bibitem{tik18}
K.S. Tikhonov, I.V. Gornyi, V.Y. Kachorovskii, A.D. Mirlin, Phys. Rev. B
  \textbf{97}, 085415 (2018).
\newblock \leavevmode\\ \doi{10.1103/PhysRevB.97.085415}

\bibitem{kong}
J.F. Kong, L.~Levitov, D.~Halbertal, E.~Zeldov, Phys. Rev. B \textbf{97},
  245416 (2018).
\newblock \leavevmode\\ \doi{10.1103/PhysRevB.97.245416}

\bibitem{prin2020}
M.~Zarenia, A.~Principi, G.~Vignale, Phys. Rev. B \textbf{102}, 214304 (2020).
\newblock \leavevmode\\ \doi{10.1103/PhysRevB.102.214304}

\bibitem{mr3}
P.S. Alekseev, A.P. Dmitriev, I.V. Gornyi, V.Y. Kachorovskii, B.N. Narozhny,
  M.~Sch\"utt, M.~Titov, Phys. Rev. B \textbf{95}, 165410 (2017).
\newblock \leavevmode\\ \doi{10.1103/PhysRevB.95.165410}

\bibitem{drag12}
R.V. Gorbachev, A.K. Geim, M.I. Katsnelson, K.S. Novoselov, T.~Tudorovskiy,
  I.V. Grigorieva, A.H. MacDonald, S.V. Morozov, K.~Watanabe, T.~Taniguchi,
  L.A. Ponomarenko, Nat. Phys. \textbf{8}(12), 896 (2012).
\newblock \leavevmode\\ \doi{10.1038/nphys2441}

\bibitem{meig1}
B.N. Narozhny, I.V. Gornyi, Frontiers in Physics \textbf{9}, 108 (2021).
\newblock \leavevmode\\ \doi{10.3389/fphy.2021.640649}

\bibitem{bistr}
R.~Bistritzer, A.H. MacDonald, Phys. Rev. Lett. \textbf{102}, 206410 (2009).
\newblock \leavevmode\\ \doi{10.1103/PhysRevLett.102.206410}

\bibitem{tse}
W.K. Tse, S.~Das~Sarma, Phys. Rev. B \textbf{79}, 235406 (2009).
\newblock \leavevmode\\ \doi{10.1103/PhysRevB.79.235406}

\bibitem{lev19}
H.Y. Xie, A.~Levchenko, Phys. Rev. B \textbf{99}, 045434 (2019).
\newblock \leavevmode\\ \doi{10.1103/PhysRevB.99.045434}

\bibitem{kash18}
O.~Kashuba, B.~Trauzettel, L.W. Molenkamp, Phys. Rev. B \textbf{97}, 205129
  (2018).
\newblock \leavevmode\\ \doi{10.1103/PhysRevB.97.205129}

\bibitem{me3}
B.N. Narozhny, Phys. Rev. B \textbf{100}, 115434 (2019).
\newblock \leavevmode\\ \doi{10.1103/PhysRevB.100.115434}

\bibitem{fl1}
G.~Falkovich, L.S. Levitov, Phys. Rev. Lett. \textbf{119}, 066601 (2017).
\newblock \leavevmode\\ \doi{10.1103/PhysRevLett.119.066601}

\bibitem{kha}
I.M. Khalatnikov, Zh. Eksp. Teor. Fiz. \textbf{29}, 253 (1956).
\newblock [Soviet Physics JETP-USSR {\bf 2}, 169 (1956)]

\bibitem{sykes}
J.~Sykes, G.A. Brooker, Ann. Phys. \textbf{56}(1), 1 (1970).
\newblock \leavevmode\\ \doi{10.1016/0003-4916(70)90002-3}

\bibitem{bur21}
V.A. Zakharov, I.S. Burmistrov, Phys. Rev. B \textbf{103}, 235305 (2021).
\newblock \leavevmode\\ \doi{10.1103/PhysRevB.103.235305}

\bibitem{read}
B.~Bradlyn, M.~Goldstein, N.~Read, Phys. Rev. B \textbf{86}, 245309 (2012).
\newblock \leavevmode\\ \doi{10.1103/PhysRevB.86.245309}

\bibitem{brad}
P.~Rao, B.~Bradlyn, Phys. Rev. X \textbf{10}, 021005 (2020).
\newblock \leavevmode\\ \doi{10.1103/PhysRevX.10.021005}

\bibitem{grom}
L.V. Delacr\'etaz, A.~Gromov, Phys. Rev. Lett. \textbf{119}, 226602 (2017).
\newblock \leavevmode\\ \doi{10.1103/PhysRevLett.119.226602}

\bibitem{ady}
T.~Holder, R.~Queiroz, A.~Stern, Phys. Rev. Lett. \textbf{123}, 106801 (2019).
\newblock \leavevmode\\ \doi{10.1103/PhysRevLett.123.106801}

\bibitem{julia1}
J.M. Link, D.E. Sheehy, B.N. Narozhny, J.~Schmalian, Phys. Rev. B \textbf{98},
  195103 (2018).
\newblock \leavevmode\\ \doi{10.1103/PhysRevB.98.195103}

\bibitem{poli16}
A.~Principi, G.~Vignale, M.~Carrega, M.~Polini, Phys. Rev. B \textbf{93},
  125410 (2016).
\newblock \leavevmode\\ \doi{10.1103/PhysRevB.93.125410}

\bibitem{bur19}
I.S. Burmistrov, M.~Goldstein, M.~Kot, V.D. Kurilovich, P.D. Kurilovich, Phys.
  Rev. Lett. \textbf{123}, 026804 (2019).
\newblock \leavevmode\\ \doi{10.1103/PhysRevLett.123.026804}

\bibitem{gal20}
Y.~Liao, V.~Galitski, Phys. Rev. B \textbf{101}, 195106 (2020).
\newblock \leavevmode\\ \doi{10.1103/PhysRevB.101.195106}

\bibitem{dragrev}
B.N. Narozhny, A.~Levchenko, Rev. Mod. Phys. \textbf{88}, 025003 (2016).
\newblock \leavevmode\\ \doi{10.1103/RevModPhys.88.025003}

\bibitem{drag}
B.N. Narozhny, M.~Titov, I.V. Gornyi, P.M. Ostrovsky, Phys. Rev. B \textbf{85},
  195421 (2012).
\newblock \leavevmode\\ \doi{10.1103/PhysRevB.85.195421}

\bibitem{ale18}
P.S. Alekseev, Phys. Rev. B \textbf{98}, 165440 (2018).
\newblock \leavevmode\\ \doi{10.1103/PhysRevB.98.165440}

\bibitem{lai92}
B.~Laikhtman, Phys. Rev. B \textbf{45}, 1259 (1992).
\newblock \leavevmode\\ \doi{10.1103/PhysRevB.45.1259}

\bibitem{gurzhi95}
R.N. Gurzhi, A.N. Kalinenko, A.I. Kopeliovich, Phys. Rev. Lett. \textbf{74},
  3872 (1995).
\newblock \leavevmode\\ \doi{10.1103/PhysRevLett.74.3872}

\bibitem{levitov19}
P.~Ledwith, H.~Guo, A.~Shytov, L.~Levitov, Phys. Rev. Lett. \textbf{123},
  116601 (2019).
\newblock \leavevmode\\ \doi{10.1103/PhysRevLett.123.116601}

\bibitem{ale20}
P.S. Alekseev, A.P. Dmitriev, Phys. Rev. B \textbf{102}, 241409 (2020).
\newblock \leavevmode\\ \doi{10.1103/PhysRevB.102.241409}

\bibitem{chus}
H.~Isobe, B.J. Yang, A.~Chubukov, J.~Schmalian, N.~Nagaosa, Phys. Rev. Lett.
  \textbf{116}, 076803 (2016).
\newblock \leavevmode\\ \doi{10.1103/PhysRevLett.116.076803}

\bibitem{ad1}
A.~Kobayashi, Y.~Suzumura, F.~Pi\'echon, G.~Montambaux, Phys. Rev. B
  \textbf{84}, 075450 (2011).
\newblock \leavevmode\\ \doi{10.1103/PhysRevB.84.075450}

\bibitem{ad2}
V.~Pardo, W.E. Pickett, Phys. Rev. Lett. \textbf{102}, 166803 (2009).
\newblock \leavevmode\\ \doi{10.1103/PhysRevLett.102.166803}

\bibitem{ad3}
S.~Banerjee, R.R.P. Singh, V.~Pardo, W.E. Pickett, Phys. Rev. Lett.
  \textbf{103}, 016402 (2009).
\newblock \leavevmode\\ \doi{10.1103/PhysRevLett.103.016402}

\bibitem{fafu}
C.~Fang, L.~Fu, Phys. Rev. B \textbf{91}, 161105 (2015).
\newblock \leavevmode\\ \doi{10.1103/PhysRevB.91.161105}

\bibitem{hua}
S.M. Huang, S.Y. Xu, I.~Belopolski, C.C. Lee, G.~Chang, T.R. Chang, B.~Wang,
  N.~Alidoust, G.~Bian, M.~Neupane, D.~Sanchez, H.~Zheng, H.T. Jeng, A.~Bansil,
  T.~Neupert, H.~Lin, M.Z. Hasan, Proc. Natl. Acade. Sci. \textbf{113}(5), 1180
  (2016).
\newblock \leavevmode\\ \doi{10.1073/pnas.1514581113}

\bibitem{kov}
P.K. Kovtun, D.T. Son, A.O. Starinets, Phys. Rev. Lett. \textbf{94}, 111601
  (2005).
\newblock \leavevmode\\ \doi{10.1103/PhysRevLett.94.111601}

\bibitem{sur19}
F.~Pe\~na Benitez, K.~Saha, P.~Sur\'owka, Phys. Rev. B \textbf{99}, 045141
  (2019).
\newblock \leavevmode\\ \doi{10.1103/PhysRevB.99.045141}

\bibitem{sur202}
C.~Hoyos, R.~Lier, F.~Pe\~na Benitez, P.~Sur\'owka, Phys. Rev. B \textbf{102},
  081303 (2020).
\newblock \leavevmode\\ \doi{10.1103/PhysRevB.102.081303}

\bibitem{lut56}
J.M. Luttinger, Phys. Rev. \textbf{102}, 1030 (1956).
\newblock \leavevmode\\ \doi{10.1103/PhysRev.102.1030}

\bibitem{ab74}
A.A. Abrikosov, Zh. Eksp. Teor. Fiz. \textbf{66}, 1443 (1974).
\newblock [Sov. Phys. JETP {\bf 39}, 709 (1974)]

\bibitem{julia20}
J.M. Link, I.F. Herbut, Phys. Rev. B \textbf{101}, 125128 (2020).
\newblock \leavevmode\\ \doi{10.1103/PhysRevB.101.125128}

\bibitem{max}
J.C. Maxwell, Phil. Trans. R. Soc. \textbf{170}, 231 (1879).
\newblock \leavevmode\\ \doi{10.1098/rstl.1879.0067}

\bibitem{falk83}
L.A. Falkovsky, Advances in Physics \textbf{32}(5), 753 (1983).
\newblock \leavevmode\\ \doi{10.1080/00018738300101601}

\bibitem{ady2}
T.~Holder, R.~Queiroz, T.~Scaffidi, N.~Silberstein, A.~Rozen, J.A. Sulpizio,
  L.~Ella, S.~Ilani, A.~Stern, Phys. Rev. B \textbf{100}, 245305 (2019).
\newblock \leavevmode\\ \doi{10.1103/PhysRevB.100.245305}

\bibitem{rai22prb}
O.E. Raichev, Phys. Rev. B \textbf{105}, L041301 (2022).
\newblock \leavevmode\\ \doi{10.1103/PhysRevB.105.L041301}

\bibitem{adam21}
A.i.e.i.f.C. Keser, D.Q. Wang, O.~Klochan, D.Y.H. Ho, O.A. Tkachenko, V.A.
  Tkachenko, D.~Culcer, S.~Adam, I.~Farrer, D.A. Ritchie, O.P. Sushkov, A.R.
  Hamilton, Phys. Rev. X \textbf{11}, 031030 (2021).
\newblock \leavevmode\\ \doi{10.1103/PhysRevX.11.031030}

\bibitem{mard}
M.P. Marder, \emph{Condensed Matter Physics} (Wiley, 2010).
\newblock \leavevmode\\ \doi{10.1002/9780470949955}

\bibitem{landauer}
R.~Landauer, IBM J. Res. Dev. \textbf{1}(3), 223 (1957).
\newblock \leavevmode\\ \doi{10.1147/rd.13.0223}

\bibitem{lucas16}
A.~Lucas, Phys. Rev. B \textbf{93}, 245153 (2016).
\newblock \leavevmode\\ \doi{10.1103/PhysRevB.93.245153}

\bibitem{falk18}
M.~Semenyakin, G.~Falkovich, Phys. Rev. B \textbf{97}, 085127 (2018).
\newblock \leavevmode\\ \doi{10.1103/PhysRevB.97.085127}

\bibitem{Giuliani}
G.~Giuliani, G.~Vignale, \emph{Quantum Theory of the Electron Liquid}
  (Cambridge University Press, 2005).
\newblock \leavevmode\\ \doi{10.1017/CBO9780511619915}

\bibitem{hill09}
A.~Hill, S.A. Mikhailov, K.~Ziegler, {EPL} (Europhysics Letters)
  \textbf{87}(2), 27005 (2009).
\newblock \leavevmode\\ \doi{10.1209/0295-5075/87/27005}

\bibitem{prin11}
A.~Principi, R.~Asgari, M.~Polini, Solid State Communications \textbf{151}(21),
  1627 (2011).
\newblock \leavevmode\\ \doi{10.1016/j.ssc.2011.07.015}

\bibitem{fei12}
Z.~Fei, A.S. Rodin, G.O. Andreev, W.~Bao, A.S. McLeod, M.~Wagner, L.M. Zhang,
  Z.~Zhao, M.~Thiemens, G.~Dominguez, M.M. Fogler, A.H.C. Neto, C.N. Lau,
  F.~Keilmann, D.N. Basov, Nature \textbf{487}(7405), 82 (2012).
\newblock \leavevmode\\ \doi{10.1038/nature11253}

\bibitem{chen12}
J.~Chen, M.~Badioli, P.~Alonso-Gonz{\'{a}}lez, S.~Thongrattanasiri, F.~Huth,
  J.~Osmond, M.~Spasenovi{\'{c}}, A.~Centeno, A.~Pesquera, P.~Godignon,
  A.~{Zurutuza Elorza}, N.~Camara, F.J.G. de~Abajo, R.~Hillenbrand, F.H.L.
  Koppens, Nature \textbf{487}, 77 (2012).
\newblock \leavevmode\\ \doi{10.1038/nature11254}

\bibitem{bas}
G.X. Ni, L.~Wang, M.D. Goldflam, M.~Wagner, Z.~Fei, A.S. McLeod, M.K. Liu,
  F.~Keilmann, B.~{\"O}zyilmaz, A.H. Castro~Neto, J.~Hone, F.M. M., B.D. N.,
  Nat. Photon. \textbf{10}, 244 (2016).
\newblock \leavevmode\\ \doi{10.1038/nphoton.2016.45}

\bibitem{kop17}
M.B. Lundeberg, Y.~Gao, R.~Asgari, C.~Tan, B.~Van~Duppen, M.~Autore,
  P.~Alonso-Gonz{\'a}lez, A.~Woessner, K.~Watanabe, T.~Taniguchi,
  R.~Hillenbrand, J.~Hone, M.~Polini, F.H.L. Koppens, Science
  \textbf{357}(6347), 187 (2017).
\newblock \leavevmode\\ \doi{10.1126/science.aan2735}

\bibitem{kop18}
D.~Alcaraz~Iranzo, S.~Nanot, E.J.C. Dias, I.~Epstein, C.~Peng, D.K. Efetov,
  M.B. Lundeberg, R.~Parret, J.~Osmond, J.Y. Hong, J.~Kong, D.R. Englund,
  N.M.R. Peres, F.H.L. Koppens, Science \textbf{360}(6386), 291 (2018).
\newblock \leavevmode\\ \doi{10.1126/science.aar8438}

\bibitem{polkop20}
P.~Novelli, I.~Torre, F.H.L. Koppens, F.~Taddei, M.~Polini, Phys. Rev. B
  \textbf{102}, 125403 (2020).
\newblock \leavevmode\\ \doi{10.1103/PhysRevB.102.125403}

\bibitem{pol20pl}
T.~Giovannini, L.~Bonatti, M.~Polini, C.~Cappelli, The Journal of Physical
  Chemistry Letters \textbf{11}(18), 7595 (2020).
\newblock \leavevmode\\ \doi{10.1021/acs.jpclett.0c02051}

\bibitem{kop20p1}
N.C.H. Hesp, I.~Torre, D.~Rodan-Legrain, P.~Novelli, Y.~Cao, S.~Carr, S.~Fang,
  P.~Stepanov, D.~Barcons-Ruiz, H.~Herzig-Sheinfux, K.~Watanabe, T.~Taniguchi,
  D.K. Efetov, E.~Kaxiras, P.~Jarillo-Herrero, M.~Polini, F.H.L. Koppens.
\newblock Observation of interband collective excitations in twisted bilayer
  graphene (2021).
\newblock \leavevmode\\ \doi{10.1038/s41567-021-01327-8}

\bibitem{kop20p2}
A.T. Costa, P.A.D. Gon{\c c}alves, D.N. Basov, F.H.L. Koppens, N.A. Mortensen,
  N.M.R. Peres, Proc. Natl. Acad. Sci. \textbf{118}(4), e2012847118 (2021).
\newblock \leavevmode\\ \doi{10.1073/pnas.2012847118}

\bibitem{mach19}
A.~Klein, D.L. Maslov, L.P. Pitaevskii, A.V. Chubukov, Phys. Rev. Research
  \textbf{1}, 033134 (2019).
\newblock \leavevmode\\ \doi{10.1103/PhysRevResearch.1.033134}

\bibitem{per20}
B.A. Ferreira, B.~Amorim, A.J. Chaves, N.M.R. Peres, Phys. Rev. A \textbf{101},
  033817 (2020).
\newblock \leavevmode\\ \doi{10.1103/PhysRevA.101.033817}

\bibitem{mach20}
A.~Klein, D.L. Maslov, A.V. Chubukov, npj Quant. Mater. \textbf{5}, 55 (2020).
\newblock \leavevmode\\ \doi{10.1038/s41535-020-0250-4}

\bibitem{falko20}
Z.M. Raines, V.I. Fal'ko, L.I. Glazman, Phys. Rev. B \textbf{103}, 075422
  (2021).
\newblock \leavevmode\\ \doi{10.1103/PhysRevB.103.075422}

\bibitem{ash}
I.L. Aleiner, B.I. Shklovskii, Phys. Rev. B \textbf{49}, 13721 (1994).
\newblock \leavevmode\\ \doi{10.1103/PhysRevB.49.13721}

\bibitem{luskin}
V.~Andreeva, D.A. Bandurin, M.~Luskin, D.~Margetis, Phys. Rev. B \textbf{102},
  205411 (2020).
\newblock \leavevmode\\ \doi{10.1103/PhysRevB.102.205411}

\bibitem{gold18}
R.~Cohen, M.~Goldstein, Phys. Rev. B \textbf{98}, 235103 (2018).
\newblock \leavevmode\\ \doi{10.1103/PhysRevB.98.235103}

\bibitem{alex17}
A.~Levchenko, H.Y. Xie, A.V. Andreev, Phys. Rev. B \textbf{95}, 121301(R)
  (2017).
\newblock \leavevmode\\ \doi{10.1103/PhysRevB.95.121301}

\bibitem{ant20}
S.~Li, A.~Levchenko, A.V. Andreev, Phys. Rev. B \textbf{102}, 075305 (2020).
\newblock \leavevmode\\ \doi{10.1103/PhysRevB.102.075305}

\bibitem{hui20}
A.~Hui, S.~Lederer, V.~Oganesyan, E.A. Kim, Phys. Rev. B \textbf{101}, 121107
  (2020).
\newblock \leavevmode\\ \doi{10.1103/PhysRevB.101.121107}

\bibitem{alex21}
S.~Li, M.~Khodas, A.~Levchenko, Phys. Rev. B \textbf{104}, 155305 (2021).
\newblock \leavevmode\\ \doi{10.1103/PhysRevB.104.155305}

\bibitem{poise}
J.L.M. Poiseuille, Annales de chimie et de physique (Series 3) \textbf{21}, 76
  (1847)

\bibitem{han20}
I.~Matthaiakakis, D.~Rodr\'{\i}guez~Fern\'andez, C.~Tutschku, E.M. Hankiewicz,
  J.~Erdmenger, R.~Meyer, Phys. Rev. B \textbf{101}, 045423 (2020).
\newblock \leavevmode\\ \doi{10.1103/PhysRevB.101.045423}

\bibitem{levch17}
A.A. Patel, R.A. Davison, A.~Levchenko, Phys. Rev. B \textbf{96}, 205417
  (2017).
\newblock \leavevmode\\ \doi{10.1103/PhysRevB.96.205417}

\bibitem{lucas20}
I.~Mandal, A.~Lucas, Phys. Rev. B \textbf{101}, 045122 (2020).
\newblock \leavevmode\\ \doi{10.1103/PhysRevB.101.045122}

\bibitem{rashba}
E.I. Rashba, Z.S. Gribnikov, V.Y. Kravchenko, Usp. Fiz. Nauk \textbf{119}(5), 3
  (1976).
\newblock [Sov. Phys. Usp. 19, 361 (1976)].
\newblock \leavevmode\\ \doi{10.3367/UFNr.0119.197605a.0003}

\bibitem{alex22}
S.~Li, A.~Levchenko, A.V. Andreev, Phys. Rev. B \textbf{105}, 125302 (2022).
\newblock \leavevmode\\ \doi{10.1103/PhysRevB.105.125302}

\bibitem{ady22}
A.~Stern, T.~Scaffidi, O.~Reuven, C.~Kumar, J.~Birkbeck, S.~Ilani (2021).
\newblock \leavevmode\\ \doi{10.48550/arXiv.2110.15369}

\bibitem{rai22}
O.E. Raichev (2022).
\newblock \leavevmode\\ \doi{10.48550/arXiv.2202.06623}

\bibitem{men11}
M.~Mendoza, H.J. Herrmann, S.~Succi, Phys. Rev. Lett. \textbf{106}, 156601
  (2011).
\newblock \leavevmode\\ \doi{10.1103/PhysRevLett.106.156601}

\bibitem{mer08}
I.~Meric, M.Y. Han, A.F. Young, B.~Ozylmaz, P.~Kim, K.L. Shepard, Nat.
  Nanotechnol. \textbf{3}, 654 (2008).
\newblock \leavevmode\\ \doi{10.1038/nnano.2008.268}

\bibitem{che}
P.A. Cherenkov, C. R. Acad. Sci. USSR \textbf{2}, 451 (1934).
\newblock [Dokl. Akad. Nauk SSSR {\bf 2}, 451 (1934)].
\newblock \leavevmode\\ \doi{10.3367/UFNr.0093.196710n.0385}

\bibitem{vav}
S.I. Vavilov, C. R. Acad. Sci. USSR \textbf{2}, 457 (1934).
\newblock [Dokl. Akad. Nauk SSSR {\bf 2}, 457 (1934)].
\newblock \leavevmode\\ \doi{10.3367/UFNr.0093.196710m.0383}

\bibitem{tam}
I.E. Tamm, I.M. Frank, C. R. Acad. Sci. USSR \textbf{14}, 107 (1937).
\newblock [Dokl. Akad. Nauk SSSR {\bf 14}, 107 (1937)].
\newblock \leavevmode\\ \doi{10.3367/UFNr.0093.196710o.0388}

\bibitem{svin19}
D.~Svintsov, Phys. Rev. B \textbf{100}, 195428 (2019).
\newblock \leavevmode\\ \doi{10.1103/PhysRevB.100.195428}

\bibitem{svin13}
D.~Svintsov, V.~Vyurkov, V.~Ryzhii, T.~Otsuji, Phys. Rev. B \textbf{88}, 245444
  (2013).
\newblock \leavevmode\\ \doi{10.1103/PhysRevB.88.245444}

\bibitem{zdy19}
T.~Zdyrski, J.~McGreevy, Phys. Rev. B \textbf{99}, 235435 (2019).
\newblock \leavevmode\\ \doi{10.1103/PhysRevB.99.235435}

\bibitem{vin10}
V.E. Dorgan, M.H. Bae, E.~Pop, Applied Physics Letters \textbf{97}(8), 082112
  (2010).
\newblock \leavevmode\\ \doi{10.1063/1.3483130}

\bibitem{ronny20}
D.~{Di Sante}, J.~Erdmenger, M.~Greiter, I.~Matthaiakakis, R.~Meyer,
  D.~{Rodriguez Fern{\'a}ndez}, R.~Thomale, E.~{van Loon}, T.~Wehling, Nat.
  Commun. \textbf{11}, 3997 (2020).
\newblock \leavevmode\\ \doi{10.1038/s41467-020-17663-x}

\bibitem{men17}
R.C.V. Coelho, M.~Mendoza, M.M. Doria, H.J. Herrmann, Phys. Rev. B \textbf{96},
  184307 (2017).
\newblock \leavevmode\\ \doi{10.1103/PhysRevB.96.184307}

\bibitem{men14}
O.~Furtmaier, M.~Mendoza, I.~Karlin, S.~Succi, H.J. Herrmann, Phys. Rev. B
  \textbf{91}, 085401 (2015).
\newblock \leavevmode\\ \doi{10.1103/PhysRevB.91.085401}

\bibitem{kel}
{Sir W. Thomson, F.R.S.}, The London, Edinburgh, and Dublin Philosophical
  Magazine and Journal of Science \textbf{42}(281), 362 (1871).
\newblock \leavevmode\\ \doi{10.1080/14786447108640585}

\bibitem{hel}
H.~Helmholz, {Monatsberichte der K\"oniglichen Preussische Akademie der
  Wissenschaften zu Berlin} \textbf{23}, 215 (1868)

\bibitem{ben}
H.~B{\'e}nard, Les tourbillons cellulaires dans une nappe liquide propageant de
  la chaleur par convection en r{\'e}gime permanent.
\newblock Ph.D. thesis, Coll{\`e}ge de France (1901)

\bibitem{ray}
{Lord Rayleigh, F.R.S.}, The London, Edinburgh, and Dublin Philosophical
  Magazine and Journal of Science \textbf{32}, 529 (1916).
\newblock \leavevmode\\ \doi{10.1080/14786441608635602}

\bibitem{baz21}
L.~Bazzanini, A.~Gabbana, D.~Simeoni, S.~Succi, R.~Tripiccione, Journal of
  Computational Science \textbf{51}, 101320 (2021).
\newblock \leavevmode\\ \doi{https://doi.org/10.1016/j.jocs.2021.101320}

\bibitem{luc18}
C.B. Mendl, A.~Lucas, Applied Physics Letters \textbf{112}(12), 124101 (2018).
\newblock \leavevmode\\ \doi{10.1063/1.5022187}

\bibitem{aiz21}
J.~Crabb, X.~Cantos-Roman, J.M. Jornet, G.R. Aizin, Phys. Rev. B \textbf{104},
  155440 (2021).
\newblock \leavevmode\\ \doi{10.1103/PhysRevB.104.155440}

\bibitem{Li17}
K.~Li, Y.~Hao, X.~Jin, W.~Lu, Journal of Physics D: Applied Physics
  \textbf{51}(3), 035104 (2017).
\newblock \leavevmode\\ \doi{10.1088/1361-6463/aa9cd0}

\bibitem{luc21}
C.B. Mendl, M.~Polini, A.~Lucas, Applied Physics Letters \textbf{118}(1),
  013105 (2021).
\newblock \leavevmode\\ \doi{10.1063/5.0030869}

\bibitem{scaf22}
J.H. Farrel, N.~Grisouard, T.~Scaffidi (2022).
\newblock \leavevmode\\ \doi{10.48550/arXiv.2112.07683}

\bibitem{oga21}
A.~Hui, V.~Oganesyan, E.A. Kim, Phys. Rev. B \textbf{103}, 235152 (2021).
\newblock \leavevmode\\ \doi{10.1103/PhysRevB.103.235152}

\bibitem{bern}
D.~Bernoulli, \emph{Hydrodynamica, sive de viribus et motibus fluidorum
  commentarii} (Strasbourg, 1738).
\newblock \leavevmode\\ \doi{10.3931/e-rara-3911}

\bibitem{eck}
C.~Eckart, Phys. Rev. \textbf{73}, 68 (1948).
\newblock \leavevmode\\ \doi{10.1103/PhysRev.73.68}

\bibitem{ray2}
{Lord Rayleigh, F.R.S.}, Philosophical Transactions \textbf{175}, 1 (1884).
\newblock \leavevmode\\ \doi{10.1098/rstl.1884.0002}

\bibitem{mal98}
J.~Maldacena, Adv. Theor. Math. Phys. \textbf{2}(2), 231 (1998).
\newblock \leavevmode\\ \doi{10.4310/ATMP.1998.v2.n2.a1}

\bibitem{hartnoll13}
A.~Donos, S.A. Hartnoll, Nat. Phys. \textbf{9}, 649 (2013).
\newblock \leavevmode\\ \doi{10.1038/nphys2701}

\bibitem{hartnoll18}
S.A. Hartnoll, A.~Lucas, S.~Sachdev, \emph{{Holographic Quantum Matter}} (MIT
  Press, 2018).
\newblock \leavevmode\\ \doi{10.48550/arXiv.1612.07324}

\bibitem{Schafer2014}
T.~Sch{\"{a}}fer, Annu. Rev. Nucl. Part. Sci. \textbf{64}(1), 125 (2014).
\newblock \leavevmode\\ \doi{10.1146/annurev-nucl-102313-025439}

\bibitem{Thomas2009}
J.E. Thomas, Nucl. Phys. A \textbf{830}(1-4), 665c (2009).
\newblock \leavevmode\\ \doi{10.1016/j.nuclphysa.2009.09.055}

\bibitem{har15}
S.A. Hartnoll, Nat. Phys. \textbf{11}(1), 54 (2015).
\newblock \leavevmode\\ \doi{10.1038/nphys3174}

\bibitem{ink20}
G.A. Inkof, J.M.C. K{\"{u}}ppers, J.M. Link, B.~Gout{\'{e}}raux, J.~Schmalian,
  Journal of High Energy Physics \textbf{2020}(11), 88 (2020).
\newblock \leavevmode\\ \doi{10.1007/JHEP11(2020)088}

\bibitem{bruin}
J.A.N. Bruin, H.~Sakai, R.S. Perry, A.P. Mackenzie, Science \textbf{339}(6121),
  804 (2013).
\newblock \leavevmode\\ \doi{10.1126/science.1227612}

\bibitem{zaa13}
J.~Zaanen, Nat. Phys. \textbf{9}, 609 (2013).
\newblock \leavevmode\\ \doi{10.1038/nphys2717}

\bibitem{zaa14}
R.A. Davison, K.~Schalm, J.~Zaanen, Phys. Rev. B \textbf{89}, 245116 (2014).
\newblock \leavevmode\\ \doi{10.1103/PhysRevB.89.245116}

\bibitem{proust19}
C.~Proust, L.~Taillefer, Annu. Rev. Cond. Matt. Phys. \textbf{10}, 409 (2019).
\newblock \leavevmode\\ \doi{10.1146/annurev-conmatphys-031218-013210}

\bibitem{legros19}
A.~Legros, S.~Benhabib, W.~Tabis, F.~Lalibert{\'{e}}, M.~Dion, M.~Lizaire,
  B.~Vignolle, D.~Vignolles, H.~Raffy, Z.Z. Li, P.~Auban-Senzier,
  N.~Doiron-Leyraud, P.~Fournier, D.~Colson, L.~Taillefer, C.~Proust, Nat.
  Phys. \textbf{15}(2), 142 (2019).
\newblock \leavevmode\\ \doi{10.1038/s41567-018-0334-2}

\bibitem{green10}
J.~Pagline, R.L. Greene, Nat. Phys. \textbf{6}, 645 (2010).
\newblock \leavevmode\\ \doi{10.1038/nphys1759}

\bibitem{john10}
D.C. Johnston, Adv. Phys. \textbf{59}(6), 803 (2010).
\newblock \leavevmode\\ \doi{10.1080/00018732.2010.513480}

\bibitem{dai}
P.~Dai, Rev. Mod. Phys. \textbf{87}, 855 (2015).
\newblock \leavevmode\\ \doi{10.1103/RevModPhys.87.855}

\bibitem{si16}
Q.~Si, R.~Yu, E.~Abrahams, Nat. Rev. Mater. \textbf{1}(4), 16017 (2016).
\newblock \leavevmode\\ \doi{10.1038/natrevmats.2016.17}

\bibitem{tbg}
Y.~Cao, V.~Fatemi, S.~Fang, K.~Watanabe, T.~Taniguchi, E.~Kaxiras,
  P.~Jarillo-Herrero, Nature \textbf{556}, 45 (2018).
\newblock \leavevmode\\ \doi{10.1038/nature23160}

\bibitem{cao20}
Y.~Cao, D.~Chowdhury, D.~Rodan-Legrain, O.~Rubies-Bigorda, K.~Watanabe,
  T.~Taniguchi, T.~Senthil, P.~Jarillo-Herrero, Phys. Rev. Lett.
  \textbf{124}(7), 076801 (2020).
\newblock \leavevmode\\ \doi{10.1103/PhysRevLett.124.076801}

\bibitem{hub00}
G.T. Horowitz, V.E. Hubeny, Phys. Rev. D \textbf{62}, 024027 (2000).
\newblock \leavevmode\\ \doi{10.1103/PhysRevD.62.024027}

\bibitem{forster2018}
D.~Forster, \emph{{Hydrodynamic Fluctuations, Broken Symmetry, and Correlation
  Functions}} (CRC Press, 2018).
\newblock \leavevmode\\ \doi{10.1201/9780429493683}

\bibitem{groz18}
S.c.v. Grozdanov, K.~Schalm, V.~Scopelliti, Phys. Rev. Lett. \textbf{120},
  231601 (2018).
\newblock \leavevmode\\ \doi{10.1103/PhysRevLett.120.231601}

\bibitem{bag18}
A.~Bagrov, B.~Craps, F.~Galli, V.~Ker\"anen, E.~Keski-Vakkuri, J.~Zaanen, Phys.
  Rev. D \textbf{97}, 086005 (2018).
\newblock \leavevmode\\ \doi{10.1103/PhysRevD.97.086005}

\bibitem{bhat09}
S.~Bhattacharyya, S.~Minwalla, J. High Energy Phys. \textbf{2009}(09), 034
  (2009).
\newblock \leavevmode\\ \doi{10.1088/1126-6708/2009/09/034}

\bibitem{deutsch}
J.M. Deutsch, Phys. Rev. A \textbf{43}, 2046 (1991).
\newblock \leavevmode\\ \doi{10.1103/PhysRevA.43.2046}

\bibitem{sred}
M.~Srednicki, Phys. Rev. E \textbf{50}, 888 (1994).
\newblock \leavevmode\\ \doi{10.1103/PhysRevE.50.888}

\bibitem{zaa19}
J.~Zaanen, SciPost Phys. \textbf{6}(5), 061 (2019).
\newblock \leavevmode\\ \doi{10.21468/SciPostPhys.6.5.061}

\bibitem{lucas19}
A.~Lucas, Phys. Rev. Lett. \textbf{122}, 216601 (2019).
\newblock \leavevmode\\ \doi{10.1103/PhysRevLett.122.216601}

\bibitem{zaa03}
D.~{van der Marel}, H.J.A. Molegraaf, J.~Zaanen, Z.~Nussinov, F.~Carbone,
  A.~Damascelli, H.~Eisaki, M.~Greven, P.H. Kes, M.~Li, Nature \textbf{425},
  271 (2003).
\newblock \leavevmode\\ \doi{10.1038/nature01978}

\bibitem{coop09}
R.A. Cooper, Y.~Wang, B.~Vignolle, O.J. Lipscombe, S.M. Hayden, Y.~Tanabe,
  T.~Adachi, Y.~Koike, M.~Nohara, H.~Takagi, C.~Proust, N.E. Hussey, Science
  \textbf{323}(5914), 603 (2009).
\newblock \leavevmode\\ \doi{10.1126/science.1165015}

\bibitem{del17}
L.~Delacr{\'{e}}taz, B.~Gout{\'{e}}raux, S.A. Hartnoll, A.~Karlsson, SciPost
  Phys. \textbf{3}(3), 025 (2017).
\newblock \leavevmode\\ \doi{10.21468/SciPostPhys.3.3.025}

\bibitem{huang17}
E.W. Huang, C.B. Mendl, S.~Liu, S.~Johnston, H.C. Jiang, B.~Moritz, T.P.
  Devereaux, Science \textbf{358}(6367), 1161 (2017).
\newblock \leavevmode\\ \doi{10.1126/science.aak9546}

\bibitem{and18}
T.~Andrade, A.~Krikun, K.~Schalm, J.~Zaanen, Nat. Phys. \textbf{14}(10), 1049
  (2018).
\newblock \leavevmode\\ \doi{10.1038/s41567-018-0217-6}

\bibitem{amo18}
A.~Amoretti, D.~{Are\'an}, B.~{Gout\'eraux}, D.~Musso, Phys. Rev. Lett.
  \textbf{120}, 171603 (2018).
\newblock \leavevmode\\ \doi{10.1103/PhysRevLett.120.171603}

\bibitem{khvesh21}
D.V. Khveshchenko, Lithuan. J. Phys. \textbf{61}, 42 (2021).
\newblock \leavevmode\\ \doi{10.3952/physics.v61i1.4406}

\bibitem{erd18}
J.~Erdmenger, I.~Matthaiakakis, R.~Meyer, D.R. Fern\'andez, Phys. Rev. B
  \textbf{98}, 195143 (2018).
\newblock \leavevmode\\ \doi{10.1103/PhysRevB.98.195143}

\bibitem{peierls}
R.~Peierls, Annalen der Physik \textbf{404}(2), 154 (1932).
\newblock [Ann. Phys. (5) {\bf 12}, 154 (1932)].
\newblock \leavevmode\\ \doi{10.1002/andp.19324040203}

\bibitem{lgu1}
L.E. Gurevich, Zh. Eksp. Teor. Fiz. \textbf{16}, 193 (1946).
\newblock [J. Phys. (USSR) {\bf 9}, 857 (1945)]

\bibitem{lgu2}
L.E. Gurevich, Zh. Eksp. Teor. Fiz. \textbf{16}, 416 (1946).
\newblock [J. Phys. (USSR) {\bf 10}, 67 (1946)]

\bibitem{gum}
Y.G. Gurevich, O.L. Mashkevich, Phys. Rep. \textbf{181}, 327 (1989).
\newblock \leavevmode\\ \doi{10.1016/0370-1573(89)90011-2}

\bibitem{gooar}
C.~Fu, T.~Scaffidi, J.~Waissman, Y.~Sun, R.~Saha, S.J. Watzman, A.K.
  Srivastava, G.~Li, W.~Schnelle, P.~Werner, M.E. Kamminga, S.~Sachdev, S.S.P.
  Parkin, S.A. Hartnoll, C.~Felser, J.~Gooth (2018).
\newblock \leavevmode\\ \doi{10.48550/arXiv.1802.09468}

\bibitem{lucas21}
X.~Huang, A.~Lucas, Phys. Rev. B \textbf{103}, 155128 (2021).
\newblock \leavevmode\\ \doi{10.1103/PhysRevB.103.155128}

\bibitem{lev20}
A.~Levchenko, J.~Schmalian, Ann. Phys. \textbf{419}, 168218 (2020).
\newblock \leavevmode\\ \doi{10.1016/j.aop.2020.168218}

\bibitem{nandi18}
N.~Nandi, T.~Scaffidi, P.~Kushwaha, S.~Khim, M.E. Barber, V.~Sunko, F.~Mazzola,
  P.D.C. King, H.~Rosner, P.J.W. Moll, M.~K{\"{o}}nig, J.E. Moore, S.~Hartnoll,
  A.P. Mackenzie, npj Quant. Mater. \textbf{3}(1), 66 (2018).
\newblock \leavevmode\\ \doi{10.1038/s41535-018-0138-8}

\bibitem{cook}
C.Q. Cook, A.~Lucas, Phys. Rev. B \textbf{99}, 235148 (2019).
\newblock \leavevmode\\ \doi{10.1103/PhysRevB.99.235148}

\bibitem{julia2}
J.M. Link, P.P. Orth, D.E. Sheehy, J.~Schmalian, Phys. Rev. B \textbf{93},
  235447 (2016).
\newblock \leavevmode\\ \doi{10.1103/PhysRevB.93.235447}

\bibitem{levy10}
N.~Levy, S.A. Burke, K.L. Meaker, M.~Panlasigui, A.~Zettl, F.~Guinea, A.H.C.
  Neto, M.F. Crommie, Science \textbf{329}(5991), 544 (2010).
\newblock \leavevmode\\ \doi{10.1126/science.1191700}

\bibitem{mnas}
K.~Mnasri, B.~Jeevanesan, J.~Schmalian, Phys. Rev. B \textbf{92}, 134423
  (2015).
\newblock \leavevmode\\ \doi{10.1103/PhysRevB.92.134423}

\bibitem{prasai}
N.~Prasai, B.A. Trump, G.G. Marcus, A.~Akopyan, S.X. Huang, T.M. McQueen, J.L.
  Cohn, Phys. Rev. B \textbf{95}, 224407 (2017).
\newblock \leavevmode\\ \doi{10.1103/PhysRevB.95.224407}

\bibitem{duine19}
C.~Ulloa, A.~Tomadin, J.~Shan, M.~Polini, B.J. van Wees, R.A. Duine, Phys. Rev.
  Lett. \textbf{123}, 117203 (2019).
\newblock \leavevmode\\ \doi{10.1103/PhysRevLett.123.117203}

\bibitem{demler18}
J.F. Rodriguez-Nieva, D.~Podolsky, E.~Demler (2018).
\newblock \leavevmode\\ \doi{10.48550/arXiv.1810.12333}

\bibitem{moo21}
C.~Zu, F.~Machado, B.~Ye, S.~Choi, B.~Kobrin, T.~Mittiga, S.~Hsieh,
  P.~Bhattacharyya, M.~Markham, D.~Twitchen, A.~Jarmola, D.~Budker, C.R.
  Laumann, J.E. Moore, N.Y. Yao, Nature \textbf{597}, 45 (2021).
\newblock \leavevmode\\ \doi{10.1038/s41586-021-03763-1}

\bibitem{fabs15}
E.~Ilievski, J.~De~Nardis, B.~Wouters, J.S. Caux, F.H.L. Essler, T.~Prosen,
  Phys. Rev. Lett. \textbf{115}, 157201 (2015).
\newblock \leavevmode\\ \doi{10.1103/PhysRevLett.115.157201}

\bibitem{langen15}
T.~Langen, S.~Erne, R.~Geiger, B.~Rauer, T.~Schweigler, M.~Kuhnert,
  W.~Rohringer, I.E. Mazets, T.~Gasenzer, J.~Schmiedmayer, Science
  \textbf{348}(6231), 207 (2015).
\newblock \leavevmode\\ \doi{10.1126/science.1257026}

\bibitem{doyon16}
O.A. Castro-Alvaredo, B.~Doyon, T.~Yoshimura, Phys. Rev. X \textbf{6}, 041065
  (2016).
\newblock \leavevmode\\ \doi{10.1103/PhysRevX.6.041065}

\bibitem{nardis16}
B.~Bertini, M.~Collura, J.~De~Nardis, M.~Fagotti, Phys. Rev. Lett.
  \textbf{117}, 207201 (2016).
\newblock \leavevmode\\ \doi{10.1103/PhysRevLett.117.207201}

\bibitem{bert20}
B.~Bertini, F.~Heidrich-Meisner, C.~Karrasch, T.~Prosen, R.~Steinigeweg,
  M.~\ifmmode \check{Z}\else \v{Z}\fi{}nidari\ifmmode~\check{c}\else
  \v{c}\fi{}, Rev. Mod. Phys. \textbf{93}, 025003 (2021).
\newblock \leavevmode\\ \doi{10.1103/RevModPhys.93.025003}

\bibitem{konik19}
J.S. Caux, B.~Doyon, J.~Dubail, R.~Konik, T.~Yoshimura, SciPost Phys.
  \textbf{6}, 70 (2019).
\newblock \leavevmode\\ \doi{10.21468/SciPostPhys.6.6.070}

\bibitem{doyon19}
M.~Schemmer, I.~Bouchoule, B.~Doyon, J.~Dubail, Phys. Rev. Lett. \textbf{122},
  090601 (2019).
\newblock \leavevmode\\ \doi{10.1103/PhysRevLett.122.090601}

\bibitem{vass21}
J.~Lopez-Piqueres, B.~Ware, S.~Gopalakrishnan, R.~Vasseur, Phys. Rev. B
  \textbf{103}, L060302 (2021).
\newblock \leavevmode\\ \doi{10.1103/PhysRevB.103.L060302}

\bibitem{tser18}
Y.~Tserkovnyak, J. Appl. Phys. \textbf{124}(19), 190901 (2018).
\newblock \leavevmode\\ \doi{10.1063/1.5054123}

\bibitem{kerr}
{John Kerr, LL.D.}, The London, Edinburgh, and Dublin Philosophical Magazine
  and Journal of Science \textbf{3}(19), 321 (1877).
\newblock \leavevmode\\ \doi{10.1080/14786447708639245}

\bibitem{jack21}
T.~{Van Mechelen}, W.~Sun, Z.~Jacob, Nat. Commun. \textbf{12}, 4729 (2021).
\newblock \leavevmode\\ \doi{10.1038/s41467-021-25097-2}

\bibitem{tser19}
Y.~Tserkovnyak, J.~Zou, Phys. Rev. Res. \textbf{1}, 033071 (2019).
\newblock \leavevmode\\ \doi{10.1103/PhysRevResearch.1.033071}

\bibitem{gal18}
V.~Galitski, M.~Kargarian, S.~Syzranov, Phys. Rev. Lett. \textbf{121}, 176603
  (2018).
\newblock \leavevmode\\ \doi{10.1103/PhysRevLett.121.176603}

\bibitem{yin19}
K.~Hattori, Y.~Hirono, H.U. Yee, Y.~Yin, Phys. Rev. D \textbf{100}, 065023
  (2019).
\newblock \leavevmode\\ \doi{10.1103/PhysRevD.100.065023}

\bibitem{cop19}
C.~Copetti, K.~Landsteiner, Phys. Rev. B \textbf{99}, 195146 (2019).
\newblock \leavevmode\\ \doi{10.1103/PhysRevB.99.195146}

\bibitem{kawa20}
R.~Toshio, K.~Takasan, N.~Kawakami, Phys. Rev. Research \textbf{2}, 032021
  (2020).
\newblock \leavevmode\\ \doi{10.1103/PhysRevResearch.2.032021}

\bibitem{has21}
E.H. Hasdeo, J.~Ekstr\"om, E.G. Idrisov, T.L. Schmidt, Phys. Rev. B
  \textbf{103}, 125106 (2021).
\newblock \leavevmode\\ \doi{10.1103/PhysRevB.103.125106}

\bibitem{sur20}
M.~Moore, P.~Sur\'owka, V.~Juri\ifmmode \check{c}\else
  \v{c}\fi{}i\ifmmode~\acute{c}\else \'{c}\fi{}, B.~Roy, Phys. Rev. B
  \textbf{101}, 161111 (2020).
\newblock \leavevmode\\ \doi{10.1103/PhysRevB.101.161111}

\bibitem{che1}
N.N. Fimin, V.M. Chechetkin, Comput. Math. Math. Phys. \textbf{58}, 449 (2018).
\newblock \leavevmode\\ \doi{10.1134/S0965542518030053}

\bibitem{che2}
A.G. Aksenov, A.V. Babakov, V.M. Chechetkin, Comput. Math. Math. Phys.
  \textbf{58}, 1287 (2018).
\newblock \leavevmode\\ \doi{10.1134/S096554251808002X}

\bibitem{aag}
I.L. Aleiner, B.L. Altshuler, M.E. Gershenson, Waves in Random Media
  \textbf{9}(2), 201 (1999).
\newblock \leavevmode\\ \doi{10.1088/0959-7174/9/2/308}

\bibitem{AltLeeWeb1991}
B.L. Altshuler, P.A. Lee, R.A. Webb (eds.), \emph{Mesoscopic Phenomena in
  Solids} (North-Holland, New York, 1991)

\bibitem{df1}
B.N. Narozhny, I.L. Aleiner, Phys. Rev. Lett. \textbf{84}, 5383 (2000).
\newblock \leavevmode\\ \doi{10.1103/PhysRevLett.84.5383}

\bibitem{df2}
B.N. Narozhny, I.L. Aleiner, A.~Stern, Phys. Rev. Lett. \textbf{86}, 3610
  (2001).
\newblock \leavevmode\\ \doi{10.1103/PhysRevLett.86.3610}

\bibitem{geimgrig}
A.K. Geim, I.V. Grigorieva, Nature \textbf{499}, 419 (2013).
\newblock \leavevmode\\ \doi{10.1038/nature12385}

\bibitem{mish03}
E.G. Mishchenko, B.I. Halperin, Phys. Rev. B \textbf{68}, 045317 (2003).
\newblock \leavevmode\\ \doi{10.1103/PhysRevB.68.045317}

\bibitem{sin07}
T.S. Nunner, N.A. Sinitsyn, M.F. Borunda, V.K. Dugaev, A.A. Kovalev, A.~Abanov,
  C.~Timm, T.~Jungwirth, J.i. Inoue, A.H. MacDonald, J.~Sinova, Phys. Rev. B
  \textbf{76}, 235312 (2007).
\newblock \leavevmode\\ \doi{10.1103/PhysRevB.76.235312}

\bibitem{sin08}
N.A. Sinitsyn, Journal of Physics: Condensed Matter \textbf{20}, 023201 (2008).
\newblock \leavevmode\\ \doi{10.1088/0953-8984/20/02/023201}

\bibitem{fabian04}
I.~\ifmmode \check{Z}\else \v{Z}\fi{}uti\ifmmode~\acute{c}\else \'{c}\fi{},
  J.~Fabian, S.~Das~Sarma, Rev. Mod. Phys. \textbf{76}, 323 (2004).
\newblock \leavevmode\\ \doi{10.1103/RevModPhys.76.323}

\bibitem{sinova15}
J.~Sinova, S.O. Valenzuela, J.~Wunderlich, C.H. Back, T.~Jungwirth, Rev. Mod.
  Phys. \textbf{87}, 1213 (2015).
\newblock \leavevmode\\ \doi{10.1103/RevModPhys.87.1213}

\end{thebibliography}

\end{document}